\documentclass[12pt]{iopart}         
\usepackage{graphicx}
\def\be{\begin{equation}}
\def\ee{\end{equation}}

\def\lambar{\lambda\llap {--}}

\def\lambar{\lambda\llap {--}}

\def\lsim{\lower 2pt \hbox{$\, \buildrel {\scriptstyle <}\over
         {\scriptstyle \sim}\,$}}
\newcommand\gsim{\buildrel > \over \sim}
\def\ba{\begin{eqnarray}}
\def\ea{\end{eqnarray}}
\def\go{\mathrel{\raise.3ex\hbox{$>$}\mkern-14mu\lower0.6ex\hbox{$\sim$}}}
\def\lo{\mathrel{\raise.3ex\hbox{$<$}\mkern-14mu\lower0.6ex\hbox{$\sim$}}}
\def\eps {{\varepsilon}}
\def\bB {{\bf B}}

\def\rkeV {{\rm keV}}

\def\br{{\bf r}}

\def\bv{{\bf v}}
\def\bR {{\bf R}}

\def\bK {{\bf K}}

\def\bPi {{\bf \Pi}}

\def\bsigma{{\mbox{\boldmath $\sigma$}}}
\def\etal {{et al.}}
\def\beps{{\mbox{\boldmath $\epsilon$}}}

\def\bmu{{\mbox{\boldmath $\mu$}}}

\def\bE{{\bf E}}
\def\bB{{\bf B}}

\def\br{{\bf r}}
\def\bk{{\bf k}}
\def\bI{{\bf I}}
\def\hatk{{\hat {\bf k}}}

\newcommand{\lp}{\left(}
\newcommand{\rp}{\right)}
\newcommand{\lb}{\left[}
\newcommand{\rb}{\right]}
\def\bj{{\bf j}}

\begin{document}
\review{Physics of Strongly Magnetized Neutron Stars}

\author{Alice K. Harding$^1$ and Dong Lai$^2$}   

\address{$^1$Code 663, NASA Goddard Space Flight Center, Greenbelt, MD 20771}
\ead{Alice.K.Harding@nasa.gov}

\address{$^2$Center for Radiophysics and Space Research, Department of Astronomy, Cornell University, 
Ithaca, NY 14853} 
\ead{dong@astro.cornell.edu}
 
\begin{abstract}
There has recently been growing evidence for the existence of neutron
stars possessing magnetic fields with strengths that exceed the
quantum critical field strength of $4.4 \times 10^{13}$ G, at which
the cyclotron energy equals the electron rest mass.  Such evidence has
been provided by new discoveries of radio pulsars having very high
spin-down rates and by observations of bursting gamma-ray sources
termed magnetars.  This article will discuss the exotic physics of
this high-field regime, where a new array of processes becomes
possible and even dominant, and where familiar processes acquire
unusual properties.  We review the physical processes that are
important in neutron star interiors and magnetospheres, including the
behavior of free particles, atoms, molecules, plasma and condensed
matter in strong magnetic fields, photon propagation in magnetized
plasmas, free-particle radiative processes, the physics of neutron
star interiors, and field evolution and decay mechanisms.  Application
of such processes in astrophysical source models, including
rotation-powered pulsars, soft gamma-ray repeaters, anomalous X-ray
pulsars and accreting X-ray pulsars will also be discussed.
Throughout this review, we will highlight the observational signatures
of high magnetic field processes, as well as the theoretical issues
that remain to be understood.

\end{abstract} 

\section{Introduction}  \label{sec:intro}

Since their theoretical conception by Baade \& Zwicky (1934) neutron
stars have been fascinating celestial objects, both for study of their
exotic environments and for their important place in stellar
evolution.  Among the first signals to be detected from neutron stars came from
radio pulsars (Hewish et al. 1968), spinning many times per second,
distinguishing themselves from the background of interplanetary
scintillation signals by their extremely regular pulsations.  Pulsars
were also soon discovered to be spinning down, their periods
increasing also very regularly.  The rotating magnetic-dipole model
(Pacini 1967, Gold 1968, Ostriker \& Gunn 1969), in which the pulsar
loses rotational energy through magnetic dipole radiation, was
dramatically confirmed with the discovery that the spin-down power
predicted for the Crab pulsar was a nearly perfect energetic match
with the radiation of its synchrotron nebula.  The rotating dipole
model also accounts for the observed rate of spin-down of the Crab and
other pulsars, with required surface magnetic fields in the range of
$10^{11} - 10^{13}$ Gauss for the first detected pulsars.  This range
has since significantly broadened, first with the discovery of a class
of pulsars having periods of several milliseconds (Backer et al. 1982), 
believed to have been spun-up by accretion torques of a
binary companion (Alpar et al. 1982), and much lower surface magnetic
fields in the range of $10^8 - 10^{10}$ Gauss.  Recent surveys have
also discovered pulsars with very high period derivatives 
(e.g. Morris et al. 2002, McLaughlin et al.~2003)
that imply surface fields up to around $10^{14}$ Gauss.

Another class of neutron stars was discovered at X-ray and
$\gamma$-ray energies and may possess even stronger surface magnetic
fields.  Such stars are now referred to as magnetars 
(Duncan \& Thompson 1992), 
since they most
probably derive their power from their magnetic fields rather than
from spin-down energy loss (see Woods \& Thompson 2005). Within the
magnetar class there are two types of sources that were originally
thought to be very different objects, although they are now believed
to be closely related.  The Anomalous X-Ray Pulsars (AXPs) were
discovered as pulsating X-ray sources in the early 1980s and were
thought at first to be an unusual type of accreting X-ray pulsar, from
which the name is derived.  The AXPs have periods in a relatively
narrow range of 5 - 11 s, are observed to be spinning down with large
period derivatives (Vasisht \& Gotthelf 1997) and have no detectable
companions or accretion disks that would be required to support the
accretion hypothesis.  Interpretation of their period derivatives as
magnetic dipole spin down imply magnetic fields in the range $10^{14}
- 10^{15}$ Gauss.  But such high fields were not widely accepted
initially, since their detected X-ray luminosities of around
$10^{35}\,\rm erg\,s^{-1}$ exceed their spin-down luminosities by
several orders of magnitude.  It was only by connection to another
subclass of magnetar, the Soft Gamma-Ray Repeaters (SGRs), that the
extremely high magnetic fields of AXPs were adopted.  SGRs were
discovered as transient sources that undergo repeated soft
$\gamma$-ray bursts, usually in widely separated episodes.  They
undergo both repeated smaller bursting of subsecond duration as well
as much more luminous superflares, lasting hundreds of seconds, which
so far have not repeated in any single source but may be repeating on
much longer timescales.  It was not until some twenty years after
their discovery that their quiescent X-ray pulsations were detected
and also very high period derivatives (Kouveliotou et al. 1998), both
with a range of values very similar to those of AXPs.  Recently,
bursts resembling the smaller bursts of SGRs were seen from several
AXPs (Kaspi et al. 2003), making it likely that SGRs and AXPs are two
variations of the same type of object 
(Thompson \& Duncan 1993), very strongly magnetized,
isolated neutron stars possibly powered by magnetic field decay.

\begin{figure} 
\includegraphics[width=15cm]{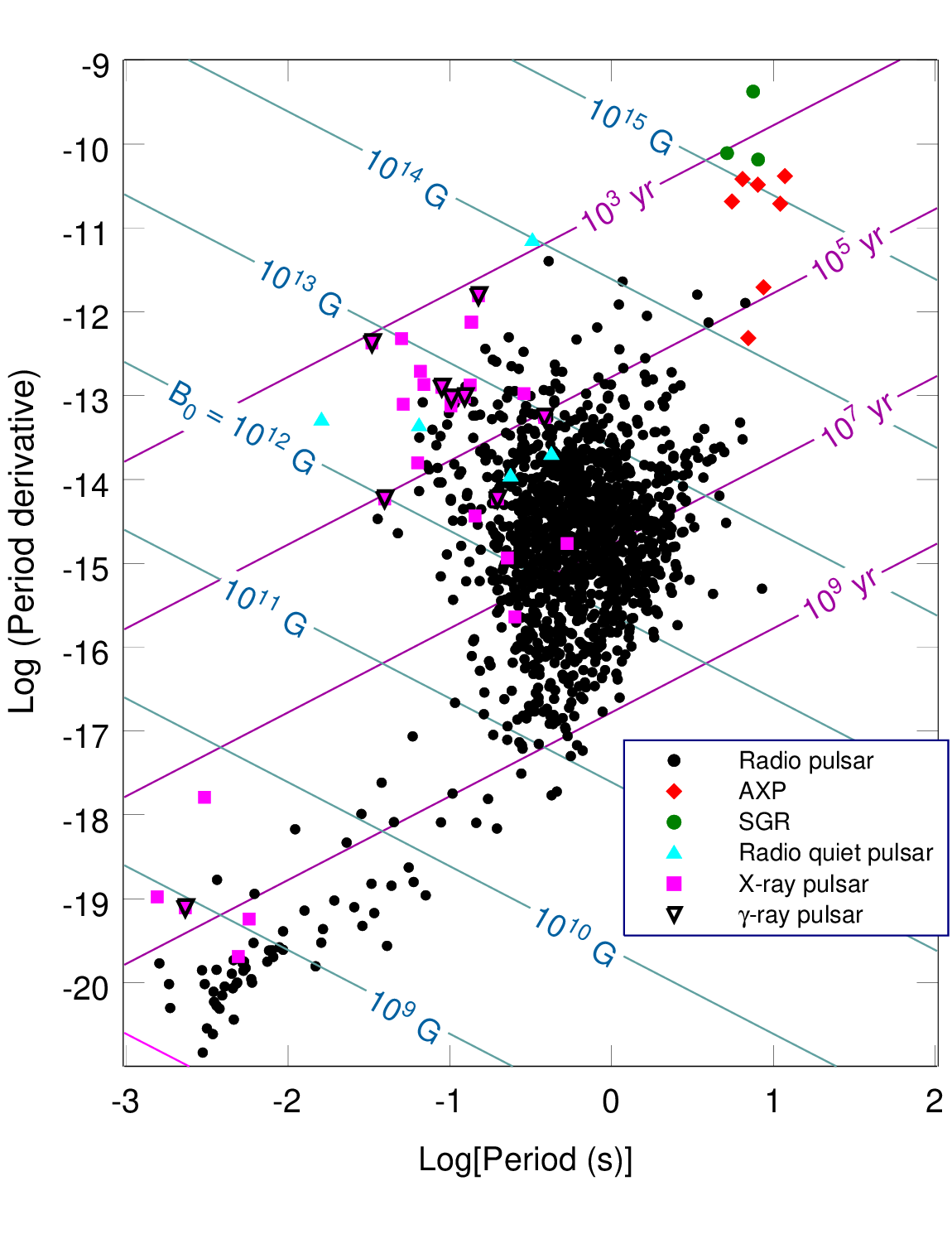}
\caption{Plot of period vs. period derivative for the presently known rotation-powered pulsars and magnetars.
Lines of constant characteristic age, $P/2\dot P$, and surface dipole field (see Eqn [\ref{eq:B0}]) are
superposed.}
\label{fig:PPdot}
\end{figure}

The periods and period derivatives of the various types of isolated
pulsars are shown in the $P$-$\dot P$ diagram of Figure
\ref{fig:PPdot}. 
Assuming that the spindown torque is due to magnetic
dipole radiation, two quantities can be defined from the measured
$P$ and $\dot P$ for each pulsar: (1) Characteristic age $P/(2\dot P)$:
From $\dot\Omega\propto -\Omega^3$ (where $\Omega=2\pi/P$),
the age of the pulsar is found to be $T=(P/2\dot P)[1-(P/P_i)^2]$,
where $P_i$ is the pulsar's initial spin period.
(2) Surface dipole magnetic field:
\be 
B_s  = \left( {\frac{{3Ic^3 P\dot P}}{{2\pi ^2 R^6 }}} \right)^{1/2} 
\simeq 2 \times 10^{12}{\rm G}\,(P \dot P_{15})^{1/2},
\label{eq:B0}\ee
where $\dot P_{15} \equiv \dot P/(10^{-15}\,\rm s\,s^{-1})$ 
(and $P$ is in units of second), and 
and $I$ ($\simeq 10^{45}$~g~cm$^2$ and $R$ ($\simeq 10^6$~cm) 
are the neutron star moment of inertia and radius.
There are presently around 1600 spin-powered radio
pulsars known, with periods from $1.5$ ms - 8 s
(Manchester 2004)\footnote{
see http://www.atnf.csiro.au/research/pulsar/psrcat/}.
Some fraction of these pulse at other wavelengths, including about 10
in $\gamma$-rays and about 30 in X-rays.  The magnetars, 
eight AXPs and four SGRs\footnote{ 
see http://www.physics.mcgill.ca/$\sim$pulsar/magnetar/main.html},
occupy the upper right-hand corner of the diagram and
curiously overlap somewhat with the region occupied by the high-field
radio pulsars.  However, the two types of objects display very
different observational behavior.  The high-field radio pulsars have
very weak or non-detectable X-ray emission and do not burst
(e.g. Kaspi \& McLaughlin 2005),
while the magnetars have no detectable radio pulsations, with the 
exception of the recent detection of radio pulsations in the transient 
AXP XTE J1810-197 (Camilo et al. 2006).  The intrinsic
property that actually distinguishes magnetars from radio pulsars is
presently not understood.

A third class of strongly magnetized neutron stars are the accreting
X-ray pulsars (Parmar 1994).  These sources are members of binary
systems with high-mass companions that either have strong stellar
winds or overflow their Roche lobes, transferring material to the
neutron star.  Inside an Alfven surface where the pressure of the
accretion flow equals the neutron star magnetic field pressure, the
accreting material is funneled along the magnetic field to the poles.
The heated accretion flow then radiates from hot spots that rotate
with the star, thereby producing pulsations.  However, since accretion
torques dominate the period derivative evolution, the surface magnetic
fields of X-ray pulsars cannot be determined using the rotating dipole
model, as they are for the rotation-powered pulsars and magnetars.
Instead, magnetic fields of these objects have been measured from the
energies of the cyclotron resonant scattering features (CRSFs) that
appear in their spectra (see Orlandini \& dal Fiume 2001, for review),
since the fundamental occurs at the electron cyclotron energy, 
$E_{ce} = \hbar eB/(m_ec)=11.58\,(B/10^{12}\,{\rm G})$~keV.  
Table 1 lists the X-ray pulsars, the energies of CRSFs that
have been detected in their spectra and the inferred magnetic field
strengths.  In most cases, the measured magnetic fields range from
$1-5 \times 10^{12}$ Gauss, with the highest fields being around $4
\times 10^{12}$Gauss.  The formation of such line features will be
discussed in section \ref{sec:XRPs}.

\begin{table}
\caption{X-Ray Pulsars with Cyclotron Resonant Scattering Features$^a$}
\begin{center}
\begin{tabular}{llc}
\hline 
Source & Energy (keV)$^b$ & B ($10^{12}$ Gauss)$^c$\\
\hline
4U 0115+63 & 12 (4) & 1.0\\
4U 1907+09 & 18 (1)& 1.6\\
4U 1538-52 & 20 & 1.7 \\
Vela X-1 & 25 (1) & 2.2 \\
V 0332+53 & 27 (2) & 2.3 \\
Cep X-4 & 28 & 2.4 \\
Cen X-3 & 28.5 & 2.5 \\
X Per & 29 & 2.5 \\
XTE J1946+274 & 36 & 3.1 \\
MX0656-072 & 36 & 3.1 \\
4U 1626-67 & 37 & 3.2 \\
GX 301-2 & 37 & 3.2 \\
Her X-1 & 41 & 3.5 \\
A 0535+26 &  47$^d$ (2) & 4.1 \\
\hline
\end{tabular}
\end{center}
{\small
$^a$ Data is from Heindl et al. (2004). 

$^b$ Numbers in parentheses are the number of cyclotron harmonics detected.

$^c$ Magnetic field strength for viewing angle along field direction.}

$^d$ Kretschmar et al. (2005)
\end{table}

Clearly, there is a broad range of astrophysical sources in which the
magnetic fields approach and exceed the quantum critical field
strength, $B_Q \equiv m_e^2c^3/(e\hbar)=4.414 \times 10^{13}$~G, 
at which the cyclotron energy equals the electron rest
mass.  In this regime, the magnetic field profoundly affects physical
processes and introduces additional processes that do not take place
in field-free environments.  From a physics point of view, strongly
magnetized neutron stars provide the only environment in which to
measure and test these effects.  From an astrophysics point of view,
it is necessary to investigate the physics of strong magnetic fields in
order to effectively model and understand the nature of the sources.
This article will attempt to address both points of view by providing
a review of the basic physical processes important in neutron star
interiors and magnetospheres, as well as a review of the source models
in which these processes play a critical role.

We begin with a description of a free electron in a magnetic field in
both classical and quantum regimes.  Subsequent sections then discuss
and review current understanding of the behavior of matter, atoms,
molecules and plasma, in strong magnetic fields, photon propagation in
magnetized plasmas, free-particle radiative processes, the physics of
neutron star interiors, and field evolution and decay mechanisms.
Then we will review models for magnetized atmospheres, non-thermal
radiation from rotation-powered pulsars, burst and quiescent radiation
from SGRs and AXPs, and emission from accreting X-ray pulsars.  Other
useful books and reviews on strongly magnetized neutron stars include
Meszaros (1992), Duncan (2000), Lai (2001) and Harding (2003).  A
complimentary work concentrating on stellar magnetism in general is
the book by Mestel (1999).

\section{Electrons in Strong Magnetic Fields} 
\label{sec:elec}

The quantum mechanics of a charged particle in a magnetic field
is presented in many texts (e.g., Landau \& Lifshitz 1977; 
Sokolov \& Ternov 1968; M\'esz\'aros 1992). Here we
summarize the basics needed for our later discussion. 

Consider first the nonrelativistic motion of a particle 
(charge $e_i$ and mass $m_i$) in a uniform magnetic field $\bB$ 
(assumed to be along the $z$-axis). In classical physics, the particle
gyrates in a circular orbit with radius and (angular) frequency given by
\be
\rho={m_icv_\perp\over |e_i|B},\quad
\omega_c={|e_i|B\over m_ic},
\label{eq:classical}\ee
where $v_\perp$ is the velocity perpendicular to the magnetic field.
In non-relativistic quantum mechanics, the kinetic energy of the transverse motion is 
quantized in Landau levels
\be
E_\perp={1\over 2}m_i\bv_\perp^2={1\over 2m_i}\bPi_\perp^2\rightarrow 
\left(n_L+{1\over 2}\right)\hbar\omega_c,\qquad n_L=0,1,2,\cdots
\label{eq:eperp}\ee
where $\bPi={\bf p}-(e_i/c){\bf A}=m_i{\bf v}$ is the mechanical momentum, 
${\bf p}=-i\hbar{\bf\nabla}$ is the canonical momentum, 
and ${\bf A}$ is the vector potential of the magnetic field. 

For an electron ($m_i\rightarrow m_e,~e_i\rightarrow -e$), the basic energy 
quantum is the cyclotron energy
\be
E_{ce}=\hbar\omega_{ce}=\hbar{eB\over m_ec}=11.577\,B_{12}~\rkeV,
\ee
where $B_{12}=B/(10^{12}\,{\rm G})$ is the magnetic field strength in 
units of $10^{12}$~Gauss. Including the kinetic energy associated with
the $z$-momentum ($p_z$) and the spin energy,
$E_{\sigma_z}=e\hbar/(2m_ec)
\bsigma\cdot\bB=\hbar\omega_{ce}\sigma_z/2$ (with $\sigma_z=\pm 1$), 
the total electron energy can be written as
\be
E_n = n\hbar\omega_{ce}+{p_z^2\over 2m_e},
\label{eqefree}\ee
where the index $n=n_L+(1+\sigma_z)/2=0,1,\cdots$.
Clearly, for the ground Landau level ($n=0$),
the spin degeneracy is one ($\sigma_z=-1$); for excited levels, 
the spin degeneracy is two ($\sigma_z=\pm 1$).  

Apart from the spin degeneracy, the Landau level associated with 
Eq.~(\ref{eq:eperp}) is degenerate by itself,
reflecting the fact that the energy is independent of the location 
of the guiding center of the gyration. To count the degeneracy, it is 
useful to define the {\it pseudomomentum} (or the generalized momentum)
\be
{\bf K}=\bPi+(e_i/c){\bf B}\times {\bf r}.
\ee
That $\bK$ is a constant of motion (i.e., it commutes with the Hamiltonian)
can be easily seen from the classical equation of motion for the particle,
$d\bPi/dt=(e_i/c)(d\br/dt)\times\bB$. 
The parallel component $K_z$ is simply the linear momentum $p_z$, 
while the perpendicular component $\bK_{\perp}$ is related to the 
position vector ${\bf R}_c$ of the guiding center by
\be
\bR_c={c\bK_{\perp}\times {\bf B}\over e_iB^2}={c\over e_iB}\bPi_\perp
\times\hat\bB+\br_\perp,
\label{eqbrc}\ee
($\hat\bB$ is the unit vector along $\bB$).
From Eqs.~(\ref{eq:eperp}) and (\ref{eqbrc}), we see that 
classical radius of gyration, $\rho=|\br_\perp-\bR_c|$ (see 
eq.~[\ref{eq:classical}]), is quantized according to
\be
|\br_\perp-\bR_c|={c\over |e_i|B}|\bPi_\perp|
\rightarrow (2n_L+1)^{1/2}\hat\rho,
\label{eqrhon}\ee
where 
\be
\hat\rho=\left({\hbar c\over |e_i|B}\right)^{1/2}
\label{eq:hatrho}\ee
is the cyclotron radius (or the magnetic length). 
Since the two components of $\bK_{\perp}$ do not commute,
$[K_x,K_y]=-i\hbar (e_i/c) B$, only one of the components can be diagonalized
for stationary states. This means that the guiding center
of the particle can not be specified.
If we use $K_x$ to classify the states,
then the wavefunction has form $e^{iK_x x/\hbar}\phi(y)$
(Landau \& Lifshitz 1977), where 
the function $\phi(y)$ is centered at $y_c=-cK_x/(e_iB)$ [see
Eq.~(\ref{eqbrc})]. 
The Landau degeneracy in an area ${\cal A}_g=L_g^2$ is thus given by 
\be
{L_g\over h}\int\! dK_x={L_g\over h}|K_{x,g}|={\cal A}_g
{|e_i|B\over h c}={{\cal A}_g\over 2\pi\hat\rho^2},
\label{eqdeg}\ee
where we have used $K_{x,g}=-e_iBL_g/c$.
Alternatively, if we choose to diagonalize $K_{\perp}^2=K_x^2+K_y^2$,
we obtain the Landau wavefunction $W_{nm}(\br_\perp)$
in cylindrical coordinates (Landau \& Lifshitz 1977),
where $m$ is the ``orbital'' quantum number (denoted by $s$ or $-s$
in some references). For the ground Landau level, this is
(for $e_i=-e$)
\be
W_{0m}(\br_\perp)\equiv
W_{m}(\rho,\phi)={1\over (2\pi m!)^{1/2}\hat\rho}
\left({\rho\over \sqrt{2}\hat\rho}\right)^m
\exp\left({\rho^2\over 4\hat\rho^2}\right)\exp(-im\phi),
\label{eqw0m}\ee
where the normalization $\int d^2\br_\perp\,|W_m|^2=1$ is
adopted. The (transverse) distance of the particle's guiding center 
from the origin of the coordinates is given by 
\be
|\bR_c|\rightarrow
\rho_m=(2m+1)^{1/2}\hat\rho,~~~~~m=0,1,2,\cdots
\label{eqrhom}\ee
The corresponding value of $K_{\perp}$ is 
$K_{\perp}^2=(\hbar |e_i|B/c)(2m+1)$. 
Note that $K_\perp^2$ assumes discrete values since $m$ is required to be an 
integer in order for the wavefunction to be single-valued. 
The degeneracy $m_g$ of the
Landau level in an area ${\cal A}_g=\pi R_g^2$ is then determined by
$\rho_{m_g} \simeq (2m_g)^{1/2}\hat\rho=R_g$, 
which again yields $m_g={\cal A}_g|e_i|B/(h c)$ as in Eq.~(\ref{eqdeg}).
Note that despite the similarity between Eqs.~(\ref{eqrhom})
and (\ref{eqrhon}), their physical meanings are quite different:
the circle $\rho=\rho_m$ does not correspond to any 
gyro-motion of the particle, and the energy is independent of $m$.
Also note that $K_{\perp}^2$ is related to the $z$-projection of angular
momentum $J_z$, as is evident from the $e^{-im\phi}$ factor in the
cylindrical wavefunction [Eq.~(\ref{eqw0m})]. In general, we have
\be
J_z=xP_y-yP_x={1\over 2e_iB}(\bK_{\perp}^2-\bPi_{\perp}^2)
= (m-n_L) {|e_i|\over e_i},
\ee
where we have used $\bPi_{\perp}^2=(\hbar |e_i|/c)B(2n_L+1)$.

For extremely strong magnetic fields such that $\hbar\omega_{ce}\go m_ec^2$, or
\be
B\go B_Q={m_e^2c^3\over e\hbar}= 4.414 \times 10^{13}~{\rm G},
\ee
the transverse motion of the electron becomes relativistic. 
The transverse momentum is again quantized according to 
Eq.~(\ref{eq:eperp}), i.e., $\bPi_\perp^2\rightarrow 
\left(2n_L+1\right)\hbar eB/c$. By solving the Dirac Equation in 
a homogeneous magnetic field, we find that Eq.~(\ref{eqefree}) for 
the energy of a free electron should be replaced by 
(e.g., Johnson \& Lippmann 1949)
\be
E_n =\left[c^2p_z^2+m_e^2c^4\left(1+2n {B\over B_Q}\right)\right]^{1/2}.
\label{eqrel}\ee
The shape of the Landau wavefunction in the relativistic theory is the
same as in the nonrelativistic theory, as seen from the fact that
$\hat\rho$ is independent of the particle mass (see Sokolov \& Ternov
1968).  Higher order corrections in $e^2$ to Eq.~(\ref{eqrel}) have
the form $(\alpha/4\pi)m_ec^2 F(B/B_Q)$, with $\alpha=1/37$ the fine 
structure constant, $F(\beta)=-\beta$,
$\beta \equiv B/B_Q$, for $\beta\ll 1$ and
$F(\beta)=\left[\ln(2\beta)-(\gamma_E+3/2)\right]^2+\cdots$ for
$\beta\gg 1$, where $\gamma_E=0.5772$ is Euler's constant (Schwinger
1988); these corrections are negligible.

Calculations based on electron wavefunctions that are solutions to the Dirac Equation 
in Cartesian and in cylindrical coordinates have both appeared in the literature. 
As is true for all quantum processes, the rates and cross sections depend on the choice of electron 
wavefunctions in a uniform magnetic field, which are dependent on spin state. 
The two most widely used relativistic wavefunctions are those of Johnson and Lippman 
(1949) and Sokolov \& Ternov (ST) (1968).  The Johnson and Lippman (JL) wavefunctions are derived in 
Cartesian coordinates and are eigenstates of the kinetic momentum operator $\pi = p + eA/c$.  
The Sokolov \& Ternov (ST) 
wavefunctions, given for the ground state in Eqn (\ref{eqw0m}), are derived in 
cylindrical coordinates and are eigenstates of the field parallel component, $\mu_z$, 
of the magnetic moment operator.  Given the different spin
dependence of the ST and JL eigenstates, one must use caution in making the appropriate choice
when treating spin-dependent processes.  Herold et al. (1982) and Melrose and Parle (1983) have 
noted that the ST eigenstates have desirable properties that the JL do not possess, such as being
eigenfunctions of the Hamiltonian including radiation corrections, having
symmetry between positron and electron states, and diagonalization of the self-energy shift operator.
As found by Graziani (1993), the ST wavefunctions also diagonalize the Landau-Dirac 
operator and are the physically correct choices for spin-dependent treatments 
and in incorporating widths in the scattering cross section.  Although the spin-averaged ST and JT 
cyclotron decay rates are equal, for example, the spin-dependent decay rates are not, except in the special case
in which the initial momentum of the electron parallel to the magnetic field vanishes.  An additional
desirable property of the ST states is that spin-dependent states are 
preserved under Lorentz transformations while Lorentz boosts mix spin states of the JL wavefunctions. 

\section{Matter in Strong Magnetic Fields} 
\label{sec:matter}

When studying matter in magnetic fields, the natural (atomic) unit
for the field strength, $B_0$, is set by equating the electron cyclotron
energy $\hbar\omega_{ce}$ to the characteristic atomic energy
$e^2/a_0=2\times 13.6$~eV (where $a_0$ is the Bohr radius), 
or equivalently by $\hat\rho=a_0$. Thus it is convenient to 
define a dimensionless magnetic field strength $b$ via 
\be
b\equiv {B\over B_0};\qquad B_0={m_e^2e^3c\over\hbar^3}=\alpha^2 B_Q=
2.3505\times 10^9\,{\rm G}
\label{eqb0}\ee
($\alpha=1/137$ is the fine structure constant).
For $b\gg 1$, the cyclotron energy $\hbar\omega_{ce}$ is much larger
than the typical Coulomb energy, so that the properties of atoms,
molecules and condensed matter are qualitatively changed by the
magnetic field.  In such a strong field regime, the usual perturbative
treatment of the magnetic effects (e.g., Zeeman splitting of atomic
energy levels) does not apply (see Garstang 1977 for a review of
atomic physics at $b\lo 1$). Instead, the Coulomb forces act as a
perturbation to the magnetic forces, and the electrons in an atom
settle into the ground Landau level. Because of the extreme
confinement ($\hat\rho\ll a_0$) of the electrons in the transverse
direction (perpendicular to the field), the Coulomb force becomes much
more effective in binding the electrons along the magnetic field
direction. The atom attains a cylindrical structure. Moreover, it is
possible for these elongated atoms to form molecular chains by
covalent bonding along the field direction. Interactions between the
linear chains can then lead to the formation of three-dimensional
condensates.

Note that when studying bound states (atoms, molecules and 
condensed matter near zero pressure) in strong magnetic fields, 
it is adequate to use nonrelativistic quantum mechanics, even for 
$B\go B_Q$. The nonrelativistic treatment of bound states
is valid for two reasons: (i) For electrons in the ground Landau level, 
eq.~(\ref{eqrel}) for the free electron energy reduces to 
$E\simeq m_ec^2+p_z^2/(2m_e)$ for $p_zc\ll m_ec^2$; 
the electron remains nonrelativistic in the $z$-direction 
(along the field axis) as long as the binding energy $E_B$ is much less 
than $m_ec^2$; (ii) As mentioned before (\S \ref{sec:elec}), the shape of the Landau 
wavefunction in the relativistic theory is the same as in the 
nonrelativistic theory.
Therefore, as long as $E_B/(m_e c^2)\ll 1$, the relativistic
effect on bound states is a small correction.
For bulk matter under pressure, the relativistic correction 
becomes increasingly important as density increases (see 
\S \ref{sec:NS}).

Note that unless specified, the expressions in this section will be in 
atomic units (a.u.), where mass and length are expressed in units 
of the electron mass $m_e$ and the Bohr radius 
$a_0=0.529\times 10^{-8}$ cm, energy in units of 
$2~{\rm Ryd}=e^2/a_0=2\times 13.6$~eV, field strength 
in units of $B_0$ [Eq.~(\ref{eqb0})]. 
A previous (more detailed) review on this subject is Lai (2001),
where more complete references can be found.

\subsection{Hydrogen Atom}
\label{subsec:hatom}

In a strong magnetic field with $b\gg 1$, the electron is confined to
the ground Landau level (``adiabatic approximation''), and the Coulomb
potential can be treated as a perturbation. Assuming infinite proton
mass (see below), the energy spectrum of the H atom is specified by
two quantum numbers, $(m,\nu)$, where $m$ measures the mean transverse
separation [Eq.~(\ref{eqrhom})] between the electron and the proton,
while $\nu$ specifies the number of nodes in the $z$-wavefunction. 
There are two distinct types of states in the energy spectrum $E_{m\nu}$.
The ``tightly bound'' states have no node in their $z$-wavefunctions 
($\nu=0$). The transverse size of the atom in the 
$(m,0)$ state is $L_\perp\sim\rho_m=[(2m+1)/b]^{1/2}$.
For $\rho_m\ll 1$, the atom is elongated with $L_z\gg L_\perp$.
We can estimate the longitudinal size $L_z$ by minimizing the
energy, $E\sim L_z^{-2}-L_z^{-1}\ln (L_z/L_\perp)$ (where the 
first term is the kinetic energy and the second term  is 
the Coulomb energy), giving 
\be
L_z\sim {1\over 2\ln(1/\rho_m)}=\left(\ln {b\over 2m+1}\right)^{-1}.
\ee
The energy of the tightly bound state is then
\be
E_{m0}\simeq -0.16A\, 
\left(\ln {b\over 2m+1}\right)^2\qquad ({\rm for}~~2m+1\ll b).
\label{eqem}\ee
Here the numerical prefactor comes from solving the Schr\"odinger equation.
The coefficient $A$ in (\ref{eqem}) is close to unity for
the range of $b$ of interest ($1\ll b\lo 10^6$) and
varies slowly with $b$ and $m$ (e.g., $A\simeq 1.01-1.3$ for $m=0-5$ when
$B_{12}=1$, and $A\simeq 1.02-1.04$ for $m=0-5$ when $B_{12}=10$.
Note that $E_{m0}$ asymptotically approaches $-0.5\,
\left[\ln {b/(2m+1)}\right]^2$ when $b\rightarrow\infty$).
Obviously the ground-state binding energy $|E_{00}|$ (e.g.,
$160$~eV at $10^{12}$~G and $540$~eV at $10^{14}$~G) is much
larger than the zero-field value (13.6~eV).
For $\rho_m\go 1$, or $2m+1\go b$ [but still $b\gg (2m+1)^{-1}$ so that the
adiabatic approximation ($|E_{m0}|\ll b$) is valid], 
we have $L_z\sim \rho_m^{1/2}$, and the
energy levels are approximated by
\be
E_{m0}\simeq -0.6\,\left({b\over 2m+1}\right)^{1/2}
\qquad [{\rm for}~~2m+1\go b\gg (2m+1)^{-1}].
\label{eqem2}\ee
Again the prefactor 0.6 comes from solution 
of the Schr\"odinger equation.
Numerical values of $E_{m0}$ for different $B$'s can be found, for 
example, in Ruder \etal~(1994). Fitting formulae for $E_{m0}$ 
are given in Potekhin (1998) and in Ho et al.~(2003).

Another type of states of the H atom has nodes in the $z$-wavefunctions 
($\nu>0$). These states are ``weakly bound'', and have energies
(for $b\gg 1$) of order $E_{m\nu}\sim -\nu^{-2}$~Ry
(see Lai 2001 and references therein; see also
Potekhin 1998 for a comprehensive set of fitting formulae).
The sizes of the wavefunctions are $\rho_m$
perpendicular to the field and $L_z\sim \nu^2a_0$ along the field. 

This simple picture of the H energy levels is modified when a finite
proton mass is taken into account.  Even for a ``stationary'' H atom,
the energy $E_{m\nu}$ should be replaced by $E_{m\nu}+m\hbar\omega_{cp}$, 
where $\hbar\omega_{cp}=\hbar eB/(m_pc)=6.3\,B_{12}$~eV 
is the proton cyclotron energy (e.g. Herold et al.~1981). 
The extra ``proton recoil'' energy 
$m\hbar\omega_{cp}$ becomes increasingly important with increasing $B$.
Moreover, the effect of center-of-mass motion is non-trivial: When
the atom moves perpendicular to the magnetic field, an electric
field is induced in its rest frame and can significantly change the
atomic structure, i.e., there is a strong coupling between the
center-of-mass motion and the ``internal'' electron motion.
It is easy to show that even including the Coulomb interaction, 
the total pseudomomentum, 
\be
\bK=\bK_e+\bK_p,
\ee
is a constant of motion. Moreover, all components of $\bK$ commute with each
other. Thus it is natural to separate the CM motion from the internal degrees
of freedom using $\bK$ as an explicit constant of motion. 
From Eq.~(\ref{eqbrc}), we find that the separation between the guiding 
centers of the electron and the proton is directly related to $\bK_\perp$:
\be
\bR_K=\bR_{ce}-\bR_{cp}={c\bB\times \bK\over eB^2}.
\ee
For sufficiently small $K_\perp$, it is convenient to think of
the effect of center-of-mass motion as the ``motional Stark effect'', and 
the kinetic energy associated with the transverse motion is 
$K_\perp^2/(2M_{\perp m})$. Here the effective mass $M_{\perp m}$ 
depends on the energy levels, and increases with increasing $B$.
For large transverse pseudomomentum $K_\perp$, the structure of 
the ``moving'' atom is qualitatively different from the 
stationary atom: The atom assumes a decentered configuration, with
transverse electron-proton separation $\sim R_K$, and longitudinal
separation $\sim R_K^{3/4}$, and 
its energy depends on $K_\perp$ as $-R_K^{-1}=-b/K_\perp$ rather than
the usual $K_\perp^2$ dependence
(see Lai 2001 and Potekhin~1998 for fitting formulae and references).

\subsection{Multielectron Atoms}

The result for H atom can be easily generalized to hydrogenic 
ions (with one electron and nuclear charge $Z$). The adiabatic
approximation (where the electron lies in the ground Landau level)
holds when $\hat\rho\ll a_0/Z$, or
$b\gg Z^2$.
For a tightly bound state, $(m,\nu)=(m,0)$, the transverse size 
is $L_\perp\sim\rho_m$, while the longitudinal size is 
\be
L_z\sim \left(Z\ln {1\over Z\rho_m}\right)^{-1}.
\ee
The energy is given by
\be
E_m\simeq -0.16\,AZ^2\,\left[\ln{1\over Z^2}\left({b\over 2m+1}\right)
\right]^2
\label{eqem3}\ee
for $b\gg (2m+1)Z^2$. Results for the weakly bound states ($\nu>0$) 
can be similarly generalized.

We can imagine constructing a multi-electron atom (with Z electrons) 
by placing electrons at the lowest available energy levels of a hydrogenic 
ion. The lowest levels to be filled are the tightly bound states with 
$\nu=0$. When $a_0/Z \gg \sqrt {2 Z-1} \hat\rho$, i.e., $b \gg 2 Z^3$,
all electrons settle into the tightly bound levels with 
$m=0,1,2,\cdots,Z-1$. The energy of the atom is approximately given
by the sum of all the eigenvalues of Eq.~(\ref{eqem3}). Accordingly, we 
obtain an asymptotic expression for $Z \gg 1$ (Kadomtsev \& Kudryavtsev
1971)
\begin{equation}
E \sim - Z^3 l_Z^2,\quad {\rm with}~~
l_Z=\ln \biggl ({a_0 \over Z \sqrt {2 Z-1} \hat \rho}\biggr )
\simeq \ln \sqrt {b \over 2 Z^3}.
\end{equation}
The size of the atom is given by
$L_\perp\sim (2Z-1)^{1/2}\hat\rho$,
$L_z\sim {a_0/(Zl_Z)}$.

For intermediate-strong fields (but still strong enough to ignore the 
Landau excitation), $Z^{4/3} \ll b \ll 2 Z^3$,
many $\nu>0$ states of the inner Landau orbitals (states with relatively 
small $m$) are populated by the electrons. In this regime a Thomas-Fermi
type model for the atom is appropriate,
i.e., the electrons can be treated as a
one-dimensional Fermi gas in a more or less
spherical atomic cell (Kadomtsev 1970; Mueller et al.~1971). 
The electrons occupy the ground Landau level, with the $z$-momentum up
to the Fermi momentum $p_F\sim n_e/b$, where $n_e$ is the 
number density of electrons inside the atom (recall that the degeneracy 
of a Landau level is $e B /hc \sim b$). The kinetic energy of electrons 
per unit volume is $\eps_k \sim b\,p_F^3\sim n_e^3/b^2$, and
the total kinetic energy is $E_k \sim 
R^3 n_e^3 /b^2 \sim Z^3 /b^2 R^6$, where $R$ is the radius of the atom.
The potential energy is $E_p \sim -Z^2/R$. Therefore the total energy of the
atom can be written as
$E \sim {Z^3/(b^2 R^6)} - {Z^2/R}$.
Minimizing $E$ with respect to $R$ yields
\begin{equation}
R \sim Z^{1/5}b^{-2/5},\quad E \sim -Z^{9/5} b^{2/5}.
\label{heavyatom}\end{equation}
For these relations to be valid, the electrons
must stay in the ground Landau level; this requires
$Z/R\ll\hbar\omega_{ce}=b$, which corresponds to $b\gg Z^{4/3}$.

Reliable values for the energy of a multi-electron atom for $b\gg 1$
can be calculated using the Hartree-Fock method, which takes into
account electron-electron direct and exchange interactions in a
self-consistent manner.  The Hartree-Fock method is approximate
because electron correlations are neglected. Due to their mutual
repulsion, any pair of electrons tend to be more distant from each
other than the Hartree-Fock wave function would indicate. In
zero-field, this correlation effect is especially pronounced for the
spin-singlet states of electrons for which the spatial wave function
is symmetrical. In strong magnetic fields, the electron spins are 
in the ground state all aligned antiparallel to the magnetic field, 
the spatial wavefunction is antisymmetric with respect to the 
interchange of two electrons.
Thus the error in the Hartree-Fock approach is expected to be
significantly smaller than the $1\%$ accuracy characteristic of
zero-field Hartree-Fock calculations 
(Neuhauser et al.~1987; Schmelcher et al.~1999).

Accurate energies of He atom as a function of $B$ in the adiabatic
approximation (valid for $b\gg Z^2$) were obtained by Virtamo (1976)
and Pr\"oschel et al.~(1982). This was extended to $Z$ up to $26$ (Fe
atom) by Neuhauser et al.~(1987) (see also Miller \& Neuhauser
1991; Mori \& Hailey 2002). 
Numerical results can be found in these papers.  Neuhauser et
al.~(1987) gave an approximate fitting formula, $E\simeq
-160\,Z^{9/5}B_{12}^{2/5}$~eV, for $0.5\lo B_{12}\lo 5$ (Comparing
with the numerical results, the accuracy of the formula is about $1\%$
for $Z\simeq 18-26$ and becomes $5\%$ for $Z\sim 10$.) For the He
atom, more accurate results (which relax the adiabatic approximation)
are given in Ruder et al.~(1994), Jones et al.~(1999) (this
paper also considers the effect of electron correlation)
and in Al-Hujaj \& Schmelcher (2003a,b).

Other calculations of heavy atoms in strong magnetic fields include
Thomas-Fermi type statistical models (see Fushiki \etal~1992; Lieb
\etal~1994a,b) and density functional theory (Jones 1985,~1986;
K\"ossl \etal~1988; Relovsky \& Ruder 1996).  The Thomas-Fermi type
models are useful in establishing asymptotic scaling relations, but
are not adequate for obtaining accurate binding energy and excitation
energies. The density functional theory can potentially give results
as accurate as the Hartree-Fock method after proper calibration is
made (see Vignale \& Rasolt 1987,1988).  Numerical results of He, C,
Fe (and the associated ions) for a wide range of field strengths
($10^{12}$-$10^{15}$~G) are given in Medin \& Lai (2006).

The effects of center-of-mass motion on multi-electron systems (heavy
atoms and molecules) in strong magnetic fields have not been studied
numerically, although many theoretical issues are discussed in Johnson
\etal~(1983) and Schmelcher \etal~(1988,1994).

While for neutral atoms the center-of-mass motion can be separated from the
internal relative motion, this cannot be done for ions (Avron et al.~1978).
Ions undergo collective cyclotron motion which depends
on the internal state.  However, the existence of an approximate
constant of motion allows an approximate pseudoseparation up to very
high fields (see Baye \& Vincke 1998).
Some numerical results for He$^+$ moving in strong magnetic
fields are obtained by Bezchastnov \etal~(1998) and 
Pavlov \& Bezchastnov (2005).

\subsection{Molecules}

In a strong magnetic field, the mechanism of forming
molecules is quite different from the zero-field case
(see Ruderman 1974; Lai \etal~1992). Consider hydrogen as an example.
The spin of the electron in a H atom is aligned anti-parallel to the 
magnetic field (flipping the spin would cost $\hbar\omega_{ce}$),
and therefore two H atoms in their ground states ($m=0$) 
do not bind together according to the exclusion principle. 
Instead, one H atom has to be excited to the $m=1$ state. The
two H atoms, one in the ground state ($m=0$), another
in the $m=1$ state then form the ground state of the H$_2$ molecule by
covalent bonding. Since the ``activation energy'' for exciting
an electron in the H atom from the Landau orbital $m$ to $(m+1)$
is small [see Eq.~(\ref{eqem})], the resulting H$_2$ molecule is stable.
Similarly, more atoms can be added to form H$_3$, H$_4,~\cdots$.

The size of the H$_2$ molecule is comparable to that of the H atom. 
The interatomic separation $a_{\rm eq}$ and the dissociation energy 
$D$ of the H$_2$ molecule scale approximately as 
\begin{equation}
a_{\rm eq}\sim {1\over \ln b},\qquad D\sim (\ln b)^2,
\end{equation}
although $D$ is numerically smaller than the ionization energy of the H atom.

Another mechanism of forming a H$_2$ molecule in strong magnetic 
fields is to let both electrons occupy the same $m=0$ Landau orbital,
while one of them occupies the tightly bound
$\nu=0$ state and the other the $\nu=1$ weakly bound state.
This costs no ``activation energy''. However, the resulting molecule 
tends to have a small dissociation energy, of order a Rydberg.
One can refer to this electronic state of the molecule as the 
weakly bound state, and to the states formed by two electrons
in the $\nu=0$ orbitals as the tightly bound states.
As long as $\ln b\gg 1$, the weakly bound states 
constitute excited energy levels of the molecule.

In the Born-Oppenheimer approximation (see Schmelcher \etal~1988,1994 for a 
discussion on the validity of this approximation in strong magnetic fields),
the interatomic potential $U(a,R_{\perp})$ is given by the total 
electronic energy $E(a,R_{\perp})$ of the system, where
$a$ is the proton separation along the magnetic field, and
$R_{\perp}$ is the separation perpendicular to the field.
Once $E(a,R_{\perp})$ is obtained, the electronic equilibrium state
is determined by locating the minimum of the $E(a,R_\perp)$ surface.
[For a given $a$, $E(a,R_\perp)$ is minimal at $R_\perp=0$].
The energy curve $E(a,R_\perp)$ can be obtained from
Hartree-Fock calculations. For large $a$, configuration interaction
must be taken into account in the Hartree-Fock scheme 
(Lai \etal~1992).

Molecular configurations with $R_\perp\neq 0$ correspond to excited
states of the molecules. To obtain $E(a,R_\perp)$, mixing of different
$m$-states in single-electron orbital needs to be taken into
account. Approximate energy surfaces $E(a,R_\perp)$ for both small
$R_\perp$ and large $R_\perp$ have been computed by Lai \& Salpeter
(1996).

Numerical results of $E(a,0)$ (based on the Hartree-Fock method) for
both tightly bound states and weakly bound states are given in Lai
\etal~(1992) and Lai \& Salpeter (1996). Quantum Monte Carlo
calculations have also been performed, confirming the validity of the
method (Ortiz \etal~1995). For example, the dissociation energy of
H$_2$ (neglecting the zero-point energy of the protons) in the ground
state is $46$ eV for $B_{12}=1$ and $150$ eV for $B_{12}=10$. By
contrast, the zero-field dissociation energy of H$_2$ is $4.75$ eV.

For the ground state of H$_2$, the molecular axis and the magnetic field 
axis coincide, and the two electrons occupy the $m=0$ and $m=1$ orbitals, 
i.e., $(m_1,m_2)=(0,1)$. The molecule can have different types of 
excitation levels (Lai \& Salpeter 1996):
(i) {\it Electronic excitations}.
The electrons occupy orbitals other than $(m_1,m_2)=(0,1)$, 
giving rise to the electronic excitations. The energy difference 
between the excited state $(m_1,m_2)$ (with $\nu_1=\nu_2=0$) 
and the ground state $(0,1)$ is of order $\ln b$, 
as in the case for atoms. Another type of electronic excitation is 
formed by two electrons in the $(m,\nu)=(0,0)$ and $(0,1)$ orbitals.  
The dissociation energy of this weakly bound state is of order a Rydberg, 
and does not depend sensitively on the magnetic field strength.
(ii) {\it Aligned vibrational excitations}.
These result from the vibration of the protons about the
equilibrium separation $a_{\rm eq}$ along the magnetic field axis.
(iii) {\it Transverse vibrational excitations}.
The molecular axis can deviate from the magnetic field direction,
precessing and vibrating around the magnetic axis.
Such an oscillation is the high-field analogy of
the usual molecular rotation; the difference is that in strong magnetic 
fields, this ``rotation'' is constrained around the magnetic field line. 
Note that in a strong magnetic field,
the electronic and (aligned and transverse) vibrational 
excitations are all comparable. This is in contrast 
to the zero-field case, where we have
$\Delta \eps_{\rm elec} \gg \hbar \omega_{\rm vib}
\gg \hbar \omega_{\rm rot}$.

Molecules of heavy elements in strong magnetic fields have not been
systematically investigated until recently. In
general, we expect that, as long as $a_0/Z\gg (2Z-1)^{1/2}\hat\rho$,
or $b\gg 2Z^3$, the electronic properties of the heavy molecule is
similar to those of H$_2$. When the condition $b\gg 2Z^3$ is not
satisfied, the molecule should be quite different and may be unbound
relative to individual atoms (e.g., at $B=10^{12}$~G bound Fe molecule
does not exist). Some Hartree-Fock results of diatomic
molecules (from H$_2$ up to C$_2$) at $b=1000$ are given in Demeur
\etal~(1994).  A recent study of molecules of H, He, C and Fe using
density functional theory for a wide range of field strengths is
given in Medin \& Lai (2006).

As more atoms are added to a molecule, the energy per atom
in a molecule saturates, becomes independent of the number of atoms
in the molecule. We then essentially
have a one-dimensional condensed matter (see the next subsection).

\subsection{Condensed Matter}
\label{sec:condense}

The binding energy of magnetized condensed matter at zero pressure can be 
estimated using the uniform electron gas model (e.g., Kadomtsev 1970).
Consider a Wigner-Seitz cell with radius $r_i=Z^{1/3}r_s$ ($r_s$ is the
mean electron spacing); the mean number density of electrons is $n_e=Z/(4\pi
r_i^3/3)$. The electron Fermi momentum $p_F$ is obtained from
$n_e=(eB/hc)(2p_F/h)$. When the Fermi energy $p_F^2/(2m_e)$ is less than the 
cyclotron energy $\hbar\omega_{ce}$, or when the electron number 
density satisfies
\be
n_e \le n_{B}
={1\over \sqrt{2}\pi^2\hat\rho^3}
=0.0716\,b^{3/2},
\label{nlandau}\ee
(or $r_i\ge r_{iB}=1.49\,Z^{1/3}b^{-1/2}$), the electrons only occupy
the ground Landau level. 
The energy per cell can be written as
\be
E_s(r_i)={3\pi^2Z^3\over 8b^2r_i^6}-{0.9Z^2\over r_i},
\label{estry}\ee
where the first term is the kinetic energy and the second term is the
Coulomb energy. For a zero-pressure condensed matter, we require
$dE_s/dr_i=0$, and the equilibrium $r_i$ and energy are then given by 
\ba
&&r_{i,0}\simeq 1.90\,Z^{1/5}b^{-2/5},\label{eqri0}\\
&&E_{s,0}\simeq -0.395\,Z^{9/5}b^{2/5}.\label{es0}
\ea
The corresponding zero-pressure condensation density is 
\be
\rho_{s,0}\simeq 561\,A\,Z^{-3/5}B_{12}^{6/5}\,{\rm g~cm}^{-3}
\label{rs0}\ee
Note that for $b\gg 1$, the zero-pressure density 
is much smaller than the ``magnetic'' density defined in 
Eq.~(\ref{nlandau}), i.e., $\rho_{s,0}/\rho_B=(r_B/r_{i,0})^3=
0.48\,Z^{2/5}b^{-3/10}$.
The uniform electron gas model can be improved by incorporating
the Coulomb exchange energy and Thomas-Fermi correction
due to nonuniformity of the electron gas (see Lai 2001).

Although the simple uniform electron gas model and its Thomas-Fermi
type extensions give a reasonable estimate for the 
binding energy for the condensed state, they are not adequate for determining
the cohesive property of the condensed matter. The cohesive energy $Q_s$
is the difference between the atomic ground-state energy and the energy per
atom of the condensed matter ground state. 
One uncertainty concerns the lattice structure of
the condensed state, since the Madelung energy can be quite different from
the Wigner-Seitz value [the second term in Eq.~(\ref{estry})]
for a non-cubic lattice. In principle, a three-dimensional
electronic band structure calculation is needed to
solve this problem. So far the only attempt to this problem has been 
the preliminary calculations by Jones (1986) for a few elements 
and several values of field strengths using density functional theory. 

Three-dimensional condensed matter can be formed by placing a pile of
parallel chains together. The energy difference 
$\Delta E_s=|E_{s,0}|-|E_\infty|$ between the 3d condensed matter and 
the 1d chain must be positive 
and can be estimated by calculating the interaction (mainly 
quadrupole--quadrupole) between the chains. Various considerations indicate
that the difference is between $0.4\%$ and $1\%$ of $|E_\infty|$
(Lai \& Salpeter 1997). Therefore, for light elements such as
hydrogen and helium, the binding of the 3d condensed matter 
results mainly from the covalent bond along the magnetic field axis, 
not from the chain-chain interaction.
The binding energies of 1D chain for some elements
have been obtained using Hartree-Fock method (Neuhauser \etal~1987;
Lai \etal~1992; Lai 2001). Density functional theory has also been used 
to calculate the structure of linear chains in strong magnetic fields 
(Jones 1985; Relovsky \& Ruder 1996; Medin \& Lai 2006).

Numerical calculations carried out so far have indicated that for
$B_{12}=1-10$, linear chains are unbound for large atomic numbers
$Z\go 6$ (Jones 1986; Neuhauser \etal~1987; Medin \& Lai 2006).  In
particular, the Fe chain is unbound relative to the Fe atom; this is
contrary to what some early calculations (e.g., Flowers \etal~1977)
have indicated.  Therefore, the chain-chain interaction must play a
crucial role in determining whether the three dimensional
zero-pressure Fe condensed matter is bound or not. The main difference
between Fe and H is that for the Fe atom at $B_{12}\sim 1$, many
electrons are populated in the $\nu\neq 1$ states, whereas for the H
atom, as long as $b\gg 1$, the electron always settles down in the
$\nu=0$ tightly bound state. Therefore, the covalent bonding mechanism
for forming molecules is not effective for Fe at $B_{12}\sim
1$. However, for a sufficiently large $B$, when $a_0/Z\gg
\sqrt{2Z+1}\hat\rho$, or $B_{12}\gg 100(Z/26)^3$, we expect the Fe
chain to be bound in a manner similar to the H chain or He chain
(Medin \& Lai 2006). The cohesive property of magnetized
condensed matter is important for understanding the physical
condition of the ``polar gap'' of the neutron star 
(see \S 9).
 
\section{Propagation of Photons in a Magnetized Plasma} \label{Sec:photons}

The magnetized plasma around a neutron star is anisotropic and
birefringent, and significantly influences the polarization state
of a photon. Many of the radiative processes discussed in 
\S \ref{sec:RadProc} are polarization-dependent (e.g., photons in 
different polarization modes often have very different opacities).
Thus it is important to understand the propagation of photons
in a magnetized plasma. The atmosphere of a neutron star (with 
density $10^{-3}-10^3$~g/cm$^3$) can be characterized as a cold plasma 
for most purposes, while the much more tenuous magnetosphere 
likely comprises relativistic electron-positron pairs. 
Even in a pure vacuum, the effect of quantum electrodynamics induces
birefringence through vacuum polarization. 

\subsection{Dielectric Tensor of Cold Plasma}

Following Ginzburg (1970), we consider a cold plasma 
composed of electrons and ions (with charge, mass and number density
given by $-e,~m_e,~n_e$ and $Ze,~m_i=Am_p,~n_i=n_e/Z$, respectively;
here $Z$ is the charge number and $A$ is the mass number of the ion)
in an external magnetic field $\bB$.
The electrons and ions are coupled by collisions, with the collision 
frequency $\nu_{ei}$. We can generalize Ginzburg (1970) by including
the radiative dampings of electrons and ions, with the
damping frequencies $\nu_{re}$ and $\nu_{ri}$, respectively. 
In the presence of an electromagnetic wave with the
electric field $\bE\propto e^{-i\omega t}$,
the equations of motion for a coupled electron-ion pair are:
\begin{eqnarray}
m_e\ddot \br_e & = & -e\bE-{e\over c}\dot\br_e\times\bB
- m_e\nu_{ei}(\dot\br_e-\dot\br_i)-m_e\nu_{re}\dot\br_e,\label{eq:emotion}\\
m_i\ddot \br_i & = & Ze\bE+{Ze\over c}\dot\br_i\times\bB
-m_e Z\nu_{ei}(\dot\br_i-\dot\br_e)-m_i\nu_{ri}\dot\br_i.
\label{eq:ionmotion}
\end{eqnarray}
Note that the drag coefficient due to e-ion collision
for electron is $m_e \langle \sigma_{ei} v_{ei}\rangle n_i
\equiv m_e\nu_{ei}$ (where $\sigma_{ei}$ is the cross-section, $v_{ei}$ is
the relative velocity), while the drag coefficient for ion is 
$m_e \langle\sigma_{ei} v_{ei}\rangle n_e= m_e Z\nu_{ei}$ since $n_e=Zn_i$.
Solving Eqs.~(\ref{eq:emotion})-(\ref{eq:ionmotion}) yields 
$\br=\br_e-\br_i$, and the polarization of the plasma 
${\bf P}=n_iZe\br_i-n_ee\br_e=
-n_ee\br$. The electric displacement vector is then ${\bf D}=\bE+4\pi{\bf P}
=\beps^{(p)}\cdot\bE$.
In the coordinate system $XYZ$ with $\bB$ along $Z$, the plasma 
dielectric tensor is given by 
\be
\lb\beps^{(p)}\rb_{{\hat Z}={\hat B}}= \lb \begin{array}{ccc}
\varepsilon & ig & 0 \\
-ig & \varepsilon & 0 \\
0 & 0 & \eta
\end{array} \rb, \label{eq:epsij0}
\ee 
where 
\ba
&&\hskip -0.8cm \varepsilon\pm g = 1-{v_e(1+i\gamma_{ri})+v_i(1+
i\gamma_{re})
\over (1+i\gamma_{re}
\pm u_e^{1/2})
(1+i\gamma_{ri}\mp u_i^{1/2})+i\gamma_{ei}},\label{eq:epsilong}\\
&&\hskip -0.8cm \eta\simeq 1-{v_e\over 1+i(\gamma_{ei}+\gamma_{re})}
-{v_i\over 1+i(\gamma_{ei}+\gamma_{ri})}.
\label{eq:eta}
\ea
In Eqs.~(\ref{eq:epsilong}) and (\ref{eq:eta}), we have
defined the dimensionless quantities
\be
u_e={\omega_{ce}^2\over\omega^2}, \quad
u_i={\omega_{ci}^2\over\omega^2}, \quad
v_e={\omega_{pe}^2\over\omega^2}, \quad
v_i={\omega_{pi}^2\over\omega^2},
\ee
where $\omega_{ce}=eB/(m_ec)$ is the electron cyclotron frequency,
$\omega_{ci}=ZeB/(m_ic)$ is the ion 
cyclotron frequency, $\omega_{pe}=(4\pi n_e e^2/m_e)^{1/2}$
is the electron plasma frequency, and $\omega_{pi}=
(4\pi n_i Z^2e^2/m_i)^{1/2}$ is the ion plasma frequency. 
The dimensionless damping rates
$\gamma_{ei}=\nu_{ei}/\omega$, $\gamma_{re}=\nu_{re}/\omega$, and
$\gamma_{ri}=\nu_{ri}/\omega$ are given by
\ba
\gamma_{ei}&=&
{Z^2n_i e^4\over\hbar\omega^2}\!\left({2\pi\over m_ekT}\right)^{1/2}
\!\!\!(1-e^{-\hbar\omega/kT})\,g_\alpha^{\rm ff},
\label{eqgammaei}\\
\gamma_{re}&=&{2e^2\omega\over 3m_ec^3},
\label{eqgammare}\\
\gamma_{ri}&=&{Z^2m_e\over Am_p}\gamma_{re}.
\ea
where $g_\alpha^{\rm ff}$ is the Gaunt factor (see Potekhin \& Chabrier 
2003 and references therein). Including these damping terms in the
dielectric tensors allows one to obtain the appropriate expressions for
the radiative opacities using the imaginary parts $\beps^{(p)}$.
If we neglect 
damping ($\gamma_{re}=\gamma_{ri}=\gamma_{ei}=0$), 
Eqs.~(\ref{eq:epsilong}) and (\ref{eq:eta}) reduce to 
\be
\varepsilon=1-{v_e\over 1-u_e}-{v_i\over 1-u_i},
\quad
g={v_e u_e^{1/2}\over 1-u_e}-{v_i u_i^{1/2}\over 1-u_i},
\quad
\eta=1-v_e-v_i.
\label{eq:cold}\ee

\subsection{Vacuum Polarization}
\label{subsec:vacuum-p}

In strong magnetic fields, vacuum polarization can significantly
influence photon propagation. According to quantum electrodynamics, a
photon may temporarily convert into virtual electron-positron pairs,
and since the pairs are ``polarized'' by the external magnetic field,
the magnetized vacuum has a nontrivial dielectric tensor $\beps$
and a nontrivial permeability tensor $\bmu$
(e.g., Heisenberg \& Euler 1936, Schwinger 1951; Adler 1971; Tsai \&
Erber 1975; see Schubert 2000 for extensive bibliography).
In the low-energy limit, $\hbar\omega\ll m_e c^2$, 
the Euler-Heisenberg effective Lagrangian can be
applied. The relevant dimensionless magnetic-field parameter is
$\beta\equiv B/B_Q$, where $B_Q=m_e^2c^3/e\hbar=4.414\times 10^{13}$~G. 
The vacuum dielectric tensor $\beps$ and inverse permeability 
tensor $\bmu^{-1}\equiv \bar\bmu$ take the form
$\beps=\bI+\Delta\beps^{(v)}$ and $\bar\bmu=\bI+\Delta\bar\bmu^{(v)}$ 
(where $\bI$ is the unit tensor), with 
\ba
&&\Delta\beps^{(v)}={\hat a}\bI +q\hat\bB\hat\bB,\\
&&\Delta\bar\bmu^{(v)}={\hat a}\bI+m\hat\bB\hat\bB,
\ea
where $a$, $q$, and $m$ are functions of $\beta$. In the limit 
$\beta\ll 1$, the vacuum polarization coefficients are given by 
(Adler 1971)
\be
\hat{a}= - \frac{2 \alpha}{45\pi}\, \beta^2,
\quad
 q= \frac{7 \alpha}{45\pi}\, \beta^2,
\quad
 m= - \frac{4 \alpha}{45\pi}\, \beta^2,
\label{vac-weak}
\ee
where $\alpha=e^2/\hbar c = 1/137$ is the fine-structure constant.
For arbitrary $B$, the vacuum polarization coefficients 
have been obtained by Heyl \& Hernquist (1997) in terms of special functions
and by Kohri \& Yamada (2002) numerically. Convenient expressions are given
by Potekhin et al.~(2004):
\ba
&&
 \hat{a} = \frac{\alpha}{2\pi} \bigg[
 \xi X(\xi) - 2 \int_1^\xi X(\xi')\, d\xi' - 0.0329199 \,\bigg],
\label{hehe-a}
\\&&
 \hat{a} + q = \frac{\alpha}{2\pi} \left[\,
 \frac{2}{9\xi^2} -\frac23 \frac{d X(\xi)}{d\xi}\,\right],
\label{hehe-q}
\\&&
 m = \frac{\alpha}{2\pi} \left[\,
 \xi X(\xi) - \xi^2 \frac{d X(\xi)}{d\xi} \right],
\label{hehe-m}
\ea
where $X(\xi)$ (with $\xi=\beta^{-1}$)
is expressed through the Gamma function $\Gamma(x)$:
\be
 X(\xi) = 2 \ln \Gamma(\xi/2) - \frac{1}{3\xi}
 - \ln \frac{4\pi}{\xi} 
 + \xi + \xi \ln \frac{2}{\xi}.
\label{hehe-xi}
\ee
The following simple fitting formulae can be used for all
values of $\beta$: 
\ba
&&
 \hat{a} \approx - \frac{2\alpha}{9\pi} \ln\bigg(
 1 + \frac{\beta^2}{5}\,\frac{ 1+0.25487\,\beta^{3/4}  }{
  1+0.75\,\beta^{5/4}}\bigg),
\label{fit-a}
\\&&
 q \approx \frac{7\alpha}{45\pi}\,\beta^2\,\frac{
 1 + 1.2\,\beta }{ 1 + 1.33\,\beta + 0.56\,\beta^2 },
\label{fit-q}
\\&&
 m \approx - \frac{\alpha}{3\pi} \, \frac{ \beta^2 }{
 3.75 + 2.7\,\beta^{5/4} + \beta^2 }.
\label{fit-m}
\ea
Equations (\ref{fit-a})--(\ref{fit-m}) exactly recover the weak-field
limits (\ref{vac-weak}) and the leading terms in 
the high-field ($b\gg1$) expansions 
(Eqs.\ [2.15]--[2.17] of Ho \& Lai (2003).
The maximum errors are 1.1\% at $\beta=0.07$ for (\ref{fit-a}), 2.3\% at 
$\beta=0.4$ for (\ref{fit-q}), and 4.2\% at $\beta=0.3$ for (\ref{fit-m}).
The derivatives of ${\hat a},~q,~m$ with respect to $\beta$ have similar
accuracies.

\subsection{Photon Modes in Cold Plasma with Vacuum Polarization}
\label{subsec:modevac}

Including vacuum polarization, the dielectric tensor
$\beps$ and inverse permeability tensor $\bmu^{-1}\equiv \bar\bmu$ 
can be written as $\beps=\beps^{(p)}+\Delta\beps^{(v)}$,
$\bar\bmu=\bI+\Delta\bar\bmu^{(v)}$ (where $\bI$ is the unit tensor).
Thus the dielectric tensor of the combined plasma+vacuum medium
still be written in the form of Eq.~(\ref{eq:epsij0}), except 
$\varepsilon\rightarrow \varepsilon'=\varepsilon+\hat a$,
$\eta\rightarrow\eta'=\eta+{\hat a}+q$.

An electromagnetic (EM) wave with a given frequency $\omega$ satisfies the
wave equation
\be
\nabla\times(\bar\bmu\cdot\nabla\times\bE)={\omega^2\over c^2}\,\beps\cdot
\bE.
\label{waveeq}\ee
For normal modes propagating along the $z$-axis, with
$\bE\propto e^{ik_\pm z}$, this reduces to
\be
-k_\pm^2\hat z\times\left[\bar\bmu\cdot (\hat z\times \bE_{\pm})
\right]={\omega^2\over c^2}\,\beps\cdot\bE_\pm,
\label{eqmode}\ee
where the subscripts ``$\pm$'' specify the two modes 
(``plus-mode'' and ``minus-mode''). 
In the $xyz$ coordinates with ${\bf k}$ along the $z$-axis and 
$\bB$ in the $x$-$z$ plane 
(such that $\hat\bB\times {\hatk}=\sin\theta_B\,\hat y$,
where $\theta_B$ is the angle between ${\bf k}$ and $\bB$), we write
the electric field of the mode as $\bE_\pm \propto (iK_\pm,1,iK_{z\pm})$, 
where the ellipticity $K_\pm=-iE_x/E_y$ is given by
\be
K_{\pm}=\beta_p\pm\sqrt{\beta_p^2+r},
\label{eqkpm}
\ee
with $r=1+(m/a)\sin^2\theta_B\simeq 1$ (where $a=1+{\hat a}$) and  
the (complex) polarization parameter $\beta_p$ is 
\be
\beta_p = -\frac{\varepsilon'^2-g^2-\varepsilon'\eta'
\lp 1+m /a\rp}{2g\eta'}\frac{\sin^2\theta_B}{\cos\theta_B}.
\label{eq:polarb2}
\ee
The  refractive index $n_\pm=ck_\pm/\omega$ is given by
\be
n_\pm^2 = \frac{g\eta'}{a\epsilon_{33}}\lp\frac{\varepsilon'}{g}
 + \frac{1}{K_\pm}\cos\theta_B\rp, \label{eq:nrefract2}
\ee
where $\epsilon_{33}=\varepsilon'\sin^2\theta_B+\eta'\cos^2\theta_B$.

The polarization parameter $\beta_p$ directly determines the
characteristics of photon normal modes in the medium. For $v_e\ll 1$,
the real part of $\beta_p$ can be written as ${\rm Re}(\beta_p)
=\beta_0\beta_V$, where 
\be
\beta_0\simeq \frac{u_e^{1/2}\sin^2\theta_B}{2\cos\theta_B}
\lp 1-u_i\rp, \label{eq:polarb0}
\ee
and 
\be
\beta_V\simeq 1+\frac{\lp q+m\rp\lp 1-u_e\rp}{u_e v_e}.
 \label{eq:polarbvp0}
\ee
In general, the modes are elliptically polarized, and the ellipticity
depends on $\beta_p$.

A special case is the magnetized vacuum with zero plasma density.
The extraordinary mode (also called $\perp$-mode) and
O-mode (also called $\parallel$-mode) are simply 
\be
\bE_\perp=(0,1,0),\quad 
\bE_\parallel \propto (1,0,K_z)
\quad {\rm with}~~K_z=\frac{q\sin\theta_{B}\cos\theta_{B}}
{a+q\cos^2\theta_{B}}.
\ee
The indices of refraction are
\be
n_\perp=\left({a\over {a+m\,\sin^2\theta_{B}}}\right)^{1/2},\quad
n_\parallel=\left({a+q\over{a+q\cos^2\theta_{B}}}\right)^{1/2}.
\label{eq:nparallel}
\ee
Clearly, acting by itself, the birefringence from vacuum polarization is 
significant (with the index of refraction differing from unity by more 
than $10\%$) for $B\go 300 B_Q$ --- for such superstrong field strengths,
one can expect appreciable magnetic lensing effect from photons emitted
near the neutron star surface (e.g., Shaviv, Heyl \& Lithwick 1999).

\begin{figure} 
\includegraphics[width=35pc]{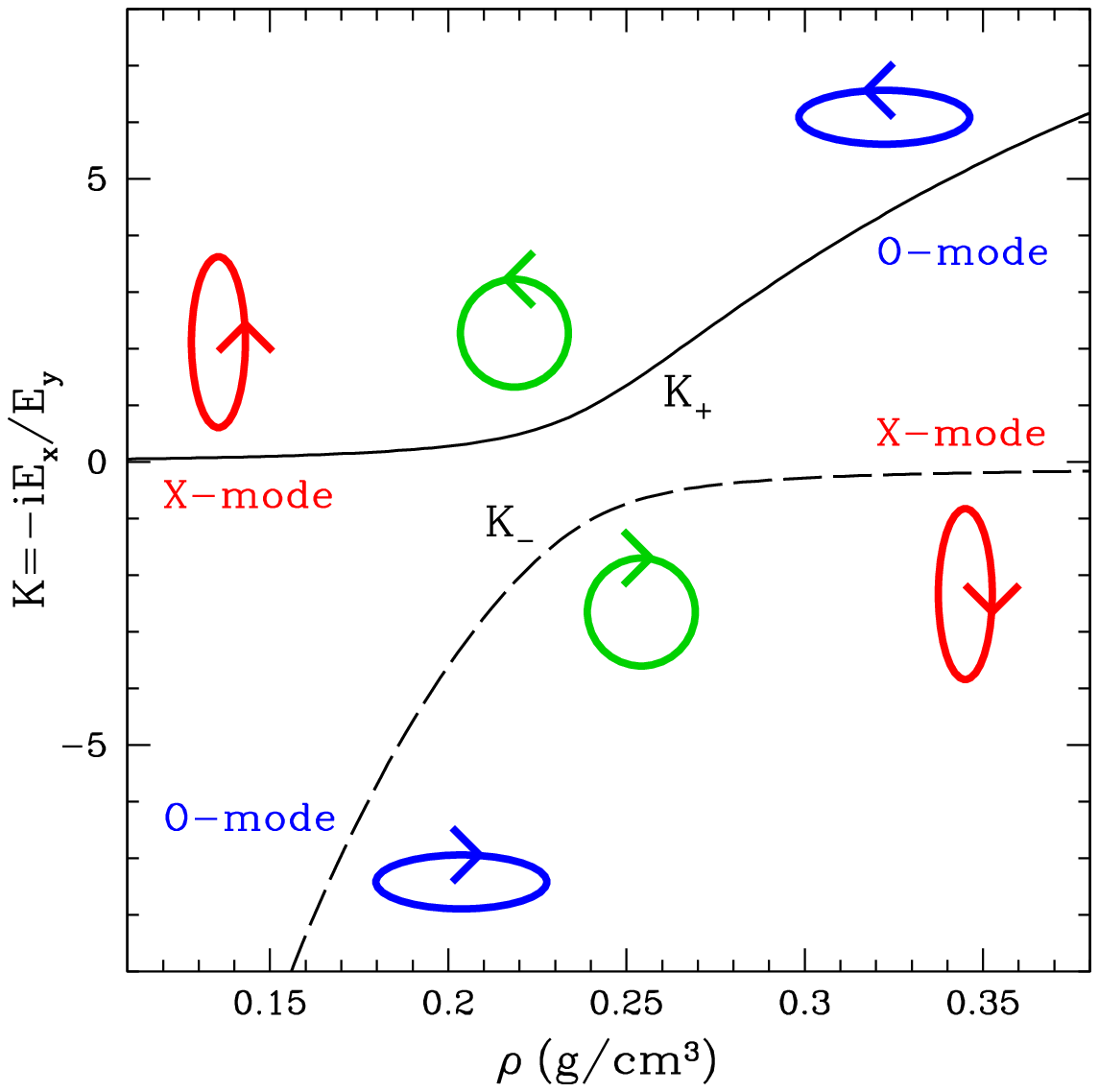}
\vskip -2.0cm
\caption{
The polarization ellipticity of the photon mode  
as a function of plasma density near the vacuum resonance. 
The two curves correspond to the two different modes.
In this example, the parameters are $B=10^{13}\,$G, 
$\epsilon=5\,$keV, $Y_e=1$, and $\theta_{kB}=45^\circ$. 
The ellipticity of a mode is specified by the ratio $K=-iE_x/E_y$, 
where $E_x$ ($E_y$) is the photon's electric field component 
along (perpendicular to) the $\bk$-$\bB$ plane. 
At densities away from the resonance density $\rho_V$, the two modes
are almost linearly polarized (with polarization ellipses orthogonal to
each other): The O-mode is characterized by $|K|\gg 1$, and the 
X-mode $|K|\ll 1$. At $\rho=\rho_V$, both modes are circularly polarized.
In the adiabatic limit, an O-mode (X-mode) photon
from the high-density region will convert to
the X-mode (O-mode) as it traverses the vacuum resonance density
$\rho_V$, with its polarization ellipse rotated by $90^\circ$.}
\label{fig:polarconv}
\end{figure}

Now consider the case of finite plasma density.
For concreteness, let us consider the regime $u_e\gg 1$, i.e., 
the photon energy $\epsilon \ll E_{ce}$. This is relevant for
propagation of thermal photons in neutron star atmospheres. 
Clearly, for generic values $\epsilon$ and $\theta_B$, $|\beta|\gg 1$, 
the two modes are almost linearly polarized:
the extraordinary mode has $|K|\ll 1$, and its $\bE$ is mostly
perpendicular to the ${\hat k}$-$\hat\bB$ plane;
the ordinary mode has $|K|\gg 1$, and is polarized
along the ${\hat k}$-$\hat\bB$ plane. This distinction of the two
modes manifests most significantly when we consider how 
they interact with matter: the ordinary-mode opacity is largely 
unaffected by the magnetic field, while the extraordinary-mode 
opacity is significantly reduced (by a factor
of order $\omega^2/\omega_{ce}^2)$ from the zero-field value.
However, this distinction becomes ambiguous when $|\beta|\lo 1$
($|K|\sim 1$). Obviously, $\beta=0$ for $\omega=\omega_{ci}$. 
But even for general energies ($\epsilon\neq \hbar\omega_{ci}$), a photon 
traveling in an inhomogeneous medium 
encounters $\beta=0$ when the condition $v_e=q+m$ is satisfied.
This is the {\it vacuum resonance} (see Gnedin et al.~1978).
Using $\hbar\omega_{pe}= \hbar (4\pi n_e e^2/m_e)^{1/2}=
28.71\,(Y_e\rho_1)^{1/2}$~eV, where $Y_e$ is the electron fraction of the
gas, and $\rho_1$ is the density in unity of 1~g~cm$^{-3}$, we find that 
for a given photon energy, vacuum resonance occurs at the density 
(Lai \& Ho 2002)
\be
\rho_V=0.964\,Y_e^{-1}(B_{14}\epsilon_1)^2 f^{-2}\,{\rm g~cm}^{-3},
\label{eqrhores}\ee
where $\epsilon_1=\epsilon/(1~{\rm keV})$ and
\be
f=\left({\alpha\beta^2/15\pi\over q+m}\right)^{1/2}
\ee
is a slow-varying function of $\beta$ [$f=1$ for $\beta\ll 1$ and
$f\rightarrow (\beta/5)^{1/2}=0.673\,B_{14}^{1/2}$
for $\beta\gg 1$; $f$ varies from $0.99$ at $B_{14}=1$ to $6.7$ at
$B_{14}=100$]. Qualitatively, the meaning of the vacuum resonance is
the following (see Fig.~\ref{fig:polarconv}):
For $\rho>\rho_V$, the plasma effect dominates the dielectric
tensor, while for $\rho<\rho_V$, vacuum polarization dominates.
Away from the resonance, the photon modes (for $\epsilon\ll E_{ce}$)
are almost linearly polarized as discussed above.
Near $\rho=\rho_V$, however, the normal modes become circularly 
polarized as a result of the ``cancellation'' of the plasma and 
vacuum effects --- both effects tend to make the mode linearly polarized, 
but in mutually orthogonal directions. 

Including the damping terms in the dielectric tensor gives rise to
the phenomenon of mode collapse (see Soffel \etal~1983; Lai \& Ho 2003a).
This occurs when the two polarization modes become
identical, i.e., $K_+=K_-$ or $n_+=n_-$. Obviously, at the point of
mode collapse, the modal description of radiative transfer 
breaks down and a rigorous treatment of radiative transfer
requires solving the transport equations for the four photon 
intensity matrix, or Stokes parameters (see Lai \& Ho 2003a).

\subsection{Wave Propagation in Inhomogeneous Medium}
\label{subsec:inhom}

Because of the strong gravity, a neutron star atmosphere has 
a rather large density gradient, with density scale height
of order a few cm's for $T=10^6$~K. When a photon propagates in such an 
inhomogeneous medium, its polarization state will evolve adiabatically 
(i.e. following the $K_+$ or $K_-$ curve in Fig.~\ref{fig:polarconv})
if the density variation is sufficiently gentle and the modes
are sufficiently distinct (i.e., $n_1-n_2$ is sufficiently large).
Away from the vacuum resonance, the adiabatic condition is 
easily satisfied. At the vacuum resonance, adiabaticity requires
(Lai \& Ho 2002)
\be
\epsilon\go \epsilon_{\rm ad}(B,\theta_B,H_\rho)
=1.49\,\bigl(f\,\tan\theta_B |1-u_i|\bigr)^{2/3}
\left({5\,{\rm cm}\over H_\rho}\right)^{1/3}~{\rm keV},
\label{condition}\ee
where $H_\rho=|dz/d\ln\rho|$ is the density scale
height (evaluated at $\rho=\rho_V$) along the ray.
For an ionized Hydrogen atmosphere, $H_\rho\simeq 2kT/(m_pg\cos\theta)
=1.65\,T_6/(g_{14}\cos\theta)$~cm, where $T=10^6\,T_6$~K is the temperature,
$g=10^{14}g_{14}$~cm~s$^{-2}$ is the gravitational acceleration, and $\theta$
is the angle between the ray and the surface normal.
In general, the probability for a nonadiabatic ``jump'' is 
\be
P_{\rm jump}=\exp\left[-(\pi/2)(\epsilon/\epsilon_{\rm ad})^3\right].
\ee
Thus for $\epsilon \go 1.3\epsilon_{\rm ad}$, the polarization evolution is 
highly adiabatic (with $P_{\rm jump}\lo 0.03$) even at the vacuum
resonance. In this case, an X-mode (O-mode) photon will be converted
into a O-mode (X-mode) as it traverses the vacuum resonance.
This resonant mode conversion is analogous to the MSW effect of neutrino
oscillation (e.g., Haxton 1995). One could also say that 
in the adiabatic limit, the photon will remain in the same plus or 
minus branch, but the character of the mode is changed across the 
vacuum resonance. Indeed, in the literature on radio wave propagation in 
plasmas (e.g., Budden 1961; Zheleznyakov et al.~1983), the nonadiabatic 
case, in which the photon state jumps across the continuous curves, 
is referred to as ``linear mode coupling''.
It is important to note that the ``mode conversion'' effect discussed here
is not a matter of semantics. The key point is that in the adiabatic limit, 
the photon polarization ellipse changes its orientation across the vacuum
resonance, and therefore the photon opacity changes significantly
(see Lai \& Ho 2003a).

The concept of adiabatic evolution of photon polarization is also
important when we consider wave/photon propagation in neutron star
magnetospheres. For example, the normal modes of
X-ray photons in the magnetosphere are determined by vacuum
polarization (since $\rho\ll \rho_V$). As a photon (of a given mode) 
propagates from the polar cap through the magnetosphere, its 
polarization state evolves adiabatically following the varying magnetic 
field it experiences, up to the ``polarization-limiting radius'' 
$r_{\rm pl}$, beyond which the polarization state is frozen.
This polarization-limiting radius is located where the adiabatic
condition breaks down (see Heyl et al.~2003; Lai \& Ho 2003b).
Another example concerns the radio wave propagation in the
magnetosphere, whose dielectric property is dominated by
highly relativistic pair plasmas. Such a propagation effect may be 
responsible for some of the observed
polarization changes in radio emission of pulsars
(e.g., Cheng \& Ruderman 1979; Barnard 1986; Melrose
\& Luo 2004).

\subsection{Warm Plasma and Relativistic Plasma}

While the dielectric tensor for cold plasma 
was derived classically, the quantum calculation, incorporating the 
quantized nature of electron motion transverse to the magnetic field, yields
the same result when the photon wavenumber $k\rightarrow 0$ 
(e.g., Canuto \& Ventura 1972; Pavlov et al.~1980). More precisely,
since the size of the Landau wavefunction is of order 
$\hat\rho=(\hbar c/eB)^{1/2}$ [see Eq.~(\ref{eq:hatrho})], this requires
$k_\perp \hat\rho\ll 1$ (where $k_\perp$ is the wave number perpendicular 
to $\bB$, or the photon energy must satisfy
$\epsilon\sin\theta_B\ll \sqrt{m_e c^2E_{ce}}$.

The cold plasma approximation is valid when the thermal
velocity of the electron, $V_T=(2k_BT/m_e)^{1/2}$, 
is much less than the phase velocity of the wave ($\omega/k\simeq c$), 
and $\omega$ is not too close to the cyclotron frequency, i.e., 
for $|\omega-\omega_{ce}|\gg (\omega/c) V_T$
(we neglect the ion component here). The cold plasma approximation
also neglects electron recoil.
Taking into account the electron motion, the cyclotron resonance 
denominator $\omega-\omega_{ce}$ in the cold
plasma dielectric should be replaced by $\omega-k_\parallel V_\parallel-
\omega_{ce}$ (where $k_\parallel$ and $V_\parallel$ are the wavenumber
and electron velocity parallel to the magnetic field), and the electron
recoil can also be included by $\omega\rightarrow \omega\pm
\hbar k_\parallel^2/(2m_e)$. Then the
dielectric tensor should be averaged over the electron velocity
distribution function. In the very strong magnetic field limit 
where electrons occupy only the lowest Landau level ($k_BT\ll E_{ce}$), 
the cold plasma expressions (\ref{eq:cold}) are replaced by (Kirk 1980)
\ba
&&\varepsilon=1-v_e\left\{1-{\omega_{ce}\over 2V_T k_\parallel}
\Bigl [W(y_-)-W(y_+)\Bigr]\right\},\\
&&g=-v_e {\omega_{ce}\over 2V_T k_\parallel}
\Bigl [W(y_-)+W(y_+)\Bigr],\label{eq:epst}\\
&&\eta=1-v_e,
\label{eq:etat}\ea
where 
\be
y_\pm={\omega\pm\omega_{ce}\over V_T k_\parallel}\pm 
{\hbar k_\parallel\over 2m_eV_T},
\ee
[the $\hbar k_\perp/(2m_eV_T)$ term arises from electron recoil],
and $W(z)$ is the plasma dispersion function
\be
W(z)={1\over\sqrt{\pi}}\int_{-\infty}^\infty\! {e^{-x^2}\,dx\over x-z},
\ee
which should be evaluated for $\omega_{ce}\rightarrow 
\omega_{ce}/(1+i\gamma_{re})$ with $\gamma_{re}>0$ 
[see Eq.~(\ref{eq:epsilong})].
Expressions (\ref{eq:epst})-(\ref{eq:etat}) are 
valid in the nonrelativistic limit, 
$\hbar\omega\ll m_ec^2$ and $k_BT\ll m_ec^2$.
One can check that for $|\omega-\omega_{ce}-\hbar k_\parallel^2/(2m_e)|\gg
|V_T k_\parallel|$, the cold plasma limit is recovered.
The net effect of finite temperature is that 
the cyclotron resonance is softened from the sharp peak 
of the cold plasma theory to a broadened one with width $\Delta\omega
\sim k_\parallel V_T$. Also, the electron recoil shifts the resonance by 
$\Delta\epsilon \simeq (E_{ce}\cos\theta_B)^2/(2m_ec^2)$, which 
becomes appreciable as $B$ approaches $B_Q$. Thus a quantitative description
of the cyclotron resonance for $B\go B_Q$ requires relativistic theory
(see \S\ref{sec:RadProc}).

Similar qualitative behavior appears in the relativistic theory
(e.g., Svetozarova \& Tsytovich (1962); Melrose 1974; Pavlov et al.~1980; Nagel 1981;
Meszaros \& Nagel 1985; see Meszaros 1992 for a review).
The polarization tensor is directly proportional to the 
forward-scattering amplitude of a photon in the plasma. Quantum
states and momentum distribution of electrons can be summed or averaged.
The advantage of such an approach is that the photon polarization modes
are calculated in a self-consistent manner as
the differential scattering cross section, which depends 
on photon modes (see \S\ref{sec:RadProc}).

The magnetosphere of a radio pulsar is believed to be filled
with ultra-relativistic electron-positron pairs (see \S \ref{sec:pulsars}). 
In the polar-cap
region, outwardly streaming pair plasmas are produced in an 
electromagnetic cascade (Sturrock 1971), with bulk Lorentz factor 
$\gamma\go 10^2$ (e.g., Daugherty \& Harding 1982; Zhang \& Harding 2000;
Hibschman \& Arons 2001).
While the polarization state
of high-energy emission is determined by the vacuum polarization effect,
radio emission generated in the inner magnetosphere can be affected 
when propagating through such a plasma 
(e.g., Cheng \& Ruderman 1979; Barnard 1986).
For example, the observed characteristics of circular polarization
(Radhakrishnan \& Rankin 1990; Han et al.~1998) are thought to develop 
as a propagation effect (e.g., Petrova \& Lyubarskii 2000;
Melrose \& Luo 2004; Petrova 2006). The most important region is associated
with the cyclotron resonance, where $\gamma\omega (1-\beta\cos\theta_B)
=\omega_{ce}$. Dispersion in a pulsar plasma has been
studied extensively, under various assumptions about the 
relative electron-positron densities and their momentum distribution,
which are uncertain (e.g., Arons \& Barnard 1986; Lyutikov 1998; 
Melrose et al. 1999; Asseo \& Riazuelo 2000). A simple model
is based on applying a Lorentz transformation to the magneto-ionic
(cold plasma) theory (Melrose \& Stoneham 1977; Melrose \& Luo 2004),
which ignores the random motions of the particles.
Such a model has been used to interpret the circular polarization in
pulsar radio emission (Melrose \& Luo 2004).
The effect of vacuum resonance in the magnetosphere plasma of pulsars
and magnetars is studied by Wang \& Lai (2006).

\section{Radiative Processes for Free Electrons} \label{sec:RadProc}

At low magnetic field strengths, many of the radiative processes we
discuss below, including cyclotron absorption and emission and Compton
scattering, may be described by classical physics.  In neutron star
magnetic fields approaching the critical field, a relativistic quantum
description is required to accurately compute the rates and the photon
spectrum of radiation, even for non-relativistic electrons.  For these
processes, we will first review the classical descriptions before
discussing the quantum description and discuss the regimes in which
each is appropriate.  Other processes we will discuss, including
one-photon pair production and annihilation, photon splitting and
bound pair creation, take place only in strong fields since they do
not conserve energy and momentum in free space.  In a magnetic field,
perpendicular momentum of particles is not conserved in transitions
between Landau states since the field effectively is able to supply or
absorb momentum. The momentum conservation along the field couples directly 
to the translational invariance of the system (see Section 2) parallel 
to B. The conservation equations for energy and parallel
momentum for transitions of electrons or positrons between initial
state $(n,s,p)$ and final state $(n',s',p')$, where $s = (\sigma_z +
1)/2$ is the electron spin state, 
$p=c p_z$ and $n$ labels the Landau level [see Eq.~(\ref{eqrel})],
read
\begin{equation} \label{eq:Econ1}
E_n - E_{n'} = \pm \epsilon
\end{equation}
\begin{equation} \label{eq:pcon1}
p' = p \mp \epsilon\cos\theta
\end{equation}
resulting in emission (upper sign) or absorption (lower sign) 
of photons at angle $\theta$ to
the magnetic field with energy $\epsilon$.  [In this Section, in
contrast to previous Sections, we will express all photon and particle
energies in units of $m_e c^2$.]  The photon energy, as determined
from the above kinematic equations, is 
\begin{equation} 
\epsilon =\pm\,\, {(E_n - p\cos\theta) - [(E_n - p\cos\theta)^2 - 2B'(n-n')
\sin^2\theta]^{1/2}\over \sin^2\theta},
\label{eq:w1}
\end{equation}
where in this section, we use $B' \equiv B/B_Q$ as the dimensionless
magnetic field parameter (equivalent to the $\beta$ parameter in
section \ref{Sec:photons}). Unless the photon is emitted or absorbed with
angle $\theta = 90^0$, the electron will experience a recoil along the
field direction, given by Eqn~(\ref{eq:pcon1}), which is needed to
determine its final energy.  The electron wavefunctions that have been
adopted in the literature for studying the processes presented in the
following subsections were discussed in \S \ref{sec:elec}.

\subsection{Cyclotron Absorption} \label{sec:CycAbs}

Cyclotron absorption, the inverse of cyclotron emission (see \S
\ref{sec:CycRad}), is a first-order process in which a photon excites
a particle to a higher Landau state.  The classical, non-relativistic
cross section for absorption of a photon by an electron at rest in the 
ground state (n=0) (or in
the rest frame of a moving electron) was derived by Blandford \&
Scharlemann (1976)
\be 
\sigma^N_{\rm abs}(\parallel, \perp)
={2\alpha\pi^2\lambar^2}\, \delta(\epsilon - NB')\left({N^2\over 2}
B'\sin^2\theta \right)^{N-1}\,{(\cos^2\theta, 1)\over (N-1)!},  
\label{eq:cycabs_cl} 
\ee
where $N$ is the harmonic number, $\alpha=1/137$, $\lambar=\hbar/(m_ec)$,
$\theta$ is the angle of photon
propagation to the magnetic field and the quantity in parentheses
refers to the two incident photon linear polarizations $(\parallel,
\perp)$ or (O,X) modes, with electric vectors parallel and perpendicular to the plane
containing the photon wavevector and the magnetic field.  The
$\delta$-function restricts the energy of the absorbed photon to be a
harmonic, $\epsilon_N = NB'$, of the cyclotron energy 
$\epsilon_B = E_{ce}/m_e c^2 =B'$.

The relativistic QED cyclotron absorption cross section was first
derived by Daugherty \& Ventura (1978) for electrons initially in the
ground state.  The required energy for excitation to state $N$ now
follows from the relativistic kinematic equations (\ref{eq:Econ1}) and
(\ref{eq:pcon1}), setting $E_n =1$, $n=0, n'=N$ and $p = 0$,

\be  \label{eq:eps_abs}
\epsilon _N  = [(1 + 2NB'\sin ^2 \theta )^{1/2}  - 1]/\sin ^2 \theta 
\ee
for a photon propagating at angle $\theta$ to the field.  Because the
electron experiences a recoil of $p' = \epsilon_N\cos\theta$ on
absorption, the cyclotron harmonics are actually anharmonic in high
magnetic fields, so that the energy difference between successive
harmonics will decrease.  The cyclotron absorption rate, summed over
final spin states of the electron, is (Harding \& Daugherty 1991)
\begin{equation}
\label{eq:cycabs_rel}
\sigma_{\rm abs}^N(\parallel, \perp) = {2\alpha \pi^2 \lambar^2 \over E_N}\,
\delta(\epsilon_N - \epsilon)\,{e^{-Z}
Z^{N-1} \over (N-1)!}\,\left[(\cos^2\theta, 1) + {Z\over N}(\sin^2\theta, 0)\right],
\end{equation}
where
\begin{equation}
Z = {\epsilon^2\sin^2\theta\over 2B'},
\end{equation}
The $\delta$-function requires that the absorbed 
photon energy have only the value given by equation (\ref{eq:eps_abs}) and effectively makes $\sigma_{\rm abs}^N$
only a function of the incident photon angle $\theta$.
The above expression reduces to the classical limit of equation
(\ref{eq:cycabs_cl}) when $NB' << 1$, where $\epsilon \approx NB'$, $Z
\approx N^2B'\sin^2\theta/2 \ll 1$ and $E_N \approx 1$.  Therefore,
the relativistic formula (\ref{eq:cycabs_rel}) should be used when
$NB'\gsim 0.1$, that is, when electron recoil becomes significant,
which is the case in fields above $\sim 10^{12}$ Gauss.

\subsection{Cyclotron and Synchrotron Radiation} \label{sec:CycRad}

Cyclotron radiation is the inverse of the cyclotron absorption process
discussed above, and results from downward transitions between Landau
levels.  The classical cyclotron radiation formula was first presented
by Schott (1912) and assumes a continuous circular orbit of the
particle in a magnetic field with negligible energy loss.  The
emission is characterized by the particle velocity $\beta$ and pitch angle, 
$\psi = \tan^{-1}(\beta_{\perp}/\beta_{\parallel})$, 
where $\beta_{\parallel}$ and $\beta_{\perp}$ are the velocity
parallel and perpendicular to the field direction.  If the particle
energy is above the cyclotron energy, the radiation is the sum over a
number of harmonics of the fundamental cyclotron energy.  The radiated
power at frequency $\nu$ and angle $\theta$ to the magnetic field is
given by (Sokolov \& Ternov 1986) \be W = {e^2 \omega^2\over
c}\,\sum_{\nu=1}^\infty \nu^2\, \int_0^\pi {\sin\theta d\theta\over
(1-\beta_{\parallel}
\cos\theta)^3}{\left\{\left({\cos\theta-\beta_{\parallel}\over
\sin\theta} \right)^2\, J_{\nu}^2(\xi) + \beta_{\perp}^2 {J'_{\nu}}^2
(\xi) \right\}} \ee 
where $\omega = eB/(\gamma m_ec)$ is the particle
gyrofrequency and 
\be \xi = {\nu\beta_{\perp}\sin\theta \over 1
-\beta_{\parallel}\cos\theta}.  \ee 
The polarized form of equation (77), in the case where $\beta_{\parallel}=0$, 
is given in Sokolov and Ternov (1986). The spectrum, given by Bekefi
(1966), Canuto and Ventura (1977) and Brainerd and Lamb (1987),
display characteristic harmonics at low energy.  When the particle
energy is relativistic and the emission is dominated by high
harmonics, it is called synchrotron radiation.  
The formula becomes somewhat simpler and the emission power per unit
photon energy is (e.g. Jackson 1975) 
\be 
I_C(\epsilon) = {\sqrt{3}\,\alpha\omega_B\sin\psi
\over 2\pi}\,\kappa\left({2\epsilon \over
3\gamma\Upsilon\sin\psi}\right),
\label{eq:IC} 
\ee 
where $\omega_B =\omega_{ce}= eB/(m_ec)$ is the cyclotron
frequency, $\Upsilon \equiv \gamma B'$, $\gamma$ is the electron
energy and the $\kappa$ function is \be \kappa(y) = y\,\int_y^{\infty}
K_{5/3}(x)\,dx.  \ee
The characteristic energy of a synchrotron photon is $\epsilon\sim
\gamma^2 B'\sin\psi$.
 
\begin{figure} 
\includegraphics[width=15cm]{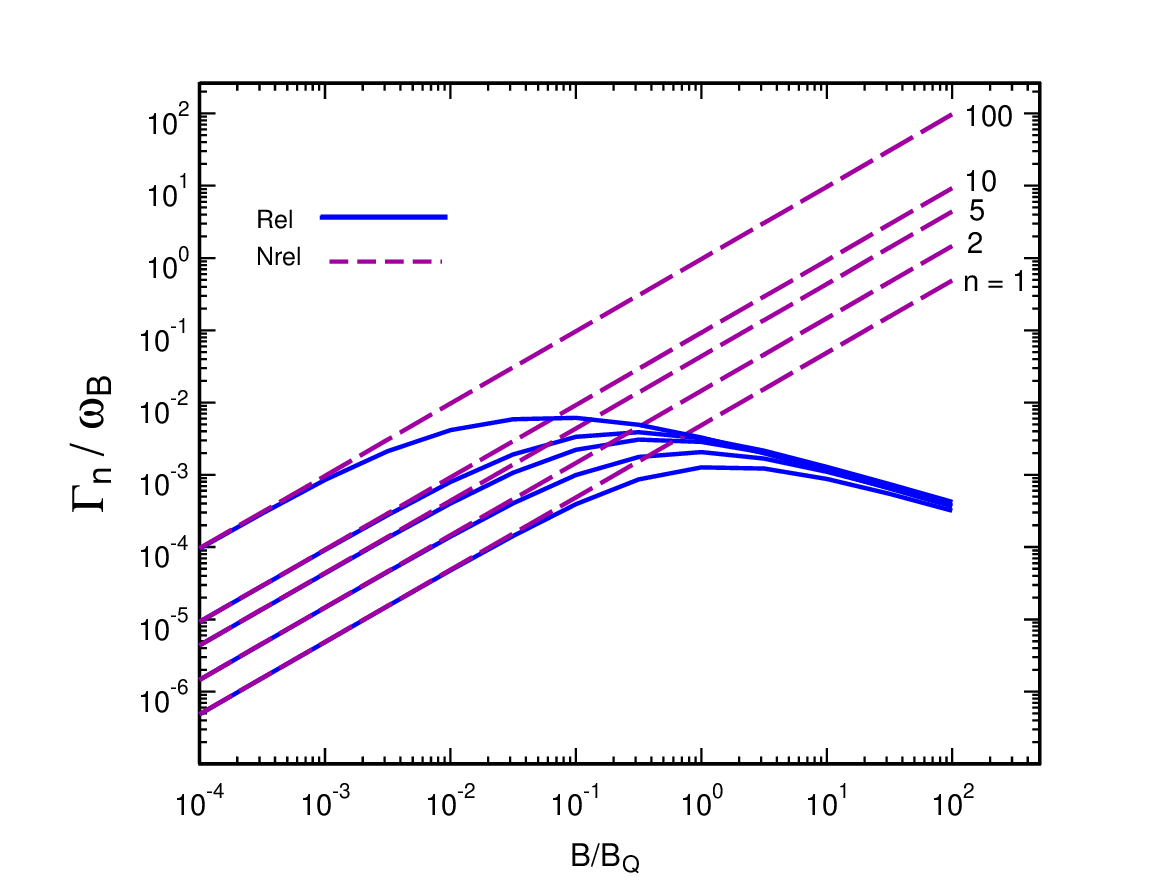}
\caption{Spin-averaged cyclotron decay rates (solid lines), in units of the cyclotron frequency $\omega_B$, 
from Landau state $n$ as a function of magnetic field 
strength $B$ in units of the critical field $B_Q$.  Also plotted are the corresponding non-relativistic
decay rates (dashed lines).}
\label{fig:CycRates}
\end{figure}

In high fields approaching the critical field, 
the classical description should be replaced with a relativistic
quantum description to accurately compute the radiative rates and
photon spectrum, even for non-relativistic electrons.  The classical 
value of the critical radiation frequency, $\gamma^2\epsilon_B$, exceeds
the electron kinetic energy, $(\gamma - 1)$ when
\begin{equation} \label{eq:ske}
B'{\gamma^2\over (\gamma-1)} > 1.
\end{equation}
The classical emissivity therefore violates conservation of energy.
The classical formula also overestimates the spectral emissivity when
\begin{equation} \label{eq:sel}
\Upsilon\sin\psi \equiv \gamma B'\sin\psi > 0.1.
\end{equation}
There is a consequent reduction of the energy loss rate for large values
of $\Upsilon$ (Erber 1966).  Finally, taking into account electron recoil gives harmonic energies
which are not simple multiples of the cyclotron frequency, as in the 
non-relativistic case.  The polarization and spin dependent transition rates for quantum synchrotron
radiation are given by Sokolov \& Ternov (1968, see also Harding \& Preece 1987) and should be used for
transitions between low Landau states in field $B' \gsim 0.1$.  The exact QED spectrum for transition
from state $n$ to the ground state $n' = 0$, averaged over electron spin, is (Latal 1986, Baring et al. 2005)
\begin{eqnarray}
{d\Gamma_{n0}\over d\epsilon} = & {\alpha c\over \lambar}{1\over B'\varepsilon_n^2}{n^{n-1}e^n\over 
\Gamma(n+1)}\left({\epsilon\over \epsilon_-} - 1\right)^{n-1} \nonumber \\
& {(\varepsilon_n - \epsilon) nB' - \epsilon\over \sqrt{(\varepsilon_n-1-\epsilon)(\varepsilon_n+1-\epsilon)}}
\exp\left({-n{\epsilon\over \epsilon_-}}\right) \\ \nonumber
\end{eqnarray}
where $\varepsilon_n \equiv \sqrt{1+2nB'}$ and $\epsilon_- \equiv nB'/\varepsilon_n$.
The rate is more complicated for transitions between higher Landau states, involving Laguerre polynomials.
The cyclotron decay rate from state $n$ is found by integrating over the radiated photon energy in each
transition, $\Gamma_{nn'}$ and then summing over the final states $n'$:
\be \label{eq:Gamma_n}
\Gamma_{n} = \sum_{n'=0}^{n-1} \Gamma_{nn'}.
\ee
The spin-averaged decay rates for some selected states, as a function of field strength, are shown in Figure 
\ref{fig:CycRates} and compared to the classical (non-relativistic) values (Herold et al. 1982).
One can see that the QED decay rate departs from the classical rate for $B'n \gsim 0.1$.  The spin-dependent
decay rates have been discussed by Daugherty \& Ventura (1978) and by Melrose \& Zheleznyakov (1981) for the
non-relativistic case, and by Herold et al. (1982) for the relativistic case.  Generally, the probability
for spin-flip transitions, those in which the electron changes its spin state, is lower than for non 
spin-flip transitions.  The ratio of the spin-flip decay rate to the non-spin-flip decay rate is 
$\propto B'^{-1}$ for the $n = 1 \rightarrow 0$ transition.

An asymptotic form of the spectrum, averaged over
spin and polarization, valid for relativistic elections, transitions between high Landau states ($n, n' \gg 1$)
and zero longitudinal momentum ($\psi = 90^\circ$), is (Sokolov \& Ternov 1968)
\be  \label{eq:IQ}
I_Q(\epsilon)  = {\sqrt{3}\alpha\omega_B\over 2\pi}\,(1-{\epsilon\over \gamma})\,
\left[\kappa(y) + y^3({3\over 2}\Upsilon)^2(1-{\epsilon\over \gamma})K_{2/3}(y)\right]
\end{equation}
where 
\begin{equation}
y = {2\epsilon\over 3\gamma\Upsilon[1-(\epsilon/\gamma)]}
\ee  
One can see that the above expression incorporates a kinematic cutoff in the spectrum at $\epsilon = \gamma$ 
that avoids energy violation.
When $\epsilon \ll \gamma$ and $B' \ll 1$, formula (\ref{eq:IQ}) reduces to the classical formula (\ref{eq:IC}).

There are several notable features of cyclotron/synchrotron radiation in high magnetic fields.  Classically
and at low field strengths, rates are highest for single harmonic number transitions and decrease monotonically
with increasing harmonic number, with transitions from high Landau states to the ground state being very
improbable.  But for field strengths $B' \gsim 0.2$, transitions from high to low states become more
probable and transitions to the ground state can actually dominate over even single harmonic number transitions
(White 1974, Harding \& Preece 1987).  The result is that electrons in very high magnetic fields radiate
energy not in small steps but often in one large transition, emitting a photon equal to its kinetic energy.  The spectrum displays an enhancement just before the kinematic cutoff, as shown in Figure \ref{fig:SynQED}.

\begin{figure} 
\includegraphics[width=15cm]{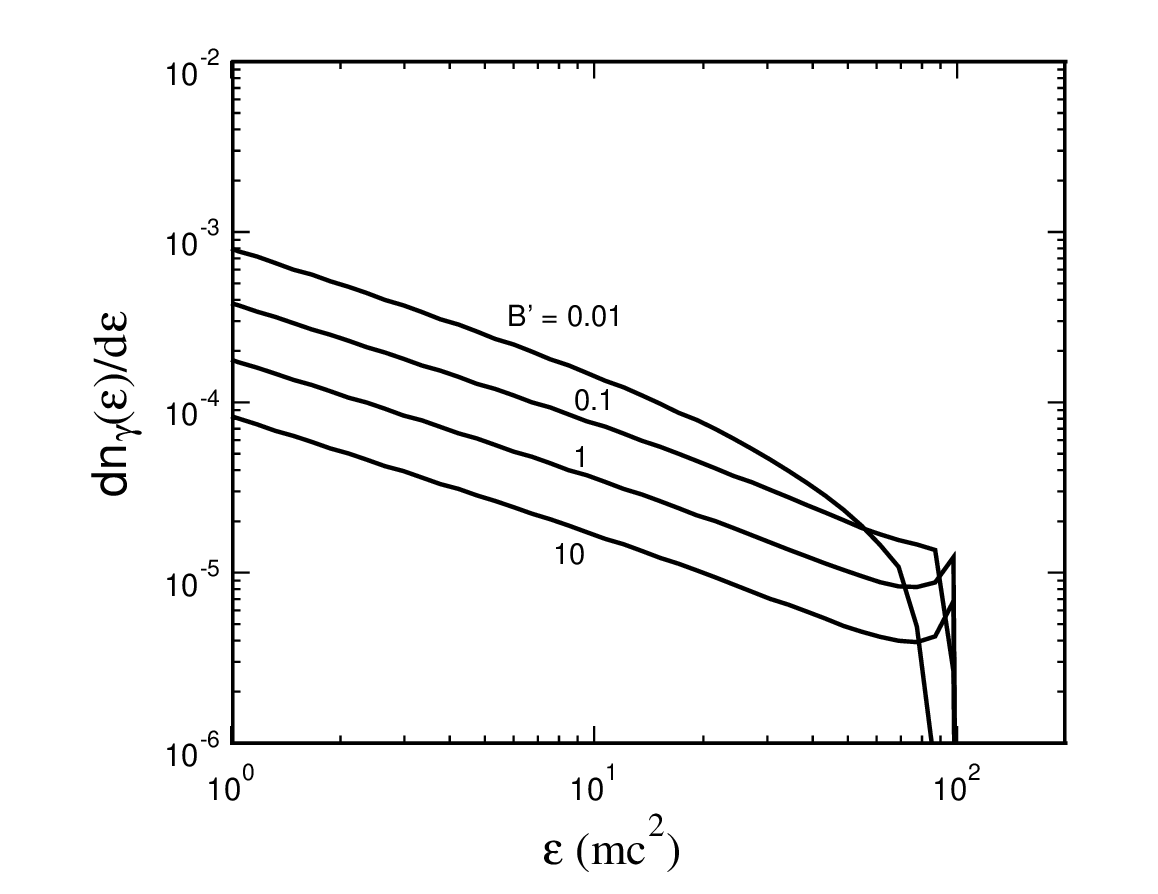}
\caption{Quantum synchrotron spectra for a particle with Lorentz factor $\gamma = 100$, initial pitch angle, $\sin\psi = 1$ and different field strengths $B'$ in units of the critical field, computed from the 
asymptotic formula in Eqn (\ref{eq:IQ}).}
\label{fig:SynQED}
\end{figure}

The energy loss rate of an electron emitting synchrotron radiation, resulting from integration of the spectrum
of Eqn (\ref{eq:IQ}) over photon energy, is shown in Figure \ref{fig:SynLossRate} as a function of the
parameter $\Upsilon$.  For low values of $\Upsilon \ll 0.1$, the synchrotron loss rate follows the 
well-known classical dependence $\dot \gamma \propto \Upsilon^2$, but begins to depart from the classical
dependence for values of $\Upsilon \gsim 0.1$.  At large values of $\Upsilon \gg 1$, the loss rate follows
a milder dependence $\dot \gamma \propto \Upsilon^{2/3}$.  This softening of the loss rate occurs because 
the total rate of emission declines at high $\Upsilon$ (Baring 1988). 

\begin{figure} 
\hskip 1cm
\includegraphics[width=12cm]{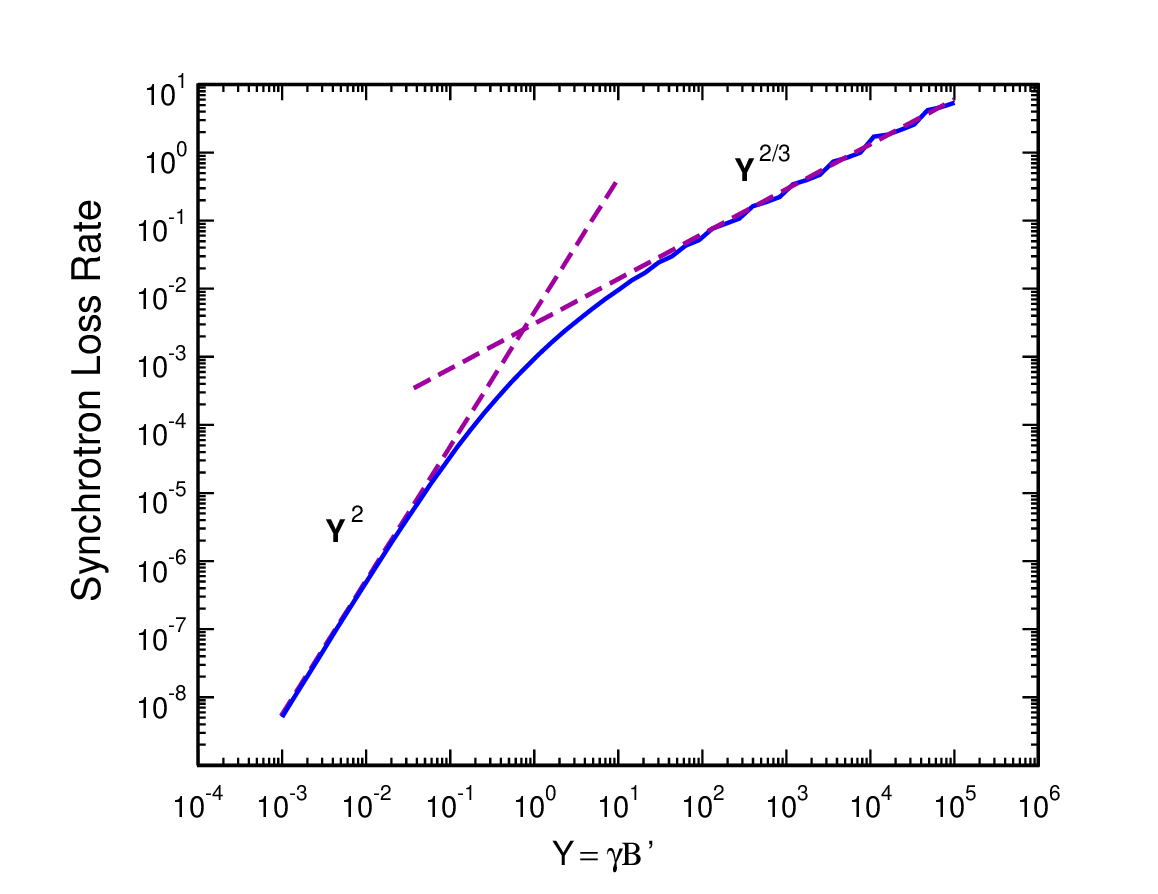}
\caption{Synchrotron energy loss rate as a function of $\Upsilon = \gamma B'$ for an electron having pitch angle
$\sin\psi = 1$.}
\label{fig:SynLossRate}
\end{figure}

\subsection{Compton Scattering} \label{sec:CompScatt}

In neutron star magnetic fields, the cyclotron decay rate is high
enough so that nearly all particles occupy the ground state. 
The cyclotron de-excitation rate $r_{cyc} \sim 3 \times 10^{15}B_{12}^{2}
\,\rm s^{-1}$, is much larger the collision rate, $r_{coll} \sim 5 \times 
10^8 (n_e/10^{21}\,
{\rm cm^{-3}})\,B_{12}^{-3/2}\,\rm s^{-1}$ (Bonazzola et al. 1979).
Radiative (rather than collisional) processes
thus control the Landau state populations.  Radiative transitions thus
dominate over collisions in astrophysical sources having strong
magnetic fields.  An electron that absorbs a cyclotron photon will
almost always de-excite by emitting another photon, rather than
collisionally de-exciting, so that the end result of the process is a
scattering of the photon rather than a true absorption. This means
that in strong fields resonant scattering dominates over absorption.
Cyclotron absorption is most accurately treated as a second-order
process, as resonances in the Compton scattering cross section, where
the excited state of the particle is a virtual state for which energy
and momentum are not strictly conserved.  The cyclotron lines are thus
broadened by the intrinsic width which is equal to the inverse of the
decay rate from that state (Harding \& Daugherty 1991, Graziani 1993),
although Doppler broadening usually dominates at most angles.  
The addition of the decay rate to the resonant denominator also renders 
the cross section finite at resonance.  The
cross section for Compton scattering in a magnetic field was first
studied in the non-relativistic limit by Canuto, Lodenquai \& Ruderman
(1971) and the full QED cross section has been computed by Herold
(1979), Daugherty \& Harding (1986) and Bussard, Meszaros \& Alexander
(1986).  Since the non-relativistic treatment is limited to dipole
radiation, only scattering at the cyclotron fundamental is allowed.
In the relativistic (QED) treatment scattering at higher harmonics is
allowed, including Raman scattering, in which the state of the
particle after scattering is higher than the initial state.

In free space, the total electron scattering cross section is just the
Thomson cross section, $\sigma_T = 6.6 \times 10^{-24}\,\rm cm^2$, for
photon energies which are non-relativistic in the electron rest frame,
or when $\epsilon\gamma \ll m_e$.  At higher energies, relativistic
effects are important both in the kinematics and in the cross section
(e.g. Rybicki and Lightman 1979).  The photon energy change in the
electron rest frame, due to electron recoil, can no longer be ignored
and the Klein-Nishina cross section (e.g. Jauch \& Rohrlich 1980) is
appropriate in the relativistic (QED) regime.

The classical, non-relativistic (Thomson) limit of the magnetized
scattering cross section (Canuto {\it et al.}  1971, Ventura 1979) has
a strong dependence on photon frequency, angle to the magnetic field
and polarization.  The non-relativistic total scattering cross section
in the electron rest frame for linearly polarized photons takes the
form (Blandford and Scharlemann 1976):
\begin{eqnarray} \label{eq:MT}
\sigma_{\parallel} & = & \sigma_T \left\{\sin^2\theta + {1 \over 2}\cos^2\theta \left[
{\epsilon^2 \over (\epsilon + \epsilon_B)^2} + {\epsilon^2 \over (\epsilon - 
\epsilon_B)^2}\right]\right\}
\nonumber \\
\sigma_{\perp} & = & {\sigma_T \over 2} \left\{{\epsilon^2 \over (\epsilon + \epsilon_B)^2} 
+ {\epsilon^2 \over (\epsilon - \epsilon_B)^2}\right\}.
\end{eqnarray}
for polarization states $(\parallel, \perp)$ where to the photon
electric vector parallel or perpendicular to the plane formed by the
photon wavevector and the field.  Here $\theta$ and $\epsilon$ are
the angle and energy of the incident photon with respect to the field
in the electron rest frame, and $\epsilon_B = \hbar eB/(m_ec)$
is the cyclotron energy.

The main effects of the magnetic field on electron scattering is the
appearance of the cyclotron resonance and a strong dependence of the
cross section on photon energy and incident angle.  For photon
energies well above the resonance, $\epsilon \gg \epsilon_B$, $\sigma
= \sigma_T$ for both polarizations.  For photon energy below the
resonance, $\epsilon \ll \epsilon_B$, $\sigma_{\parallel} \simeq
\sigma_T[\sin^2\theta + \cos^2\theta(\epsilon/\epsilon_B)^2]$ and
$\sigma_{\perp} \simeq \sigma_T(\epsilon/\epsilon_B)^2$.  A magnetized
plasma becomes quite optically thin for propagation parallel to the
magnetic field for photon frequencies below the cyclotron frequency.

As in the case of cyclotron absorption, the non-relativistic cross
section for scattering is not accurate for 
$B'\gsim 0.1$ 
and furthermore describes scattering only in the fundamental and so cannot
be used to treat scattering in higher harmonics.  As we will discuss
in more detail in \S \ref{sec:XRPs}, cyclotron line modeling at low
field strengths is still possible using the absorption cross section
for harmonics above the fundamental, where the scattering is treated
as cyclotron absorption followed by cyclotron emission of the photon.
Even at high field strengths, 
use of the relativistic absorption cross
section can give a reasonable approximation to the scattering cross
section for low harmonics.  But this treatment neglects the
contribution from non-resonant scattering which is important for
photon energies away from resonance.

The relativistic magnetic scattering cross section is quite
complicated even for the simplest case of ground state-to-ground state
scattering (Herold 1979), and is even more unwieldy for the case of
scattering to arbitrary Landau states (Daugherty \& Harding (1986) and
Bussard, Meszaros \& Alexander 1986).  However, a number of
interesting features appear in the relativistic scattering cross
section (Daugherty \& Harding 1986).  One is the possibility of ground
state-to-ground state scattering through higher intermediate Landau
states (i.e. $0 \rightarrow n \rightarrow 0$, producing resonant
contributions to the cross section at higher harmonics.  Another
possibility is a $0 \rightarrow 0 \rightarrow n$ scattering where a
photon of energy $\epsilon \simeq nB'$ is almost completely `absorbed'
producing a scattered photon having very low energy.  The inverse
process in which an electron in an excited state scatters to the
ground state, can convert an X-ray photon to a $\gamma$-ray photon
(Brainerd 1989).  Another possibility that arises solely in QED is
two-photon emission, where an electron makes a transition between two
Landau states with the emission of two photons (Alexander \& Meszaros
1991a, Semionova \& Leahy 1999, Melrose \& Kirk 1986) Gonthier et al. (2000) 
have derived a
simplified analytic approximation to the relativistic Compton
scattering cross section for the case where the incident photon is
parallel to the field, in which case there is a resonance only at the
fundamental and scattering to higher Landau states can effectively be
neglected.  Such a situation would apply to scattering by a
relativistic electron moving along the magnetic field, where photons
would primarily appear predominatly in a Lorentz cone ($1/\gamma$) centered 
on the field direction in the electron rest
frame.  The approximate expression for the total relativistic,
polarization-dependent scattering cross section for the case $\theta =
0^\circ$ is

\begin{eqnarray}  \label{QEDscatApp}
\sigma_{||\to ||}&=&\sigma _{\perp \to ||}={{3\sigma _T} \over {16}}\left\{
{g(\epsilon )-h(\epsilon )} \right\}\left[ {{1 \over {(\epsilon -B')^2}}+{1 \over
{(\epsilon +B')^2}}} \right],\nonumber \\
\sigma _{||\to \perp }&=&\sigma _{\perp \to \perp
}={{3\sigma _T} \over {16}}\left\{ {f(\epsilon )-2\epsilon h(\epsilon )}
\right\}\left[
{{1 \over {(\epsilon -B')^2}}+{1 \over {(\epsilon +B')^2}}} \right] \
\end{eqnarray}
\noindent
where
\begin{eqnarray}
g(\epsilon )&=&{{\epsilon^2 (3+2\epsilon )+2\epsilon} \over {\sqrt {\epsilon
(2+\epsilon )}}}\ln \left( {1+\epsilon -\sqrt {\epsilon (2+\epsilon )}}
\right)+{\epsilon  \over 2}\ln (1+4\epsilon )\nonumber \\ & & +\epsilon
(1+2\epsilon )\ln (1+2\epsilon )+2\epsilon ,\nonumber \\ f(\epsilon )&=&-\epsilon
^2\ln (1+4\epsilon )+\epsilon (1+2\epsilon )\ln (1+2\epsilon ),\ {\rm and} \\
h(\epsilon )&=&\left\{
\begin{array}{ll}
{{\epsilon ^2} \over {\sqrt {\epsilon (2-\epsilon )}}}\tan ^{-1}\left(
{{{\sqrt {\epsilon (2-\epsilon )}} \over {1+\epsilon }}} \right), & \mbox{for
$\epsilon <2$, and} \\ {{\epsilon ^2} \over {2\sqrt {\epsilon (\epsilon
-2)}}}\ln \left( {{{\left( {1+\epsilon +\sqrt {\epsilon (\epsilon -2)}}
\right)^2} \over {1+4\epsilon }}} \right), & \mbox{for $\epsilon >2$.}
\end{array}
\right. \nonumber
\end{eqnarray}
This expression reduces to the nonrelativistic limit for small
$\epsilon$.  Figure \ref{fig:Gonthier2000_Fig2} shows this
approximation compares to the exact QED cross section for the same
case of incident photon angle $\sin\theta = 0$.  For incident photon
energies above the cyclotron energy, the electron can be excited to
higher Landau states and this contribution (for states up to $n' =
500$) is included in the exact cross section.  Even though the
approximation assumed only $n' = 0$, it does surprisingly well.  For
photon energies well above $\epsilon_B$, the cross section tends
toward the field-free relativistic (Klein-Nishina) cross section.  The
suppression of the relativistic magnetic cross section is due
primarily to Klein-Nishina effects, i.e. electron recoil.  The
approximation for the differential scattering cross section for this
case is given in Gonthier et al.  (2000).

\begin{figure} 
\hskip -1cm1
\includegraphics[width=17cm]{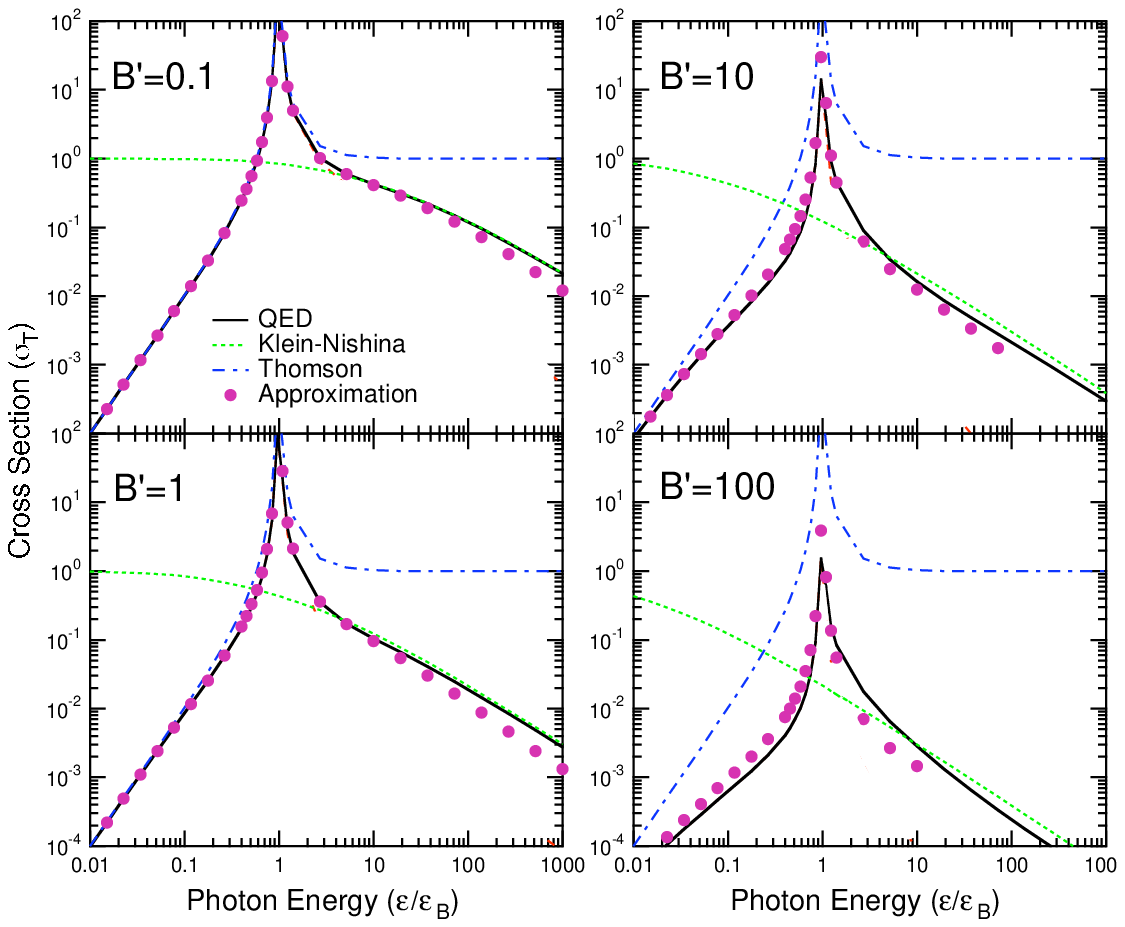}
\vskip -8cm
\caption{Total polarization-averaged Compton scattering cross section for an electron at rest in the ground state, in units of the Thompson cross section, as a function of incident photon energy in units of cyclotron 
energy, for the case of incident photon angle $\sin\theta = 0$.   Solid line: relativistic QED cross section 
(e.g. Bussard et al. 1986, Daugherty \& Harding 1986) summed over final electron Landau states up to 500, dot-
dashed line: magnetic Thompson cross section, Eqn. (\ref{eq:MT}), dotted line: Klein-Nishina (non-magnetic) 
cross section, dots: approximate QED cross section in Eqn (\ref{QEDscatApp}).  Adapted from Gonthier et al.
(2000).}
\label{fig:Gonthier2000_Fig2}
\end{figure}

\subsection{Pair Production and Annihilation}  \label{sec:PairProd}

In a strong magnetic field, single photons as well as two or more
photons may convert into electron-positron pairs.  One-photon pair
production cannot conserve both energy and momentum in field-free
space, but a magnetic field can absorb the extra momentum of a
photon with the energy required to created a pair.

\subsubsection{One-photon pair creation and annihilation} \label{sec:1pho}

A photon with energy $\epsilon$, traveling at angle $\theta$ to the
magnetic field, can produce an electron with parallel momentum $p$ and
a positron with parallel momentum $q$ only in the discrete Landau
states that are kinematically allowed by the energy and momentum
conservation equations,
\be \label{eq:Econ2}
E_n + E_{n'} = \epsilon
\ee
\be \label{eq:pcon2}
p + q = \epsilon\cos\theta
\ee
where $E_n = (1 + p^2 + 2nB')^{1/2}$ and $E_{n'} = (1 + q^2 +
2n'B')^{1/2}$ are the energies of the electron and positron.  The
threshold, $\epsilon = 2/\sin\theta$, is the photon energy needed to
produce a pair with momenta $p = q = \epsilon\cos\theta / 2$ in the
ground state ($n = n' = 0$)\footnote{This applies only to the $\parallel$ mode;
for the $\perp$ mode, we must have $n=0,n'=1$ or $n=1,n'=0$, so the
threshold is replaced by $\epsilon\sin\theta=1+\sqrt{1+2B'}$.}.
In the case of high photon energies and
low magnetic fields, $\epsilon^2\sin^2\theta/B' \gg 1$, where the pair
is produced far above threshold in high Landau states, the
polarization-averaged pair production attenuation coefficient can be
expressed in the asymptotic form (Klepikov 1954, Erber 1966)
\be  \label{eq:1gpair}
R_{1\gamma}^{\rm pp}(\chi) ={\alpha\over 2\lambar} B'\sin\theta\,\,\, T(\chi),
\ee
\be 
T(\chi) \approx 4.74 \chi^{-1/3} {\rm Ai}^2(\chi^{-2/3}) = \left\{
\begin{array}{lr} 
0.377 \exp \mbox{\Large $(-{4\over 3\chi})$} & \chi \ll 1 
\\ \\ 
0.6 \chi^{-1/3} & \chi \gg 1 
\end{array} \right. 
\ee
where $\chi \equiv \epsilon B'\sin\theta/2$, Ai is the Airy function and 
$\lambar$
is the electron Compton wavelength.  
The probability of one-photon pair production thus rises exponentially
with increasing photon energy and transverse field strength.
A rule-of-thumb is that magnetic pair production will be important for this regime when the
argument of the exponential in Eqn~(\ref{eq:1gpair}) approaches unity, or when
$\chi \ge 0.1$.  When the magnetic field $B' \gsim 0.1$, this condition will
be satisfied below the threshold, so that in such high fields pair production will occur near
threshold.  The pair will then be produced in low Landau states and the attenuation coefficient 
will exhibit resonances at the threshold for each pair state (Toll 1952, Daugherty \& Harding 1983).   
Eqn~(\ref{eq:1gpair}) will not be valid
in this case, but may be corrected for near-threshold effects by making the substitution, 
$\chi \rightarrow \chi/F$, where $F = 1 + 0.42(\omega\sin\theta)^{-2.7}$ in Eqn~(\ref{eq:1gpair}) 
(Daugherty \& Harding 1983),
or by using the approximate expression (Baring 1988, see corrections in Baring 1991)

\begin{eqnarray} \label{eq:1gpair_nt}
R_{1\gamma}^{\rm pp} (\epsilon, \theta) \simeq {\alpha \over \lambar}\,{B'\sin\theta\over \epsilon_{\perp}^2}\,{3\epsilon_{\perp}^2-4\over
2(\epsilon_{\perp}+2)^2}\,\left[\log{(1/\zeta)}\right]^{-1/2}\,\left[{4\epsilon_{\perp}\over \epsilon_{\perp}^2-4} +
\log{\zeta}\right]^{-1/2}
\nonumber \\
~~~~~~~~~~~~~\exp\left\{{-{\epsilon_{\perp}\over B'} - {\epsilon_{\perp}^2-4\over 4B'}\,\log{\zeta}}\right\},
~~~~~~~~~~~~~~2 < \epsilon_{\perp} \ll 1/B'.
\end{eqnarray}
where $\epsilon_{\perp} \equiv \epsilon\sin\theta$, $\zeta = (\epsilon_{\perp}-2)/(\epsilon_{\perp}+2)$.  
Either of the above prescriptions will approximate the decrease in $R_{1\gamma}$ near threshold (see Figure 
\ref{fig:1gpairRate}).
Semionova \& Leahy (2001) have derived the one-photon pair production attenuation coefficient for all
cases of photon polarization and electron and positron spin states, making use of the proper spin
eigenstates defined by Sokolov \& Ternov (1986).

\begin{figure} 
\hskip 1cm
\includegraphics[width=12cm]{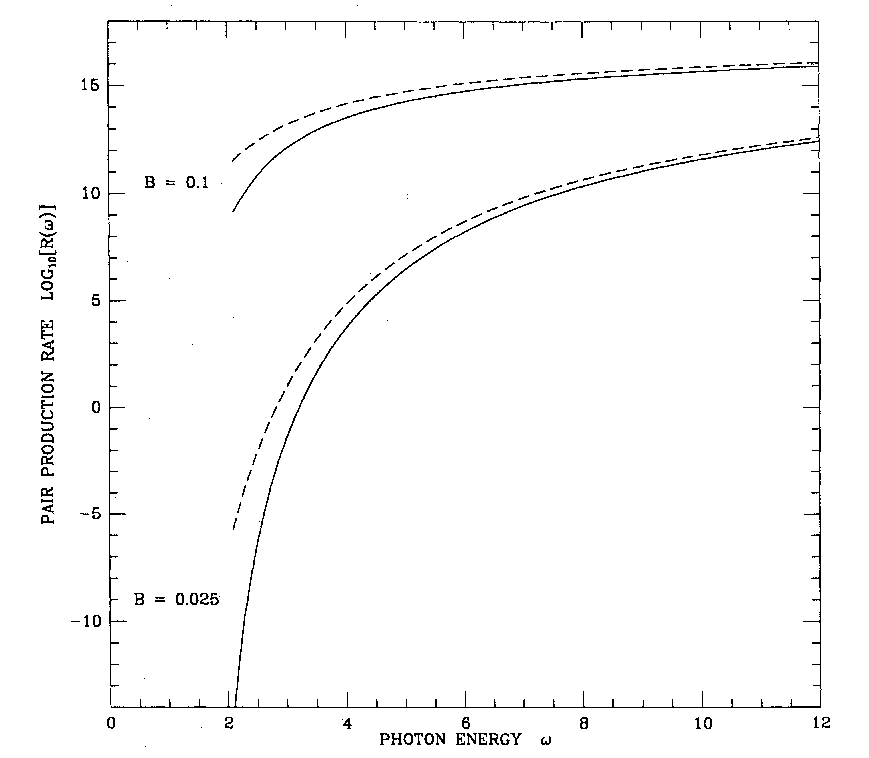}
\caption{One-photon pair production rate as a function of photon energy (in units of $mc^2$), for two
different magnetic field strengths (in units of the critical field).  The solid lines are the approximation 
of Eqn (\ref{eq:1gpair_nt}) to the exact rate and the dashed lines are the asymptotic formula of Eqn 
(\ref{eq:1gpair}). From Baring (1988).}
\label{fig:1gpairRate}
\end{figure}

In the case where an electric field is present perpendicular to the magnetic field, the pair production attenuation coefficient can be obtained by a Lorentz transformation perpendicular to the magnetic field (Daugherty 
\& Lerche 1975).  When an electric field of strength $E$ parallel to the magnetic field is present, 
Daugherty \& Lerche (1976) determined that the pair attenuation coefficient is increased by an additional
amount of order $E^2/B^2$, which is usually small in neutron star magnetospheres since $E \ll B$ is always true for
rotationally induced electric fields.

The inverse process to one-photon pair creation is one-photon pair
annihilation, in which an electron-positron pair annihilates into a
single photon and is likewise only permitted in a strong field.  The
kinematic equations are the same as those for one-photon pair
creation, Eqns (\ref{eq:Econ2}) and (\ref{eq:pcon2}), with the
electron and positron initially occupying Landau states $n$ and $n'$
with momenta $p$ and $q$.  The annihilation rate for pairs in the
ground state, $n = n' = 0$ has been calculated by Wunner (1979) and
Daugherty \& Bussard (1980) and the expression in the
center-of-momentum frame (where $p = -q$) is particularly simple:
\be  \label{eq:1gann}
R_{1\gamma}^{\rm ann}(E_p) = n_o^+n_o^-\,{\alpha\lambda^2 c\over
B'}{1\over E_p^2} \exp\left({-2E_p^2\over B'}\right), 
\ee 
where $\lambda \equiv 2\pi \lambar$ is the electron Compton wavelength, 
$E_p =(p^2 + 1)^{1/2}$ and $n_o^+$ and $n_o^-$ are the number
densities of positrons and electrons.  The annihilation rate in an
arbitrary frame can be obtained by a Lorentz transformation along the
magnetic field direction.  Since the rate in Eqn (\ref{eq:1gann})
increases exponentially with increasing field strength, one-photon
annihilation overtakes the two-photon annihilation rate for pairs at
rest at a field strength around $10^{13}$ Gauss.  One-photon
annihilation from the ground state results in a line at $2m$,
broadened asymmetrically toward higher energies by the parallel
momenta of the pairs.  Unlike in the case of two-photon annihilation
(cf. \S \ref{sec:2pho}), Doppler broadening results only in a
blueshift here, because the photon must take all of the kinetic energy
of the pair in addition to the rest mass.  The annihilation photons
are emitted in a fan beam transverse to the field, which is broadened
if the pairs have nonzero parallel momenta.  Pairs annihilating from
excited states produce additional lines above 1 MeV which at high
energies blend together into a continuum.  The one-photon annihilation
rate of pairs from excited states can proceed at a rate orders of
magnitude faster than from the ground state in fields below $10^{13}$
G (Harding 1986, Wunner {\it et al.} 1986).  The one-photon rate
therefore becomes comparable to the two-photon rate (which does not
rise rapidly in excited states) at lower field strengths for pairs in
excited states.  However, the effectiveness of synchrotron cooling may
keep the densities of excited states low enough relative to the ground
state to cancel the increase in the annihilation rate.  The one-photon
annihilation rate for different photon polarizations and particle spin
states has been derived by Wunner et al. (1986) and Semionova \& Leahy
(2000).

\subsubsection{Two-photon pair creation and annihilation in a strong magnetic field}  \label{sec:2pho}

Pair creation by two photons is significantly modified by strong
magnetic fields from its field-free behavior.  In field-free,
the two-photon pair production cross section near threshold in terms
of the photon energy in the center-of-momentum frame, $\epsilon_{_{\rm
CM}} = [\epsilon_1\epsilon_2(1-\cos\theta_{12})/2]^{1/2}$, is
(Svensson 1982)
\begin{equation}
\sigma_{2\gamma} \simeq {3\over 8}\,\sigma_T \left\{
\begin{array}{lr}
(\epsilon_{_{\rm CM}}^2 - 1)^{1/2}  & (\epsilon_{_{\rm CM}} - 1) \ll 1 \\
{\left[2 \ln(2\epsilon_{_{\rm CM}}) - 1 \right]} / \epsilon_{_{\rm CM}}^6,  & \epsilon_{_{\rm CM}} \gg 1 \\
\end{array} \right.
\end{equation}
where $\epsilon_1$ and $\epsilon_2$ refer to the energies of the
photons and $\cos\theta_{12}$ is the cosine of the angle between
their propagation directions.  The full relativistic cross section can
be found in Jauch and Rohrlich (1980).  In a magnetic field, the
kinematic equations for this process impose conservation of energy and
parallel momentum only:
\begin{equation} \label{eq:Econ3}
\epsilon_1 + \epsilon_2 = E_n + E_{n'}
\end{equation}
\begin{equation} \label{eq:pcon3}
\epsilon_1\cos\theta_1 + \epsilon_2\cos\theta_2 = p + q,
\end{equation}
where $\theta_1$ and $\theta_2$ are their angles with respect to the field.  
The threshold depends on photon polarization
direction with respect to the field, with the threshold for producing a pair
in the ground state ($n=n'=0$), taking the form (Daugherty and Bussard 1980):
\begin{equation}  \label{eq:2gppth}
(\epsilon_1\sin\theta_1 + \epsilon_2\sin\theta_2)^2 + 
2\epsilon_1\epsilon_2[1 - \cos(\theta_1-\theta_2)] \ge 4.
\end{equation}
The second term is 
similar to the field-free threshold
condition and the first term appears as a result of non conservation of
perpendicular momentum.  Thus it is possible for photons traveling
parallel to each other ($\theta_1 = \theta_2 \ne 0^0$) to produce a
pair, an event not permitted in field-free space.

The two-photon pair production cross section in a strong magnetic field,
like the one-photon pair production cross section,
has resonances near threshold due to the discreteness of the pair 
states.  The two-photon cross section in a strong magnetic field
has been calculated by Kozlenkov and Mitrofanov (1987) for photon
energies below one-photon pair production threshold and shows the same
sawtooth behavior as the one-photon process.  Near threshold, the magnetic 
field decreases the cross section below its free-space value, due to the
decreased phase space available to the pair.  Above its threshold, one-photon pair 
creation will dominate since it is a lower order process than two-photon pair creation.
A comparison of the relative importance of the one-photon and two-photon processes 
(Burns and Harding 1984) shows that  
the one-photon process will generally dominate in magnetic fields above $\approx 10^{12}$ Gauss. 

The free-space cross section for two-photon pair annihilation in the
non-relativistic limit is $\sigma_{2\gamma} = (3/8)\sigma_T/\beta_r$,
where $\beta_r$ is the relative velocity of the positron and electron.
The annihilation rate for unpolarized positrons and electrons with densities $n_+$
and $n_-$ is then
\begin{equation} \label{eq:R2g}
R_{2\gamma} = {3\over 8}\sigma_Tc n_+ n_-.
\end{equation}
The two-photon annihilation rate in a strong magnetic field has been calculated
for pairs in the ground Landau state (Daugherty and Bussard 1980, Wunner 1979) and is unchanged from
the field-free rate below $B' \sim 0.1$.  At $B' \gsim 0.2$, the rate decreases sharply due to the
smaller phase-space of the virtual pair states, just as the one-photon annihilation rate is increasing
exponentially.  Two-photon annihilation of non-relativistic pairs results in a line at 511 keV as
in free space, but the relaxation of transverse momentum conservation in a magnetic field causes
a broadening mostly on the red side of the line at viewing angles other than $90^{\circ}$.  At a
viewing angle of $90^{\circ}$ to the field direction, the broadening is symmetric for $B' \lsim 0.1$
but becomes asymmetrically broadened on toward the blue side for $B' \gsim 0.1$, as shown in Figure
\ref{fig:BaringHarding1992}.  The magnetic broadening
can be approximated as
\be  \label{eq:2glinebr}
(\Delta \epsilon)^B \sim 
\left\{ \begin{array}{lr}
B'/2,  &  \sin\theta < \sqrt{2B'} \\
\sin\theta\sqrt{B'/2} , & \sin\theta > \sqrt{2B'} \\
\end{array}
\right.
\ee

There is an increasing tendency in very high fields for one of the
photons to be produced with almost all of the pair energy, so that
two-photon annihilation behaves more like one-photon annihilation.
The angular distribution of the photons from annihilation at rest also
becomes more anisotropic with increasing field strength, with the peak
of emission perpendicular to {\bf B} , again similar to one-photon
annihilation.  Similar broadening occurs when the pairs have non-zero
momenta parallel to the magnetic field (Kaminker et al. 1987, Baring
\& Harding 1992).

\begin{figure} 
\hskip -1.0cm
\includegraphics[width=17cm]{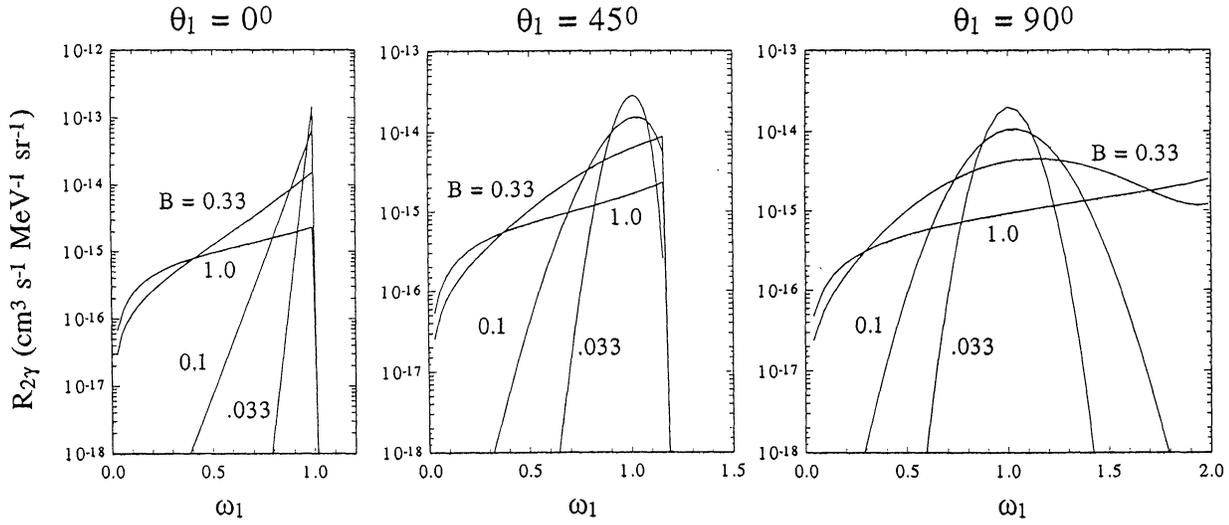}
\caption{Two-photon annihilation spectra for pairs at rest in a magnetic field in the ground Landau state, 
for different field strengths (in units of the critical field) and viewing angles to the field, $\theta_1$.  
The kinematic cutoffs occur when $|\cos\theta_2| = 1$, where $\theta_2$ is the angle of the unseen photon.
From Baring \& Harding (1992).}
\label{fig:BaringHarding1992}
\end{figure}

\subsection{Bound Pair Creation} \label{sec:BoundPair}

As was first noted by Shabad \& Usov (1982), the dispersion curve (as shown in Figure \ref{fig:ShabadUsov85})
of a photon propagating 
along a curved magnetic field, with $\epsilon < 2/\sin\theta$ initially below the threshold for 
one-photon pair creation, will cross the lowest bound state of positronium prior to reaching the 
threshold for creating a free pair in the ground state ($n=n'=0$ for $\parallel$ polarization, 
$n=0, n'=1$ or $n=1, n'=0$ for $\perp$ polarization).  This situation was originally interpreted as
a ``capture" of the photon by the magnetic field for $B' \gsim 0.1$, and a complete 
prevention of free pair creation until the photon could travel to a region of lower field strength. 
But Herold et al. (1985) pointed out that as the photon approaches the crossing point,
it will adiabatically convert into positronium, after evolving through a mixed photon-positronium
state (Shabad \& Usov 1985, 1986), if the magnetic field is strong enough.  
The positronium state produced in this way from a single photon is stable to
annihilation, since positronium cannot annihilate from the ground state in a magnetic field 
(Wunner et al. 1981).  
In this case, creation of a free pair is suppressed unless the positronium is unbound and several
possible mechanisms have been discussed.  Positronium may be ionized by an electric field
parallel to the magnetic field, whose strength is sufficient when combined with the Coulomb
field to form a potential barrier low enough for the electron and positron to tunnel through to 
become a free pair.  The probability of ionization by a field of strength $E_{\parallel}$ is
(Usov \& Melrose 1995) 
\be  \label{eq:Pfionpos}
W_{\rm E} = {eE_{\parallel} \over 2mc (\Delta \varepsilon_{00})^{1/2}}\,
\exp\left[{-4(\Delta \varepsilon_{00})^{3/2}\over 3 (eE_{\parallel}/mc^2)\lambar} \right]
\ee
where $\Delta \varepsilon_{00}$ is the positronium binding energy in the ground state.
On the scale of a neutron star radius, R, the critical ionizing field strength, below which 
the ionization probability is negligible, is 
\be  \label{eq:Efionpos}
E_{\parallel}^{\rm ion} \approx {2\over 3} {mc^2(\Delta \varepsilon_{00})^{3/2}\over {e\lambar}}
\ee 
and is a slowly varying function of $B'$.  For $B' \sim 0.2$, $eE_{\parallel}^{\rm ion} \sim  
10^{10}\,\rm Volts\,cm^{-1}$.  
Alternatively, the bound pair could be photoionized by a UV radiation field of sufficient intensity
(Herold et al. 1985).  For example, in a blackbody radiation field of temperature $T_6 = T/10^6$ K 
the mean-free path for photoionization of positronium moving with Lorentz factor $\Gamma$ 
in a magnetic field $B' \sim 0.1$ is approximately (Bhatia et al. 1992)
\be  \label{eq:mphionpos}
\ell_{\rm ph} \simeq 5 \times 10^4\,{\rm cm}\,\left({\Gamma\over 10^2}\right)^3\,T_6^{2}
\ee
For $T > 10^5$ K, the photoionization mean-free path of positronium will be short enough to 
free a bound pair in a fraction of a neutron star radius.

\begin{figure} 
\hskip 1.0cm
\includegraphics[width=15cm]{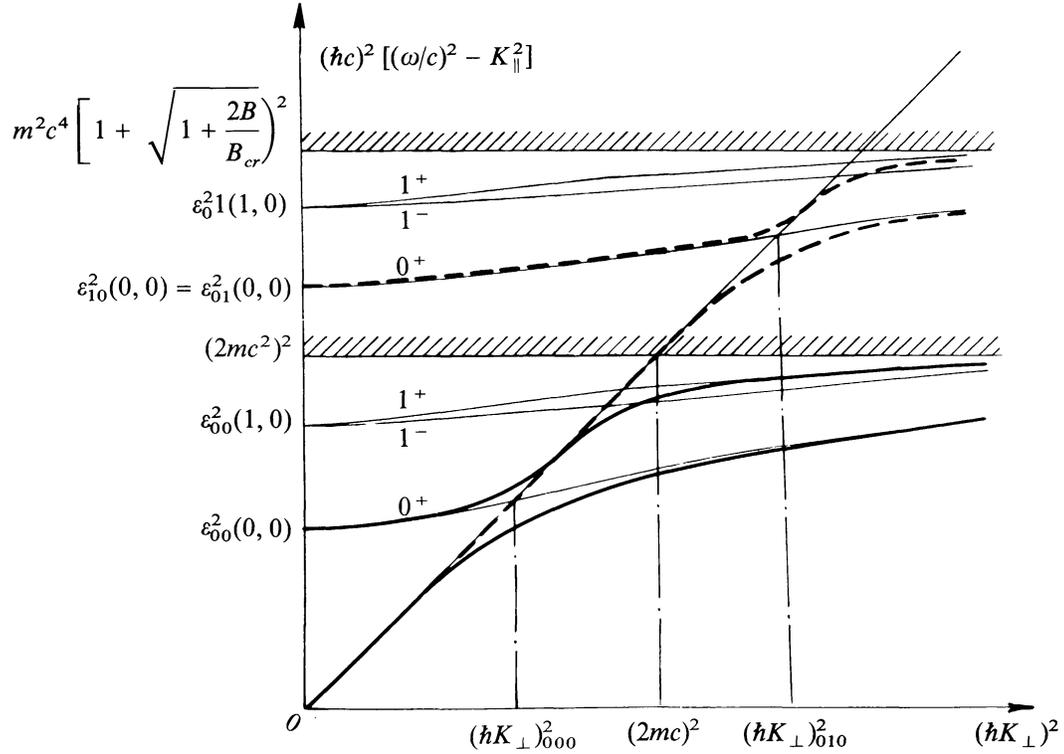}
\caption{Dispersion curves for photons propagating in a strong magnetic field.  The sloped straight line
is the unperturbed dispersion curve $(\omega/c)^2 - K^2_{\parallel} = K^2_{\perp}$.  
$\epsilon_{n,n'}(n_c, p_c)$ are the energies of the positronium states with quantum number $n_c$ 
and parallel momentum $p_c$.  The thin solid curves, labeled by $n_c^{\pm}$, are
the positronium dispersion curves for the lowest states having $n = n' = 0$ and $n + n' = 1$ 
for the electron and positron Landau states, where $\pm$ are the parities.  The hatched borders are the 
thresholds for production of a free pair in the two lowest states, which are the same as the positronium continua for each series of states.  The thick solid curves are the lowest dispersion curves 
of the mixed photon-positronium states with $n = n' = 0$.  The thick dashed lines are the lowest
dispersion curves for the $n + n' = 1$ mixed states. From Shabad \& Usov (1985).}
\label{fig:ShabadUsov85}
\end{figure}

\subsection{Photon Splitting} \label{sec:PhoSplit}

Photon splitting is a third-order process in which one photon divides
into two or more photons in the presence of a strong external field.
This process can be most easily understood in the context of the
vacuum polarization effect on photon propagation in a strong magnetic
field, which is due to the interaction of the photon virtual pairs
with the field.  To lowest order, photon splitting occurs when one
member of a vacuum polarization pair radiates.  While it is
kinematically allowed in field-free space, it is forbidden by Furry's
theorem (Furry 1937, Jauch \& Rohrlich 1980), a QED charge conjugation
symmetry stating that the S-matrix vanishes for any closed diagram
with an odd number of vertices having only external photon lines.  The
symmetry is broken by a strong magnetic field.  The photon splitting
rate in a magnetic field was computed by Bialynicka-Birula \&
Bialynicki-Birula (1970), and Adler et al. (1970, 1971) in the weakly
dispersive limit.  In this limit, there are three photon polarization
modes permitted by QED (i.e., they do not violate CP conservation):
$\perp \rightarrow \parallel\parallel$, $\parallel \rightarrow
\perp\parallel$ and $\perp \rightarrow \perp\perp$.  A simplified expression 
(for $B'\lsim 1$) for the polarization-dependent attenuation coefficient for
photon splitting was derived by Baring (1991):
\be  \label{eq:pspol}
R^{sp}_{\perp \rightarrow \parallel\parallel} = {1\over 2}R^{sp}_{\parallel \rightarrow \perp\parallel}
= \left({13\over 24}\right)^2 R^{sp}_{\perp \rightarrow \perp\perp} = {1\over 6}
\left({26\over 19}\right)^2 R_{sp}
\ee
and
\be  \label{eq:psatten}
R_{sp} \approx {\alpha^3\over 10\pi^2\lambar}\left({19\over 315}\right)^2\,B'^6\epsilon^5
\sin^6\theta
\ee
where $\epsilon$ is the incident photon energy propagating at angle $\theta$ to the magnetic 
field of strength $B'$.  Since there is no threshold for photon splitting the expression above is
valid for arbitrarily small photon energies.  For large values of $B' > 1$, the $B'$ dependence in
Eqn (\ref{eq:psatten}) saturates (i.e., the factor $B'^6 \rightarrow 1$ for $B' \gg 1$).  The full
expression for the photon splitting attenuation coefficient is given by Adler (1971) and Stoneham (1979).

Since photon splitting is a third order process, its rate is smaller
than that of one-photon pair production above $\epsilon = 2/\sin\theta$.  
However, it can compete with one-photon pair
production in neutron star magnetospheres, where photons are below
pair creation threshold at their emission points (see \S
\ref{sec:pulsars}).  As the photon 
travels in a region of curved field lines, 
its angle to the field increases and the photon may split before
reaching pair creation threshold (Baring 1991, 1995, Harding et
al. 1997).  Since the photon splitting rate is very sensitive to field
strength, its influence grows quickly with increasing field strength
with the switch from pair creation to photon splitting being dependent
on field geometry.  For photons propagating in neutron star dipole
magnetic field geometry, photon splitting becomes dominant as an
attenuation mechanism for $B' \gsim 1$ (Baring \& Harding 1998).

Adler (1971) examined the kinematic selection rules for photon
splitting in the weakly dispersive limit and found that only the
$\perp \rightarrow \parallel\parallel$ mode conserved energy and
momentum in the magnetized vacuum  
[another channel, $\perp\rightarrow \parallel \perp$, is also 
allowed kinematically, but the rate is suppressed by a factor of order
$n_\parallel -1$ (see Eq.~\ref{eq:nparallel})
compared to the $\perp \rightarrow \parallel\parallel$ channel,
and thus is negligible in the weakly dispersive limit].
Usov (2002), including only the
linear positronium contribution to the polarization tensor, has shown
that Adler's kinematic selection rules hold in arbitrarily high
magnetic fields.  However, the moderate vacuum dispersion present in
magnetar fields (see \S \ref{sec:magnetars}) may critically depend on
higher order (non-linear) contributions from the polarization tensor
(e.g. Melrose and Parle 1983).  If Adler's selection rules are in fact
ubiquitous it has profound implications, since it means that the only
polarization mode permitted to split produces only photons of the mode
that cannot split.  Such a situation will prevent pure photon
splitting cascades (Baring 1995, Harding \& Baring 1996), where the
two product photons of an initial splitting split into two more
photons, each of which split further, reprocessing the spectrum to
lower and lower energies.  Such a selection rule also leads to an
extremely strong polarizing mechanism, since only one mode is
attenuated and produces photon of the other mode.  Bulik (1998) has
derived the photon splitting rate including the effect of a plasma.
He finds that the presence of a plasma changes the kinematic selection
rules, allowing addition splitting modes to become important.
However, these changes seem to operate only in a small region of phase
space in neutron star atmospheres, between $B' \sim 0.1 - 2$, $\theta < 30^{\circ}$ and plasma
density $\rho > 1\,\rm g\,cm^{-3}$.

\section{Physics of Neutron Star Interiors and Envelopes}
\label{sec:NS}

\subsection{Overview}

Many observable properties of neutron stars, such as the mass-radius
relation, the maximum rotation rate and the thermal evolution, 
depend on the properties of the star's interior (e.g., Baym \& Pethick 1979; 
Shapiro \& Teukolsky 1983; Glendenning 2000; Heiselberg 2002;
Lattimer \& Parkash 2005). An important goal of neutron star astrophysics
is to use various observations of neutron stars to probe the property 
of matter under extreme conditions, particularly at super-nuclear densities.

A neutron star can be divided into several different regions: the
atmosphere/envelope, the crust and the core.  The atmosphere/envelope
contains a negligible amount of mass, but plays an important role in
determining the emergent photon spectrum and flux (see \S
\ref{sec:atm}). The crust, constituting $1-2\%$ of the stellar mass
and extending $\sim 1$~km below the surface, contains atomic nuclei
embedded in an electron Fermi sea.  Since a neutron star is formed as a
collapsed, hot ($k_BT\go 10$~MeV) core of a massive star in a
supernova explosion, the neutron star matter may be assumed to be
fully catalyzed and in the absolute lowest energy state (e.g.,
Salpeter 1961; Baym, Pethick \& Sutherland 1971).  The dominant nuclei
in the crust vary with density, and ranges from $^{56}$Fe at $\rho\lo
8\times 10^6$~g~cm$^{-3}$ to neutron-rich nuclei with $A\sim 200$ and
$Z/A\sim 0.1$ near the core-crust interface at $\rho\sim
10^{14}$~g~cm$^{-3}$ (see Douchin \& Haensel 2001 for recent work on
neutron star crust composition)\footnote{Neutron stars that have 
undergone accretion are expected to have different crust 
compositions because of nuclear reactions and weak interactions
during the accretion phase (e.g., Haensel \& Zdunik 1990; 
Blaes \etal~1990; Schatz \etal~1999).}. 
This ``standard'' picture of neutron
star crust, in which one particular kind of nucleus is present at a
given density and these nuclei form a near-perfect crystal lattice,
has been challenged by Jones (1999), who argued that thermodynamic
fluctuations at the time of freezing would ensure the presence of
several kinds of nuclei at a given pressure, yielding an impure
solid. This would substantially increase the resistivity of the crust,
particularly at low temperatures, and thus greatly affect magnetic
field evolution and heat conduction in the crust (Jones 2004).  At
densities above the $4\times 10^{11}$~g~cm$^{-3}$, neutrons ``drip''
out of nuclei and permeate the crust (this region is called ``inner
crust''). At temperatures less than $\sim 0.1$~MeV, these neutrons are
expected to form Cooper pairs and turn superfluid, largely decoupling
dynamically from the rest of the star. This has been invoked to
explain pulsar glitches, quick increases of the observed rotation rate
the star, which might be caused by a sudden transfer of angular
momentum from the more rapidly rotating neutron superfluid to the
crust.  Around the highest density region in the crust, the nuclei may
assume rod- and plate-like shapes (``nuclear pasta''; Pethick \&
Ravenhall 1995), before an eventual transition to uniform nuclear
matter of the liquid core.

The liquid core (with densities $\go 2\times 10^{14}$~g~cm$^{-3}$)
consists of a mixture of neutrons, protons and
electrons, which at progressively higher densities are joined by 
muons and other exotic particles such as strangeness-bearing hyperons
and/or pion or kaon Bose condensates. Chemical equilibrium among 
these particles is established by weak interactions such as
neutron beta decay ($n\rightarrow p+e^-+\bar\nu$) and electron
capture ($e^-+p\rightarrow n+\nu$), and the nuclear symmetry 
energy plays an important role in determining the relative abundance of
neutrons and protons. It is possible that both neutrons and
protons in the core are in a superfluid/superconducting state, in which the
neutron vorticity is concentrated in quantized vortex lines, and magnetic
flux may be similarly concentrated in proton flux tubes. The transition
temperatures for these superfluid states are highly
uncertain. Such a superfluid would alter the specific heat and neutrino
emissivities of the core, thereby affecting the thermal evolution
of the neutron star (Yakovlov \& Pethick 2004). 

At sufficiently high densities, free quarks may appear in the stellar core.
The so-called quark stars may come in several forms depending on 
the details of the nuclear to quark matter phase transition (Weber 2005).
A pure quark star (also called ``strange'' star) consists of up, down and
strange quarks with electrons to fulfill charge neutrality; the star may be
bare or may be enveloped in a thin nuclear crust (Alcock et al.~1986).
A hybrid star has a core of quark matter and a mantle of nuclear matter
(e.g., Alford et al.~2005). A mixed star has mixed phase of 
nuclear and deconfined quark matter over a range of density and radius
(Glendenning 1992). It is likely that a condensate of quark Cooper pairs
may appear at low temperatures characterized by a BCS gap usually referred 
to as color superconductivity (e.g., Alford et al 2001). Such a phase 
is predicted to have a unique property (e.g., the densest QCD phase,
the so-called color-flavor locked phase, is a color superconductor but 
an electromagnetic insulator).

An important probe of the neutron star interior is the mass-radius relation.
Neutron star masses have been measured accurately from timing
observations of binary radio pulsars, and less accurately from
binary X-ray pulsars (Stairs 2004). 
Of particular interest is PSR J0751+1807 (with white dwarf companion
in a 6 hour orbit), which has a mass of $2.1\pm 0.2M_\odot$ 
(Nice et al.~2005). This measurement, together with that of
Vela X-1 ($1.86\pm 0.16M_\odot$), rules out a class of soft equations of
state. The neutron star radius can be measured from
observations of the star's surface emission, including possible spectral
line identifications. This requires detailed modeling of neutron
star atmospheres (\S \ref{sec:atm}). Recent observations of
quiescent neutron star binaries in globular clusters (with known distances)
yield an effective radius $R_\infty=R(1+z)$ (where $z$ is the 
gravitational redshift at the neutron star surface) in the range
of 13 to 16~km (Rutledge et al.~2002; Heinke et al.~2003), 
consistent with many nuclear equations of state. 

The study of neutron star cooling can potentially provide
useful information on the neutron star interior, such as
superfluidity, direct vs. modified URCA processes (which depend
on the symmetry energy of nuclear matter). Theoretical cooling curves
can be compared with observations if the neutron star age can be estimated 
and the atmosphere emission spectrum can be properly interpreted
(Yakovlev \& Pethick 2004). Observations of pulsar glitches and 
possible precession also provide indirect constraints
on the interior physics (e.g. Link 2003; Akgun et al.~2006). 
In the future, measurement of neutron-star moments of inertia (in the double
pulsar system PSR J0737-3039; Lyne et al.~2004) and 
detection of gravitational waves from coalescing neutron star binaries 
will lead to useful information of neutron star equation 
of state (e.g., Lai \& Wiseman 1996; Faber et al.~2002;
Shibata et al.~2005).

Because of its high density, the equation of state of the
bulk interior of a neutron star is not strongly affected 
by the magnetic field (e.g., Broderick et al.~2000), unless one considers 
field strengths approaching the maximum ``virial'' value,
\be
B_{\rm max}\simeq 10^{18}\left({M\over 1.4M_\odot}\right)
\left({R\over 10\,{\rm km}}\right)^{-2}\,{\rm G},
\ee
which is set by equating the magnetic energy 
$(4\pi R^3/3)(B^2/8\pi)$ to the gravitational binding energy
$GM^2/R$. A strong magnetic field, however, can significantly
influence the property of neutron star envelopes, 
which we discuss in the following.

\subsection{Free Electron Gas in Strong Magnetic Fields}

Before discussing various properties of neutron star envelopes,
we summarize the basic thermodynamical properties
of a free electron gas in strong magnetic fields at finite 
temperature $T$ (see, e.g., Landau \& Lifshitz 1980 for the $B=0$ 
case).

The number density $n_e$ of electrons is related to the chemical potential
$\mu_e$ by 
\be
n_e={1\over (2\pi\hat\rho)^2\hbar}\sum_{n=0}^\infty g_{n}
\int_{-\infty}^\infty\! f\,\,dp_z,
\label{eqne}\ee
where $g_{n}$ is the spin degeneracy of the Landau level 
($g_0=1$ and $g_{n}=2$ for $n\ge 1$), and
$f$ is the Fermi-Dirac distribution
\be
f=\left[1+\exp\left({E_n-\mu_e\over k_BT}\right)\right]^{-1},
\ee
with $E_n$ given by Eq.~(\ref{eqrel}). The electron 
pressure is given by
\be
P_e={1\over (2\pi\hat\rho)^2\hbar}\sum_{n=0}^\infty g_{n}
\int_{-\infty}^\infty\! f\,\,{p_z^2c^2\over E_n}\,dp_z.
\label{eqpe}\ee
Note that the electron pressure is isotropic
\footnote{The transverse kinetic pressure $P_{e\perp}$ 
is given by an expression similar to Eq.~(\ref{eqpe}), except that 
$p_z^2c^2$ is replaced by $\langle p_\perp^2c^2\rangle
= n\beta (m_ec^2)^2$. Thus the kinetic pressure is anisotropic,
with $P_{e\parallel}=P_e=P_{e\perp}+{\cal M}B$, where ${\cal M}$ is the
magnetization. When the electron gas is compressed perpendicular to $\bB$,
work must be done against the Lorentz force density
$(\nabla\times {\bf\cal M})\times\bB$ involving the magnetization current.
Thus there is a magnetic contribution to the perpendicular 
pressure, with the magnitude ${\cal M}B$. The composite pressure tensor 
is therefore isotropic, in agreement with the thermodynamic result 
$P_e=-\Omega/V$ (Blandford and Hernquist 1982). 
For a nonuniform magnetic field, the net force (per unit volume) 
on the stellar plasma is 
$-\nabla P_e-\nabla (B^2/8\pi)+(\bB\cdot\nabla)\bB/(4\pi)$.}. 
The grand thermodynamic potential is 
$\Omega=-P_e V$, from which all other
thermodynamic quantities can be obtained. Note that for 
nonrelativistic electrons (valid for $E_F\ll m_ec^2$ and $k_BT\ll m_ec^2$),
we use Eq.~(\ref{eqefree}) for $E_n$, and the expressions for 
the density $n_e$ and pressure $P_e$ can be simplified to
\ba
&&n_e={1\over 2\pi^{3/2}\hat\rho^2\lambda_{Te}}\sum_{n=0}^\infty
g_{n}I_{-1/2}\left({\mu_e-n\hbar\omega_{ce}\over k_BT}\right),\label{eqne2}\\
&&P_e={k_BT\over \pi^{3/2}\hat\rho^2\lambda_{Te}}\sum_{n=0}^\infty
g_{n}I_{1/2}\left({\mu_e-n\hbar\omega_{ce}\over k_BT}\right),
\ea
where $\lambda_{Te}\equiv (2\pi\hbar^2/m_ek_BT)^{1/2}$ is the thermal wavelength
of the electron, and $I_\eta$ is the Fermi integral:
\be
I_\eta(y)=\int_0^\infty\!\! {x^\eta\over \exp(x-y)+1}\,dx.
\ee

First consider degenerate electron gas at zero temperature. The Fermi energy
(excluding the electron rest mass) $E_F=\mu_e(T=0)-m_e c^2=(m_e c^2)
\epsilon_F$ is determined from
\be
n_e={\beta\over 2\pi^2\lambda_e^3}
\sum_{n=0}^{n_{\rm max}}g_{n}\,x_F(n),
\label{ef0}\ee
with
\be
x_F(n)={p_F(n)\over
m_ec}=\left[(1+\epsilon_F)^2-(1+2n\beta)\right]^{1/2},
\ee
where $\lambda_e=\hbar/(m_ec)$ is the electron Compton wavelength, 
$\beta=B/B_Q$, and $n_{\rm max}$ is set by the condition 
$(1+\epsilon_F)^2\ge (1+2n_{\rm max}
\beta)$. The electron pressure is given by
\be
P_e={\beta\, m_ec^2\over 2\pi^2\lambda_e^3}\sum_{n=0}^{n_{\rm max}}
g_{n}(1+2n\beta)\,\Theta\!\left[{x_F(n)\over (1+2n\beta)^{1/2}}\right],
\label{eqpe0}\ee
where
\be
\Theta(y)={1\over 2}y\sqrt{1+y^2}-{1\over 2}\ln\left(y+\sqrt{1+y^2}\right),
\ee
which approaches $y^3/3$ for $y\ll 1$.
The critical ``magnetic density'' below which only the ground Landau
level is populated ($n_{\rm max}=0$) is determined by
$(1+\epsilon_F)^2=1+2\beta$, which gives [see Eq.~(\ref{nlandau})]
\be
\rho_B=0.802\,Y_e^{-1}
b^{3/2}~{\rm g~cm}^{-3}=7.04\times 10^3\,Y_e^{-1} B_{12}^{3/2}
~{\rm g~cm}^{-3},
\label{eqrhob}\ee
where $Y_e=Z/A$ is the number of electrons per baryon. Similarly, the
critical density below which only the $n=0,1$ levels are occupied
($n_{\rm max}=1$) is $\rho_{B1}=(2+\sqrt{2})\rho_B=3.414\,\rho_B$.
For $\rho<\rho_B$, equation (\ref{ef0}) simplifies to 
\be
\rho=3.31\times 10^4\,Y_e^{-1}
B_{12}\left[(1+\epsilon_F)^2-1\right]^{1/2}~{\rm g~cm}^{-3}.  \ee 
For nonrelativistic electrons ($\epsilon_F\ll 1$), the Fermi temperature
$T_F=E_F/k_B=(m_e c^2/k_B)\epsilon_F$ is given by 
\be T_F=2.70\, B_{12}^{-2}\left({Y_e\,\rho}\right)^2~{\rm K} \qquad ({\rm
for}~~\rho<\rho_B),\label{eqtf}\ee
where $\rho$ is in units of 1~g~cm$^{-3}$.
For $\rho\gg\rho_B$, many Landau levels are filled by the electrons, 
Eqs.~(\ref{eqne}) and (\ref{eqpe}) reduce to the zero-field expressions. 
In this limit, the Fermi momentum $p_F$ is given by
\be 
x_F={p_F\over m_ec}={\hbar\over m_ec}(3\pi^2n_e)^{1/3}=
1.009\times 10^{-2}\left({Y_e\,\rho}\right)^{1/3},\qquad (B=0)
\label{eqxf}\ee
and the Fermi temperature is 
\be
T_F={m_ec^2\over k_B}\left(\sqrt{1+x_F^2}-1\right)\simeq 3.0\times 10^5
\left({Y_e\,\rho}\right)^{2/3}~{\rm K},\qquad (B=0)
\label{eqtf0}\ee
where the second equality applies to nonrelativistic electrons
($x_F\ll 1$). Comparison between Eq.~(\ref{eqtf}) and
Eq.~(\ref{eqtf0}) shows that the magnetic field lifts the degeneracy
of electrons even at relatively high density (see Fig.~\ref{fig:t_rho}).

\begin{figure} 
\includegraphics[width=35pc]{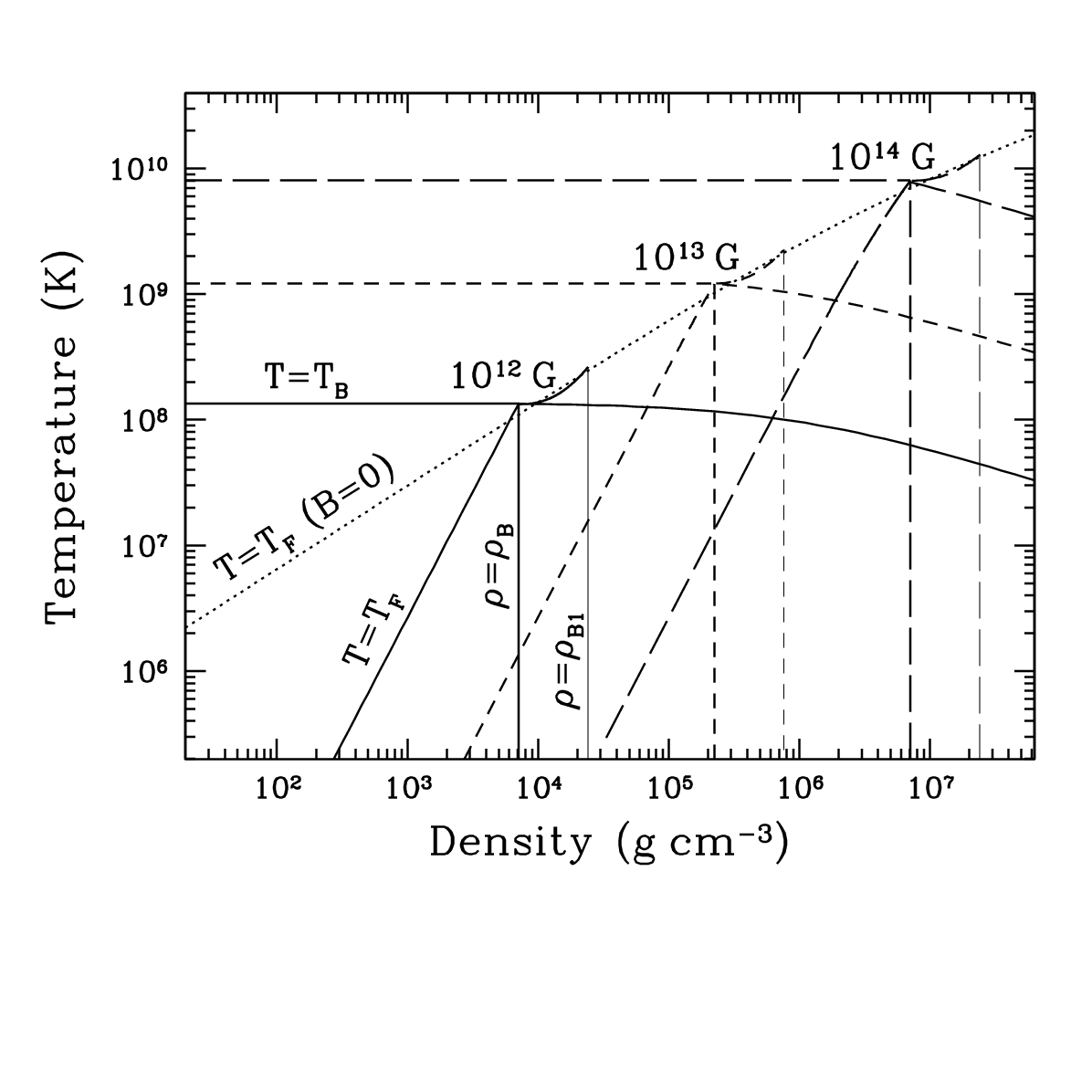}
\vskip -1.5cm
\caption{ 
Temperature-density diagram illustrating the different
regimes of magnetic field effects on the thermodynamic properties of a
free electron gas. The solid lines are for $B=10^{12}$~G, short-dashed
lines for $B=10^{13}$~G, and long-dashed lines for $B=10^{14}$~G. For
each value of $B$, the vertical lines correspond to $\rho=\rho_B$ (the
density below which only the ground Landau level is occupied by the
degenerate electrons) and $\rho=\rho_{B1}$ (the density below which
only the $n=0,1$ levels are occupied); the Fermi temperature is shown
for $\rho\le\rho_B$ and for $\rho_B<\rho\le\rho_{B1}$; the line marked
by ``$T=T_B$'' [see Eq.~(\ref{eqtb})] corresponds to the temperature
above which the Landau level effects are smeared out.  The dotted line
gives the Fermi temperature at $B=0$. The magnetic field is strongly
quantizing when $\rho\lo \rho_B$ and $T\lo T_B$, weakly quantizing
when $\rho\go \rho_B$ and $T\lo T_B$, and non-quantizing when $T\go
T_B$ or $\rho\gg\rho_B$. }
\label{fig:t_rho}
\end{figure}

Finite temperature tends to smear out Landau levels. 
Let the energy difference between the $n=n_{\rm max}$ level and the
$n=n_{\rm max}+1$ level be $\Delta E_B$. We can define a ``magnetic
temperature'' 
\be
T_B={\Delta E_B\over k_B}={m_e c^2\over k_B}\left(\sqrt{1+2n_{\rm max}\beta
+2\beta}-\sqrt{1+2n_{\rm max}\beta}\right).
\label{eqtb}\ee
Clearly, $T_F=T_B$ at $\rho=\rho_B$ (see Fig.~\ref{fig:t_rho}). 
The effects due to Landau quantization are diminished when $T\go T_B$. 
For $\rho\le \rho_B$, we have
$T_B=\left(\sqrt{1+2\beta}-1\right)(m_ec^2/k_B)$, which reduces to
$T_B\simeq \hbar\omega_{ce}/k_B$ for $\beta=B/B_Q \ll 1$. 
For $\rho\gg \rho_B$ (or $n_{\rm max}\gg 1$), equation (\ref{eqtb})
becomes
\be
T_B\simeq {\hbar\omega_{ce}\over k_B}\left({m_e\over m_e^\ast}\right) 
=1.34\times 10^8\,B_{12}\,(1+x_F^2)^{-1/2}~{\rm K},
\ee
where $m_e^\ast=\sqrt{m_e^2+(p_F/c)^2}=m_e\sqrt{1+x_F^2}$, with
$x_F$ given by Eq.~(\ref{eqxf}). 

There are three regimes characterizing the effects of Landau
quantization on the thermodynamic properties of the electron gas 
(see Fig.~\ref{fig:t_rho}; see also Yakovlev \& Kaminker 1994):

(i) $\rho\lo\rho_B$ and $T\lo T_B$: In this regime, the electrons
populate mostly the ground Landau level, and the magnetic field
modifies essentially all the properties of the gas. The field is
sometimes termed ``strongly quantizing''. For example, for degenerate,
nonrelativistic electrons ($\rho<\rho_B$ and $T\ll T_F\ll
m_ec^2/k_B$), the internal energy density and pressure are 
\ba
&&u_e={1\over 3}n_eE_F,\\ &&P_e=2u_e={2\over 3}n_eE_F\propto
B^{-2}\rho^3.
\label{eqpe1}\ea
These should be compared with the $B=0$ expression 
$P_e=2u_e/3\propto\rho^{5/3}$.
Note that for nondegenerate electrons ($T\gg T_F$), the classical ideal gas
equation of state, 
\be
P_e=n_ek_BT,
\ee
still holds in this ``strongly quantizing''
regime, although other thermodynamic quantities are 
significantly modified by the magnetic field.

(ii) $\rho\go\rho_B$ and $T\lo T_B$: In this regime, the electrons are
degenerate (note that $T_F>T_B$ when $\rho>\rho_B$; see
Fig.~\ref{fig:t_rho}), and populate many Landau levels but the level
spacing exceeds $k_BT$. The magnetic field is termed ``weakly
quantizing''.  The bulk properties of the gas (e.g., pressure and
chemical potential), which are determined by all the electrons in the
Fermi sea, are only slightly affected by such magnetic
fields. However, the quantities determined by thermal electrons near
the Fermi surface show large oscillatory features as a function of
density or magnetic field strength. These de Haas - van Alphen type
oscillations arise as successive Landau levels are occupied with
increasing density (or decreasing magnetic field).  The oscillatory
quantities are usually expressed as derivatives of the bulk quantities
with respect to thermodynamic variables; examples include heat
capacity, magnetization and magnetic susceptibility, adiabatic index
$(\partial\ln P_e/\partial\ln\rho)$, sound speed, and electron
screening length of an electric charge in the plasma (e.g., Ashcroft
\& Mermin 1976; Blandford \& Hernquist 1982; Lai \& Shapiro 1991;
Yakovlev \& Kaminker 1994).  With increasing $T$, the oscillations
become weaker because of the thermal broadening of the Landau levels;
when $T\go T_B$, the oscillations are entirely smeared out, and the
field-free results are recovered.
 
(iii) $T\go T_B$ or $\rho\gg\rho_B$: In this regime, many
Landau levels are populated and the thermal widths of the Landau
levels ($\sim k_BT$) are higher than the level spacing. The magnetic
field is termed ``non-quantizing'' and does not affect the
thermodynamic properties of the gas.

\subsection{Magnetized Crusts, Effects on Thermal Structure and 
Cooling of Neutron Stars}
\label{subsec:crust}

As discussed above, the effect of Landau quantization on the
density-pressure relation is important
only for $\rho\lo \rho_B$ [see Eq.~(\ref{eqrhob})]. Deeper in the
neutron star envelope (and the interior), we expect the magnetic field
effect on the bulk equation of state to become negligible as more 
Landau levels are filled. In general, we can use the condition
$\rho_B\go \rho$, or $B_{12}\go 27\,(Y_e\rho_6)^{2/3}$, to estimate
the critical value of $B$ above which Landau quantization will affect
physics at density $\rho$. For example, at $B\go 10^{14}$~G, the
neutronization transition from $^{56}$Fe to $^{62}$Ni (at
$\rho=8.1\times 10^{6}$~g~cm$^{-3}$ for $B=0$; Baym, Pethick 
\& Sutherland 1971) in the crust can be significantly 
affected by the magnetic field (Lai \& Shapiro 1991).

The ions in the neutron star envelope form a one-component plasma and
are characterized by the Coulomb coupling parameter 
\be
\Gamma={(Ze)^2\over r_i k_BT}=22.75\,{Z^2\over T_6}\left({\rho_6\over
A} \right)^{1/3}, \ee 
where $r_i=(3/4\pi n_i)^{1/3}$ is the
Wigner-Seitz cell radius, $n_i=\rho/m_i =\rho/(Am_p)$ is the ion
number density, $\rho_6=\rho/(10^6\,{\rm g~cm}^{-3})$ and
$T_6=T/(10^6\,{\rm K})$. For $\Gamma\ll 1$, the ions form a classical
Boltzmann gas whose thermodynamic property is unaffected by the
magnetic field. For $\Gamma\go 1$, the ions constitute a strongly
coupled Coulomb liquid. The liquid freezes into a Coulomb crystal at
$\Gamma=\Gamma_m\simeq 175$, corresponding to the classical melting
temperature $T_m$ (e.g., Nagara \etal~1987; Potekhin \& Chabrier
2000). The quantum effects of ion motions (zero-point vibrations)
tends to increase $\Gamma_m$ (Chabrier et al.~1992; Chabrier 1993) or
even suppress freezing (e.g., Ceperley \& Alder 1980; Jones \&
Ceperley 1996). At zero-field, the ion zero-point vibrations have
characteristic frequency of order the ion plasma frequency $\Omega_p$,
with \be \hbar\omega_{pi}=\hbar\left({4\pi Z^2e^2n_i\over
m_i}\right)^{1/2}=675\, \left({Z\over A}\right)\,\rho_6^{1/2}~{\rm
eV}.  \ee For $T\ll T_{\rm Debye}\sim \hbar\omega_{pi}/k_B$, the ion
vibrations are quantized. The effects of magnetic field on strongly
coupled Coulomb liquids and crystals have not been systematically
studied (but see Usov \etal~1980).  The cyclotron frequency of the ion
is given by $\hbar\omega_{ci}=\hbar
(ZeB/Am_pc)=6.3\,(Z/A)\,B_{12}~{\rm eV}$.  The ion vibration frequency
in a magnetic field may be estimated as
$(\omega_{pi}^2+\omega_{ci}^2)^{1/2}$. Using Lindeman's rule, we
obtain a modified melting criterion: \be
\Gamma\,\left(1+{\omega_{ci}^2\over\omega_{pi}^2}\right)\simeq 175.
\ee For $\omega_{ci}\ll \omega_{pi}$, or $B_{12}\ll
100\,\rho_6^{1/2}$, the magnetic field does not affect the melting
criterion and other properties of ion vibrations.

A strong magnetic field can significantly affect the transport
properties and thermal structure of a neutron star crust.
Even in the regime where the magnetic quantization effects are small
($\rho\gg\rho_B$), the magnetic field can still greatly modify 
the transport coefficients (e.g., electric conductivity and heat
conductivity). This occurs when the effective
gyro-frequency of the electron, $\omega_{ce}^\ast=eB/(m_e^\ast c)$,
where $m_e^\ast=\sqrt{m_e^2+(p_F/c)^2}$, is much larger than the
electron collision frequency, i.e., 
\be \omega_{ce}^\ast\tau_0\simeq
1.76\times 10^3\,{m_e\over m_e^\ast}\,B_{12}\,\left({\tau_0\over
10^{-16}\,{\rm s}}\right), \ee 
where $\tau_0$ is the effective
electron relaxation time. In a degenerate Coulomb plasma with a 
nonquantizing magnetic field, $\tau\simeq 
3\pi\hbar^3/(4Zm_e^\ast e^4\ln\Lambda)
\simeq 6\times 10^{-17}m_e/(m_e^\ast Z\ln\Lambda)$~s.
When $\omega_{ce}^\ast\tau_0\gg 1$, the electron
heat conductivity perpendicular to the magnetic field,
$\kappa_\perp$, is suppressed by 
a factor $(\omega_{ce}^\ast\tau_0)^{-2}$. In this classical 
regime, the heat conductivity along the field, $\kappa_\parallel$, 
is the same as the $B=0$ value. In a quantizing magnetic field,
the conductivity exhibits oscillatory behavior of the
de Haas - van Alphen type. On average, the longitudinal
conductivity is enhanced relative to the $B=0$ value.
The most detailed calculations of the electron transport 
coefficients of magnetized neutron star envelopes
are due to Potekhin (1999), where earlier references
can be found (see also Hernquist 1984, Yakovlev \& Kaminker 1994).

The thermal structure of a magnetized neutron star envelope 
has been studied by many authors. Hernquist (1985) and
Van Riper (1988) considered the cases where $\bB$ is normal
or tangential to the stellar surfaces, while
Schaaf (1990), Heyl \& Hernquist (1998,2001), 
Potekhin \& Yakovlev (2001) (for a Fe crust)
and Potekhin et al.~(2003) (for an accreted magnetized
envelope) analysed arbitrary magnetic orientations and 
used increasingly more accurate transport coefficients.
In general, a normal magnetic field reduces the thermal insulation
as a result of the (on average) increased $\kappa_\parallel$ due to 
Landau quantization of electron motion, while 
a tangential magnetic field (parallel to the stellar surface)
increases the thermal insulation of the envelope because the
Larmor rotation of the electron significantly reduces
the transverse thermal conductivity $\kappa_\perp$.
A consequence of the anisotropic heat transport is that for a given
internal temperature of the neutron star, the surface temperature 
is nonuniform, with the magnetic poles hotter and the magnetic
equator cooler (see Fig.~\ref{fig:tprofile}).
Additional heating and surface temperature inhomogeneity comes from
the bombardment of high-energy particles from the magnetosphere
on the polar cap.  Models for particle acceleration in neutron star
magnetospheres predict that some fraction of electron-positron pairs that are 
produced in pair cascades (see \S \ref{sec:pulsars}) decelerate, 
turn around and accelerate downward to the neutron-star surface. 
In space-charge limited flow models (e.g. Arons \& Scharlemann 1979, Harding
\& Muslimov 2001), the trapped positrons needed to screen the 
$E_{\parallel}$ are only a small fraction of the number of primary electrons. 
Such models predict moderate heating, so that the heated polar caps have 
X-ray luminosities that are small compared to cooling luminosities for
typical ages less than about $10^6$ yr.
In polar vacuum gap (e.g. Ruderman \& Sutherland 1975) and outer gap
(Cheng et al. 1986) models, roughly half of the produced particles return to 
heat the surface, causing a much higher degree of polar cap heating.

\begin{figure} 
\vskip -5cm
\hskip -1cm
\includegraphics[width=42pc]{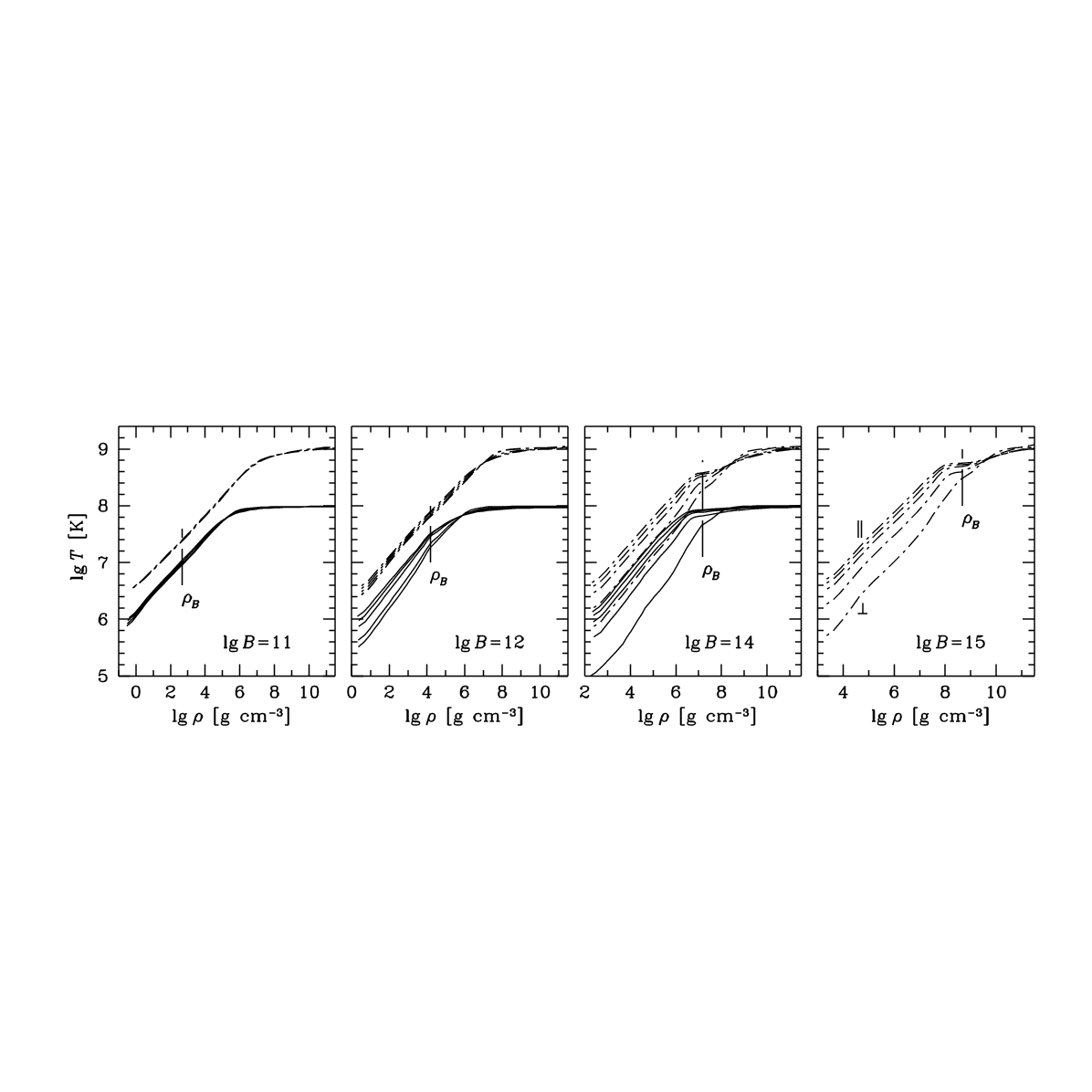}
\vskip -6cm
\hskip -2cm
\caption{Temperature profiles through the iron envelope of a neutron 
star at (from left to right) $B=10^{11},\,10^{12}\,10^{14}$ and
$10^{15}$~G. The internal temperature is fixed to $10^8$~K (solid
lines) or $10^9$~K (dot-dashed lines). The lines of
each group correspond to $\cos\theta=1$ (the lowest line), 0.7, 0.4,
0.1, and 0 (the highest line). From Potekhin \& Yakovlev (2001).}
\label{fig:tprofile}
\end{figure}

A superstrong magnetic field ($B\go 10^{14}$~G) also 
affects the cooling curve of a neutron star
(see Potekhin \& Yakovlev 2001; Potekhin et al.~2003).
This is because for a given core temperature
(whose time evolution largely depends on neutrino
emission from the core, and thus is unaffected by 
the field strength unless $B\gg 10^{15}$~G), an enhanced
radiation flux emerges from the magnetic polar regions,
and such enhancement more than compensates for the reduced
flux from the equatorial region where the heat conductivity
is suppressed. Figure \ref{fig:nscoolb}
shows some examples of theoretical cooling curves
for neutron stars with mass $1.5M_\odot$
and $1.3M_\odot$, and with different surface
field strengths (assuming dipole field structure).
Note that in these models, the $1.5M_\odot$ star has
a sufficiently high central density for the direct
URCA process to operate, while only the slower, modified 
URCA process operates in the $1.3M_\odot$ star (see
Yakovlev \& Pethick 2004); thus the $1.5M_\odot$ star
cools much faster than the $1.3M_\odot$ star. We also see
that for $t\lo 10^5$~years, a superstrong magnetic field
($B\go 10^{14}$~G) makes the neutron star brighter, while at later 
times the star cools down more quickly compared to the $B=0$ case.

\begin{figure} 
\includegraphics[width=36pc]{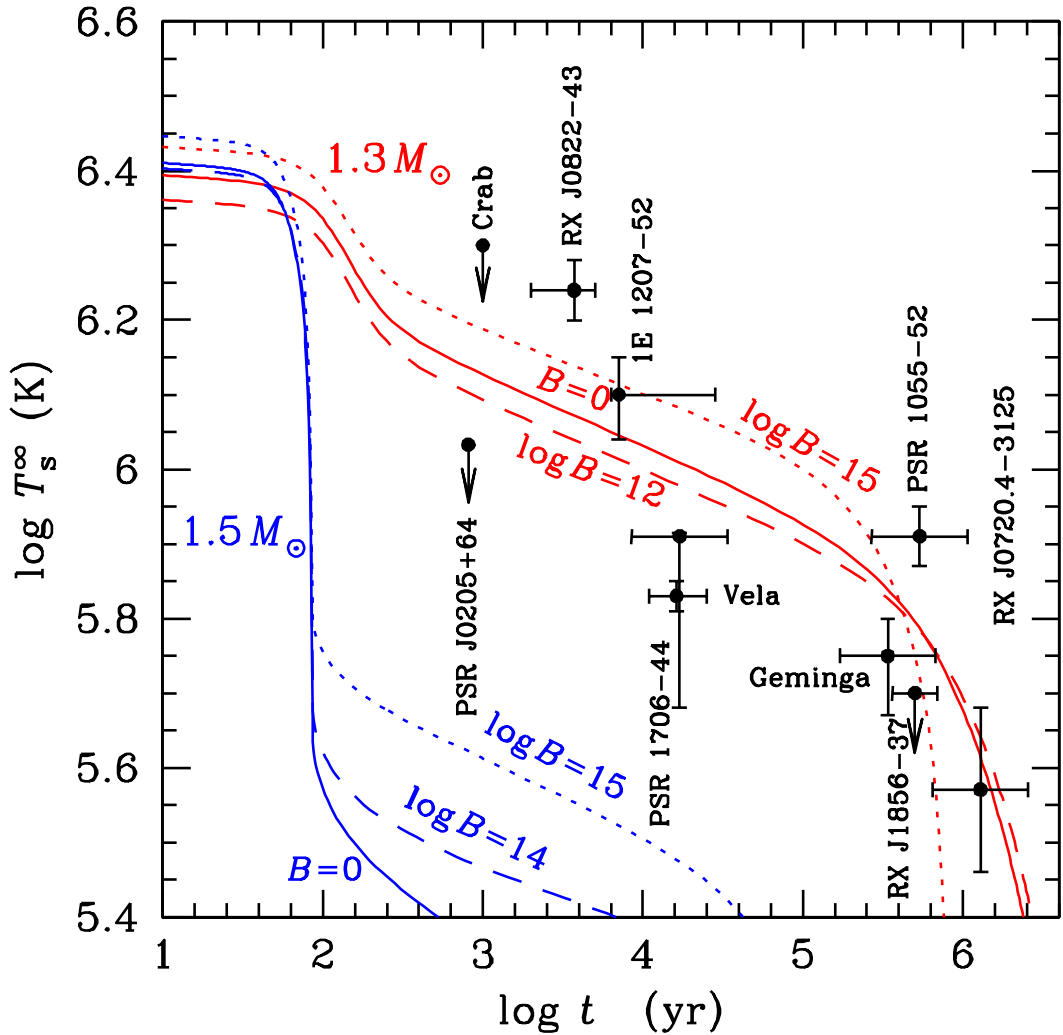}
\caption{Effective surface temperature (as seen by distant
observer) versus neutron star age for assumed mass $1.3M_\odot$
and $1.5M_\odot$. The dots with error bars show estimates of neutron
star ages and effective temperatures from various observations;
the dots with errors indicate observational upper limits.
The different curves show the cooling of neutron stars with 
iron envelope for different magnetic field strengths
(log $B$ in Gauss). From Chabrier, Saumon \& Potekhin (2006).}
\label{fig:nscoolb}
\end{figure}


\subsection{Outermost Envelopes}
\label{subsec:outermost}

The outermost layer of a neutron star is of great importance since it
mediates the emergent radiation from the stellar surface to 
the observer. The chemical composition of this layer is unknown. 
The surface of a fully catalyzed neutron star is expected to consist
of $^{56}$Fe formed at the star's birth, although C, O may also be
present. This may be the case for young radio pulsars that have 
not accreted any gas. Once the neutron star accretes material, 
or has gone through a phase of accretion, either during 
supernova fallback, or from the interstellar medium or 
from a binary companion, the surface composition can be quite different.
A H layer is expected even if a small amount of fallback/accretion 
occurs after neutron star formation. A He layer results if H 
is completely burnt out. The atmosphere composition may 
also be affected by the (slow)  nuclear burning on the NS surface 
layer (Chang, Arras \& Bildsten 2004) and by the bombardment 
of high-energy particles on the surface (Thompson \& Beloborodov 2005).

Despite the uncertainties in the surface composition mentioned 
in the last paragraph, a great simplification arises due 
to the efficient gravitational separation of light and heavy elements.
As an example, consider a trace amount of ions 
(with charge number $Z_i$ and mass number $A_i$) embedded in 
the ionized hydrogen plasma (with the electron number density $n_e$)
of a neutron star atmosphere.
The downward drift speed of the ion relative to the H plasma is 
$v_{\rm drift}\simeq g\,t_f$, where $g$ is the gravitational acceleration.
The friction timescale $t_f$ can be estimated from
${A_im_p/t_f}\approx n_e m_p \sigma_{pi}v_{p}$,
where the ion-proton Coulomb collision cross-section is $\sigma_{pi}
\sim (Z_ie^2/k_BT)^2\ln\Lambda$ (here 
$\ln \Lambda$ is the Coulomb logarithm),
and the thermal velocity of the proton is $v_p\sim (k_BT/m_p)^{1/2}$. 
This gives
\be
t_f\sim {A_i\over Z_i^2\ln\Lambda}{m_p^{1/2}(k_BT)^{3/2}\over e^4n_e}.
\ee
Using typical parameters ($g\sim 2\times 10^{14}$~cm~s$^{-2}$, 
$T\sim 10^6$~K, $n_e\sim 10^{24}$~cm$^{-3}$), we find that the
drift speed is of order $10\, (A_i/Z_i^2\ln\Lambda)$~cm~s$^{-1}$.
The timescale for gravitational settling over the pressure scale
height $H\sim k_BT/(m_pg)$ is of order a second.
Thus the lightest elements, H and He, if present, are 
the most important chemical species in the outermost layer of the star.

The hydrogen envelope of a neutron star may assume different forms,
depending on the temperature and magnetic field.  For ``modest'' field
strengths ($B\lo {\rm a~few}\times 10^{13}$~G) and temperatures
($T\go 10^5$~K), conditions satisfied by most observable neutron stars,
the atmosphere is largely nondegenerate, and consists mainly of
ionized hydrogen, H atoms and small H$_N$ molecules, and 
the condensed phase can be neglected in the photosphere.
As density increases, the matter gradually transform into a degenerate
Coulomb plasma (see above). 
The atmosphere (with density $\rho\sim 10^{-3}-10^4$~g~cm$^{-3}$)
constitutes a non-ideal, partially ionized, Coulomb plasma.
An important issue concerns the ionization equilibrium 
of atoms in such a plasma. This is nontrivial 
because in a strong magnetic field, an atom cannot move
free across the field, and there is strong coupling between
the internal atomic structure and the center-of-mass motion
(see \S \ref{subsec:hatom}). 
Lai \& Salpeter (1995) gave an approximate analytic solution for 
a limited magnetic-field--temperature--denisty regime 
(see also Khersonskii 1987; Pavlov \& Meszaros 1993). 
To date the most complete treatment of the equation of state 
(including Saha equilibrium) of partially ionized hydrogen plasma
is that of Potekhin et al.~(1999) (for $10^{11}$~G$\lo B \lo 
10^{13}$~G) and Potekhin \& Chabrier (2004) (for 
$10^{13}$~G$\lo B \lo 10^{15}$~G). They used the numerical energy 
levels and fitting formulae of a moving H atom as obtained by 
Potekhin (1994,1998), and their free energy model takes into account
the Coulomb plasma nonideality, the interactions of bound species 
with one another and with the electrons and protons. 
Ionization equilibrium is given by minimization of the free energy 
with respect to particle numbers under the stoichiometric constraints
(keeping the volume and the total number of free and bound protons
constant). The derivatives of the free energy 
with respect to $\rho$ and $T$ and their combinations provide
the other thermodynamic functions.
Figure \ref{fig:phase} shows the domains of partial ionization
in the $\rho-T$ plane for different values of $B$. The higher $B$, 
the greater $T$ at which the bound species are important. 
The calculated atomic fractions 
are needed to obtain the radiative opacities in the atmosphere.
Note that the Potekhin et al. model becomes less reliable 
at relatively low $T$
where larger molecules can be important (Lai 2001).
Generally, this occurs within the $\rho-T$ domain where
$x_\mathrm{H}\go 0.1$ (i.e., to the left of the solid lines
in Fig.~\ref{fig:phase}).

\begin{figure} 
\includegraphics[width=38pc]{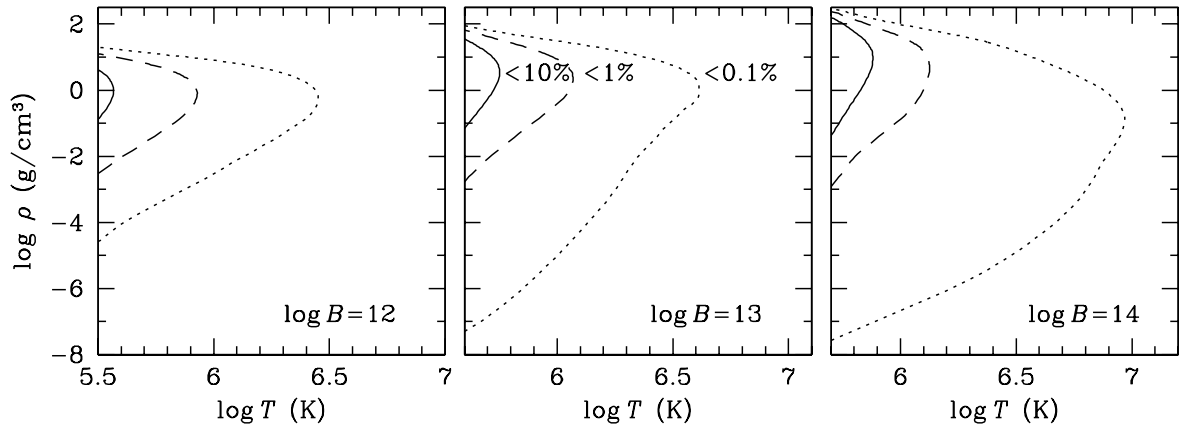}
\caption{Domains of partial ionization in the $\rho-T$ plane
for $B=10^{12}$, $10^{13}$ and $10^{14}$~G.
The contours delimit the domains where the atomic fraction
$x_\mathrm{H}<0.1$\% (to the right of the dotted lines),
0.1\%$<x_\mathrm{H}<1$\% (between the dashed and dotted lines),
1\%$<x_\mathrm{H}<10$\% (between the solid and dashed lines)
or $x_\mathrm{H}>10$\% (to the left of the solid lines).
From Potekhin et al.~(2006).}
\label{fig:phase}
\end{figure}

For sufficiently strong magnetic fields and/or low temperatures, the
hydrogen layer may be in a condensed, metallic form, since the binding
energy of the condensed hydrogen increases as a power-law function of
$B$, while the binding energies of atoms and small molecules increase
only logarithmically.  Lai \& Salpeter (1997) studied the phase
diagram of H under different conditions and showed that in strong
magnetic fields, there exists a critical temperature $T_{\rm crit}$
below which a phase transition from gaseous to condensed state occurs,
with $kT_{\rm crit}$ about $10\%$ of the cohesive energy of the
condensed hydrogen. Thus, $T_{\rm crit}\sim 8\times 10^4,\, 5\times
10^5,\,10^6$~K for $B=10^{13},\,10^{14},\,5\times 10^{14}$~G (Lai
2001). An analogous ``plasma phase transition'' was also obtained in
an alternative thermodynamic model for magnetized hydrogen plasma
(Potekhin et al.~1999). While this model is more restricted than Lai
\& Salpeter (1997) in that it does not include H$_N$ molecules, it
treats more rigorously atomic motion across the strong $B$ field and
Coulomb plasma nonideality. In the Potekhin et al. model, the density
of phase separation is roughly the same as in Lai \& Salpeter (1997),
i.e., $\rho_{s,0}\sim 10^3\,B_{12}^{6/5}$~g~cm$^{-3}$ [see
Eq.~(\ref{rs0})], but the critical temperature is several times
higher. Thus, although there is a factor of a few uncertainty
in $T_{\rm crit}$, it is almost certain that for $T\lo T_{\rm
crit}/2$, the H surface of a neutron star is in the form of 
the condensed metallic state, with negligible vapor above it. 
The radiative property of such condensed phase is 
discussed in \S\ref{sec:atm}.

For heavy elements such as Fe, detailed study of the phase diagram is
not available since our current knowledge of the various states/phases
in strong magnetic fields is not as complete as for H (see \S
\ref{sec:condense}). Calculations so far have shown that at
$B=10^{12}-10^{13}$~G, a linear chain is unbound relative to individual
atoms for $Z\go 6$, and thus chain-chain interactions play an important
role in determining whether 3D zero-pressure condensed matter is bound
or not. For larger field strengths, $B\gsim 10^{14}(Z/26)^3$~G, a linear
chain becomes bound (Medin \& Lai 2006). If the condensed Fe is
unbound with respect to the Fe atom, then the outermost Fe layer of
the neutron star is characterized by gradual transformation from
nondegenerate gas at low densities, which includes Fe atoms and ions,
to degenerate plasma as the pressure (or column density) increases.
On the other hand, even a weak cohesion of the Fe condensate 
can give rise to a phase transition at sufficiently low temperatures.
Crude numerical results by Jones (1986)
together with approximate scaling relations suggest an
upper limit of the cohesive energy (for $Z\go 10$) $Q_s\lo
Z^{9/5}B_{12}^{2/5}$~eV. Thus for
Fe, the critical temperature for phase transition $T_{\rm crit}\lo
0.1Q_s/k\lo 10^{5.5}B_{12}^{2/5}$~K.

The iron surface layers of magnetic neutron stars can also studied
using Thomas-Fermi type models (e.g., Fushiki, Gudmundsson \& Pethick 
1989; Thorolfsson et al.~1998). While these models are too crude to 
determine the cohesive energy of the condensed matter, they provide 
a useful approximation to the gross properties of the neutron star
surface layer. At zero temperature, the pressure is zero 
at a finite density which increases with increasing $B$. 
This feature is qualitatively the same as in the uniform electron gas 
model. Neglecting the exchange-correlation energy and 
the nonuniformity correction, we can write the pressure of a 
zero-temperature uniform electron gas as
\be
P=P_e-{3\over 10}\left({4\pi\over 3}\right)^{1/3}\!\!\!\!
(Ze)^2\!\left({\rho\over Am_p}\right)^{4/3},
\label{puniform}\ee
where the first term is given by Eq.~(\ref{eqpe0}) [or by
Eq.~(\ref{eqpe1}) in the strong field, degenerate limit], and the
second term results from the Couloumb interactions among the electrons
and ions. Setting $P=0$ gives the condensation density $\rho_{s,0}$ as
in Eq.~(\ref{rs0}). At finite temperatures, the pressure does not
go to zero until $\rho\rightarrow 0$, i.e., an atmosphere is
present. Note that the finite-temperature Thomas-Fermi model only
gives a qualitative description of the dense atmsophere; important
features such as atomic states and ionizations are not captured in
such a model.

\section{Neutron Star Magnetic Field Evolution} \label{sec:Bfield}

Magnetic field is perhaps the single most important quantity
that determines the various observational manifestations of 
neutron stars. Thus it is natural that a large amount of work 
has been devoted to the study of neutron star magnetic field 
evolution. Recent reviews include Bhattacharya \& Srinivasan (1995),
Ruderman (2004), Reisenegger et al.~(2005).

\subsection{Observations}

As discussed in \S \ref{sec:intro}, with a few exceptions, our current
knowledge of neutron star magnetic fields are largely based on
indirect inferences:

(i) For radio pulsars, of which about 1600 are known today (see
Manchester 2004), the period $P$ and period derivative $\dot P$ can be
measured by timing the arrivals of radio pulses. If we assume that the
pulsar spindown is due to magnetic dipole radiation, 
we obtain an estimate of the neutron star surface field $B$
[see Eq.~(\ref{eq:B0})].  For most radio
pulsars, the inferred $B$ lies in the range of
$10^{11}-10^{14}$~G. For a smaller population of older, millisecond
pulsars, $B\sim 10^{8-9}$~G -- Such field reduction is thought to be
related to the recycling process that turns a dead pulsar into an
active, millisecond pulsar.

(ii) For Anomalous X-ray pulsars and Soft Gamma Repeaters (Woods \&
Thompson 2005), superstrong magnetic fields ($B\sim
10^{14}-10^{15}$~G) are again inferred from the measured $P$ (in the
range $5-12$~s) and $\dot P$ (based on X-ray timing study) and the
assumption that the spin-down is due to the usual magnetic dipole
radiation or Alfv\'{e}n wave emission. In addition, strong theoretical
arguments have been put forward (Thompson \& Duncan 1995,1996,2001)
which suggest that a superstrong magnetic field is needed to explain
various observed properties of AXPs and SGRs (e.g., energetics of SGR
bursts/flares, including the spectacular giant flares from three SGRs:
the March 5 1979 flare of SGR 0525-66 with energy $E\go 6\times
10^{44}$~erg, the August 27 1998 flare from SGR 1900+14 with $E\go
2\times 10^{44}$~erg, and the December 27 2004 flare from SGR 1806-20
with $E\sim 4\times 10^{46}$~erg; quiescent luminosity $L_x\sim
10^{34-36}$~erg/s, much larger than the spindown luminosity
$I\Omega\dot\Omega$) (see \S \ref{sec:magnetars}).
Tentative detections of spectral features during SGR/AXP bursts have
been reported in several systems (e.g., Gavriil et al.~2003; Ibrahim
et al.~2003; Rea et al.~2003), which, when interpreted as proton
cyclotron lines, imply $B\sim 10^{15}$~G.

(iii) For a number of thermally emitting neutron stars (``dim isolated
neutron stars''), spectral features with energy $0.2-1$~keV have been
detected (see \S\ref{subsec:obs}). While the identification of these
features is unclear, if one assumes that they are due to proton
cyclotron resonance or atomic transitions in light elements, one would
infer $B\sim 10^{13}-10^{14}$~G.
 
(iv) For about a dozen accreting X-ray pulsars in binary
systems, electron cyclotron features have been detected,
implying $B\sim 10^{12}-10^{13}$~G (see \S 1). For many other 
X-ray pulsars with no detectable cyclotron features,
we can use the measured spin period together with theoretical ideas 
of spin equilibrium (i.e., spinup due
to accretion of matter is balanced by magnetic braking -- loss of
stellar angular momentum via magnetic fields) to estimate the surface
magnetic field of the neutron star. For the $\sim 100$ 
neutron stars in high-mass X-ray binaries, this typically 
gives $B\sim 10^{12}$~G (e.g., Bildsten et al.~1997).
In the last few years, half a dozen or so accreting
millisecond pulsars have been discovered (e.g., Chakrabarty 2005),
with the estimated field strength $B\sim 10^9$~G --- these systems 
are thought to be neutron stars undergoing the recycling process.

Several lines of observations/arguments point toward the possibility
of an evolving magnetic fields in neutron stars.

(i) From the discussion above, it appears that young neutron stars
have strong magnetic fields $\sim 10^{11}-10^{15}$~G (most radio
pulsars, magnetars, high-mass X-ray binaries), whereas old neutron
stars have weak fields $\lo 10^9$~G (millisecond pulsars, low-mass
X-ray binaries). If these two groups have an evolutionary connection,
then the field must be significantly reduced on timescale of
$10^9$~yr. Most likely, this reduction is associated with the
accretion process that recycled the neutron star.

(ii) Magnetic field evolution in isolated pulsars has long been
studied using the statistical distribution of pulsars on 
the $P-\dot P$ diagram, as well as pulsars' 
spatial and velocity distributions. Such study probes field 
evolution on the timescale of the ages of 
radio pulsars ($\lo 10^7$~yr). Unfortunately, the conclusions of such 
population study have often been conflicting. For example,
Narayan \& Ostriker (1990) and  
Gonthier et al.~(2004) suggested that field decay occurs on timescale
$\lo 5$~Myr. Similar studies by Bhattacharya et al.~(1992), 
Lorimer et al.~(1997), Faucher-Gigu\'er\'e \& Kaspi (2006) found no
evidence for field decay during the radio pulsar lifetime, implying a decay 
time constant $\go 100$~Myr. Such divergent results illustrate 
the difficulties of controlling various systematic uncertainties
(e.g. selection effects, luminosity evolution law, dependence of
beaming fraction on period) in pulsar population studies.

(iii) Magnetar emission is most likely powered by magnetic energy,  
implying the dissipation of superstrong magnetic fields
(see Thompson \& Duncan 1995,1996; Woods \& Thompson 2005).

(iv) One explanation for the ``anomalous'' braking indices $n<3$ in
young pulsars is that their magnetic dipole field increases with time
(but see Melatos 1997).
It has also been pointed out that following each pulsar glitch, the
apparent magnetic field $\propto \sqrt{P\dot P}$) increases, which
might imply that some pulsars (Crab and Vela), may evolve into
magnetars (Lyne 2004; Lin \& Zhang 2004). Since many AXPs and SGRs are
observed to be associated with supernova remnants or clusters of
massive stars, most magnetars are expected to be younger than radio pulsars,
rather than the other way around.

\subsection{Origin of Neutron Star Magnetic Field}
\label{subsec:origin}

The magnetic fields of neutron stars were most likely already present
at birth. The traditional fossil field hypothesis suggests that the
magnetic field is inherited from the progenitor, with magnetic flux
conserved and field amplified ($B\propto R^{-2}$) during core
collapse.  In the case of magnetic white dwarfs (with measured fields
in the range $\sim 3\times 10^4 - 10^9$~G), there is strong evidence
that the fields are the remnants from a main-sequence phase (Ap/Bp
stars, with $B\sim 200$~G --25~kG) (see, e.g., Ferrario \&
Wichramasinghe 2005). Neutron stars descend from main-sequence stars
with mass $\go 8M_\odot$ (i.e., O and early B stars).  Only recently
have the large-scale magnetic fields (with $B\sim 1$~kG) of O stars
been detected (in two stars so far; see Donati et al.~2006).
It is interesting that the magnetic flux of such O stars $\Phi\sim
10^5\pi R_\odot$~G (for $R\sim 10R_\odot$; of course, not
all of this flux threads the inner $1.4M_\odot$ core) is of the same order 
of magnitude as the flux of a $10^{15}$~G neutron star 
($R\sim 10^{-5}R_\odot$) as well as the fluxes of the most strongly magnetic
Ap/Bp stars and white dwarfs. It is also of interest to note that 
magnetic white dwarfs ($B\go 1$~MG) tend to be more massive 
(mean mass $\sim 0.93M_\odot$) than their non-magnetic counterparts 
(mean mass $\sim 0.6M_\odot$). Since white dwarfs with $M\go 0.7M_\odot$
and neutron stars form exclusively from the material that belongs to the
convective core of a main-sequence star (Reisenegger 2001), this suggests
that the magnetic field may be generated in the convective core 
of the main-sequence progenitor. It has also been 
suggested that magnetized neutron stars could be produced, in principle, 
by accretion-induced collapse of magnetic white dwarfs (Usov 1992).

Alternatively, it has been argued that magnetic field  may be
generated by a convective dynamo in the first $\sim 10$ seconds of 
a proto-neutron star (Thompson \& Duncan 1993). In principle,
the maximum field achievable is either $B\sim (4\pi\rho)^{1/2}v_{\rm con}
\sim 4\times 10^{15}$~G (for convective eddy speed 
$v_{\rm con}\sim 10^3$~km~s$^{-1}$) or $B\sim (4\pi\rho)^{1/2}
R\Delta\Omega\sim 2\times 10^{17}$~G [for differential rotation $\Delta\Omega
\sim 2\pi/(1~{\rm ms})$]. How much dipole field can be generated is
more uncertain. It could be that a large-scale field of $\sim 10^{15}$~G
is generated if the initial spin period of the neutron star is 
comparable to the convective turnover time ($\sim H_p/v_{\rm con}\sim 1$~ms
for pressure scale height of 1~km), whereas in general only a small-scale
($\sim H_p$) field is produced, resulting in mean field of order 
$10^{12}-10^{13}$~G typical of radio pulsars. We note that  
studies of presupernova evolution of massive stars, including magnetic fields,
typically find that a newly formed neutron star has rotation rate appreciably
slower than the breakup rate (Heger et al.~2005).

In either the fossil field or dynamo scenario, the magnetic field 
of a proto-neutron star is likely ``messy'' and unstable. 
Before the crust forms (around $100$~s afetr collapse),
the field will evolve on the Alfven crossing time $t_A\sim (4\pi \rho)^{1/2}
R/B$ ($\sim 0.1$~s for $B\sim 10^{15}$~G) into a stable configuration 
in which a poloidal field coexists with a toroidal field of the same
order of magnitude (see Braithwaite \& Spruit 2006).

Gradual field generation due to a thermomagnetic effect 
in the crust has been discussed (Urpin \& Yakovlev 1980; 
Blandford et al.~1983; Wiebicke \& Geppert 1996), but it seems
unlikely to be capable of producing large-scale fields stronger than
$10^{12}$~G.

\subsection{Physics of Field Evolution in Isolated Neutron Stars}

The magnetic field in a neutron star evolves through a series
of quasi-equilibrium states, punctuated by the release of 
elastic stress in the crust and hydrodynamic motions in its liquid core.
The physics of the quasi-equilibrium field evolution was
discussed by Goldreich \& Reisenegger (1992) (see also
Reisenegger et al.~2005). The bulk region of a neutron star comprises 
a liquid core of mostly neutrons with a small fraction ($Y_e\sim 
{\rm a~few}\%$) of protons and electrons. The medium is stably stratified
due to the $Y_e$ gradient, thus the magnetic field cannot force
the bulk neutron fluid to move (e.g. due to buoyancy) unless
$B^2/(8\pi)\go Y_e P$ (or $B\go 10^{17}$~G) or the fluid can change its
composition as it rises (which takes place on a timescale longer than the
neutrino cooling time for $B\lo 10^{17}$~G). The magnetic field
evolves according to
\be
{\partial\bB\over\partial t}=-\nabla\times
\left({c^2\over 4\pi\sigma}\nabla\times
\bB\right)+\nabla\times\left(-{\bj\over n_ee}\times\bB\right)+
\nabla\times\left(\bv_a\times\bB\right),
\ee
where the three terms on the right-hand side represent three different
effects:

(i) The first term represents Ohmic diffusion of the magnetic field
($\sigma$ is the zero-field conductivity defined by $\bj=\sigma{\bf
E}$), with timescale 
\be 
t_{\rm Ohmic}\sim {4\pi \sigma L^2\over c^2}
\ee 
for fields which vary on lengthscale $L$. In the core, the
conductivity is large since all the particles are degenerate, leading to
$t_{\rm Ohmic}\sim 2\times 10^{11}(L_{\rm km}/T_8)^2 (\rho/\rho_{\rm
nuc})^3$~yr (Baym, Pethick \& Pines 1969), where $\rho_{\rm
nuc}=2.8\times 10^{14}$~g~cm$^{-3}$, $L_{\rm km}=L/(1~{\rm km})$.  
Of course, currents which are confined to the crust would have a much
shorter decay time (e.g., Sang \& Chanmugam 1987; Cumming et
al.~2004).

(ii) The second term describes advection of the field by Hall drift:
The magnetic field is carried by the electron fluid, which drifts with
respect to the ions with velocity $\bv_e=-\bj/(n_ee)$. This term is
non-dissipative, but can change the field structure on timescale \be
t_{\rm Hall}\sim {4\pi n_e eL^2\over cB}\sim 5\times 10^8 {L_{\rm
km}^2\over B_{12}}\left({\rho\over\rho_{\rm nuc}}\right)~{\rm yr}.
\ee 
Goldreich \& Reisenegger (1992) suggested that the nonlinear Hall
term may give rise to a turbulent cascade to small scale, thus
enhancing the Ohmic dissipation rate of the field. This ``Hall
cascade'' has been confirmed by recent simulations of electron MHD
turbulence (also known as whistler turbulence; see Cho \& Lazarian
2004). Several studies have also demonstrated magnetic energy transfer 
between different scales (e.g., Urpin \& Shalybkov 1999; 
Hollerbach \& R\"udiger 2002). 
A specific example of the Hall-drift enhanced Ohmic
dissipation can be seen by considering a purely toroidal field (see
Reisenegger et al.~2005): In this case, the evolution equation for the
field reduces to the Burgers equation (see also Vainshtein et
al.~2000), where the field tends to develop a cusp (and the associated
current sheet) and fast Ohmic decay. Recently, Rheinhardt \& Geppert
(2002) and Rheinhardt et al.~(2004) discussed a ``Hall drift
instability'' that can lead to nonlocal transfer of magnetic energy to
small scales, leading to enhanced crustal field dissipation.

(iii) The third term describes ambipolar diffusion, which involves
a drift of the combined magnetic field and the bulk electron-proton fluid
relative to the neutrons. The drift velocity $\bv_a$ is determined by
force balance $m_p\bv_a/\tau_{pn}={\bf f}_B-\nabla (\Delta\mu)$,
where $\tau_{pn}$ is the proton-neutron collision time,  
${\bf f}_B=\bj\times\bB/(cn_p)$ is the magnetic force per proton-electron
pair, and $\nabla(\Delta\mu)$ (with $\Delta\mu=\mu_p+\mu_e-\mu_n$) is 
the net pressure force due to imbalance of $\beta$-equilibrium (Goldreich
\& Reisenegger 1992). There are two modes of ambipolar diffusion:
The solenoidal mode ($\nabla\cdot \bv_a=0$) is noncompressive,
and does not perturb $\beta$-equilibrium, thus $\bv_a=\tau_{pn}પ
\tau_{pn}{\bf f}_B/m_p$, and the associated diffusion time 
$t_{\rm Amb}^{\rm s}\sim L/v_a$ is
\be
t_{\rm Amb}^{\rm s}\sim {4\pi n_p m_p L^2\over \tau_{pn}B^2}
\sim 3\times 10^9\,{T_8^2L_{\rm km}^2\over B_{12}^2}~{\rm yr},
\ee
where we have used $n_p\simeq 0.05\rho/m_p$ and $\tau_{pn}
\simeq 2\times 10^{-17}T_8^{-2}(\rho/\rho_{\rm nuc})^{1/3}$~s.
The irrotational mode ($\nabla\times\bv_a=0$) is compressive,
and is impeded by chemical potential gradients $\nabla (\Delta\mu)$,
and is possible only if weak interaction re-establishes 
$\beta$-equilibrium during the drift. This gives a 
timescale $t_{\rm Amb}^{\rm ir}\sim t_{\rm cool}/B_{17}^2\sim
5\times 10^{15}T_8^{-6}B_{12}^{-2}$~yr
(where $t_{\rm cool}$ is the neutrino cooling time and modified
URCA rate has been used). Like Ohmic decay,
ambipolar diffusion is dissipative, which leads to 
heating of the core and deep crust of a magnetar; 
this may be the power source for the persistent X-ray emission 
of magnetars (Thompson \& Duncan 1996; Heyl \& Kulkarni 1998; 
Arras et al.~2004; see \S\ref{sec:magnetars}).

In addition the the ``steady'' field evolution discussed above,
crust fracture can also lead to sudden change of the magnetic field
(Thompson \& Duncan 1996). The crust has a finite shear 
modulus $\mu$, and when the yield strain $\theta_{\rm max}$ 
is exceeded the lattice will fracture.
The characteristic yield field strength is of order 
$\sim (4\pi\theta_{\rm max}\mu)^{1/2}=2\times 10^{14}
(\theta_{\rm max}/10^{-3})^{1/2}$~G. For example, Hall drift of the
magnetic field causes the stresses in the crust to build up, 
and irregularities in the field can be damped quickly by 
crustal yielding -- this may be responsible for magnetar bursts
(Thompson \& Duncan 1995; see \S\ref{sec:magnetars}).

\subsection{Accretion-Induced Field Reduction}

The weak ($10^8-10^9$~G) magnetic fields found in millisecond pulsars
and low-mass X-ray binaries have long led to the suggestion that the
dipole field is reduced during accretion, perhaps through diamagnetic
screening of the magnetic flux by the accreted plasma (e.g.,
Bisnovati-Kogan \& Komberg 1975; Shibazaki et al.~1989;
Romani 1990).  Exactly how this is
achieved (if at all) is not clear, and a number of idealized problems
have been studied. To give some recent examples: 
Cumming et al.~(2001) considered a 1D
model in which matter accretes onto horizontal field lines, and showed
that inward field advection can compete with outward Ohmic diffusion
only when the accretion rate is larger than $10\%$ of the Eddington
rate; Payne \& Melatos (2004) constructed global hydromagnetic
equilibrium sequence of a neutron star accreting at the magnetic poles
(see also Melatos \& Phinney 2001);
Lovelace et al.~(2005) considered a model in which the magnetic field
is screened by the current in the infalling plasma outside the star.
In any case, it is not clear that such field burial is effective or
permanent, as the field tends to resurface due to various
instabilities (e.g. buoyancy). If the original field is sustained 
by currents in the crust, then accelerated Ohmic decay due to
accretion-induced heating may be a viable mechanism 
(e.g., Urpin \& Geppert 1995; Konar \& Bhattacharya 1998). 

A different type of accretion-induced field reduction mechanism
relies on the interaction between superfluid neutron vortices 
and superconducting magnetic fluxoids in the stellar interior
(see Srinivasan et al.~1990). 
When the star spins up, the fluxoids are dragged together with 
the vortices toward the rotation axis, which leads to a reduction of
the effective dipole moment even though the local field strength at 
the magnetic poles actually increases (Ruderman 1991a,b).
Various observational consequences of such vortices migration
have been reviewed by Ruderman (2004).

\section{Thermal Radiation from Magnetized Neutron Stars} 
\label{sec:atm}

\subsection{Observational Background}
\label{subsec:obs}

It has long be recognized that 
thermal, surface emission from isolated neutron stars
(NSs) can potentially provide invaluable information on the physical
properties and evolution of NS (equation of state at super-nuclear
densities, superfluidity, cooling history, magnetic field, surface
composition, different NS populations, etc. See, e.g., Prakash et
al.~2001; Yakovlev \& Pethick 2004 for review). In the last few years,
considerable observational resources (e.g.~{\it Chandra} and {\it
XMM-Newton}) have been devoted to such study.
For example, the spectra of a number of radio pulsars (e.g.,
PSR~B1055-52, B0656+14, Geminga and Vela) have been observed to
possess thermal components that can be attributed to emission from NS
surfaces and/or heated polar caps (e.g., Becker \& Pavlov
2002). Phase-resolved spectroscopic observations are becoming
possible, revealing the surface magnetic field geometry and emission
radius of the pulsar (e.g., Caraveo et al.~2004; De Luca et
al.~2005; Jackson \& Halpern 2005). 
{\it Chandra} has also uncovered a number of compact
sources in supernova remnants with spectra consistent with thermal
emission from NSs (see Pavlov et al.~2003), and useful constraints on
NS cooling physics have been obtained (e.g., Slane et al.~2002;
Yakovlev \& Pethick 2004).

Surface X-ray emission has also been detected
from a number of soft gamma-ray repeaters (SGRs) and anomalous X-ray
pulsars (AXPs) --- these are thought to be magnetars, whose radiation
is powered by the decay of superstrong ($B\go 10^{14}$~G) magnetic
fields (see Thompson \& Duncan 1995,1996; Woods \& Thompson 2005).  
Fits to the quiescent
magnetar spectra with blackbody or with crude atmosphere models
indicate that the thermal X-rays can be attributed to magnetar surface
emission at temperatures of (3--7) $\times 10^6$~K (see, e.g., Juett
et al 2002; Tiengo et al.~2002; Patel et al.~2003; Kulkarni et
al.~2003; Tiengo et al.~2005). One of the intriguing puzzles is the
absence of spectral features (such as ion cyclotron line around 1~keV
for typical magnetar field strengths) in the observed thermal
spectra. Clearly, detailed observational and theoretical studies of
surface emission can potentially reveal much about the physical
conditions and the nature of magnetars.

Also of great interest are the seven or so isolated, radio-quiet NSs
(so-called ``dim isolated NSs''; see Haberl 2005).  These NSs share
the common property that their spectra appear to be entirely thermal,
indicating that the emission arises directly from the NS atmospheres,
uncontaminated by magnetospheric processes.  Thus they offer the best
hope for inferring the precise values of the temperature, surface
gravity, gravitational redshift and magnetic field strength. The true
nature of these sources, however, is unclear at present: they could be
young cooling NSs, or NSs kept hot by accretion from the ISM, or
magnetars and their descendants (e.g., van Kerkwijk \& Kulkarni 2001,
Mori \& Ruderman 2003; Haberl 2005; Kaspi et al.~2005).  Given their
interest, these isolated NSs have been intensively studied by deep
{\it Chandra} and {\it XMM-Newton} observations.
While the brightest of these, RX J1856.5-3754, has a featureless
spectrum remarkably well described by a blackbody (Drake et al.~2002;
Burwitz et al.~2003), absorption lines/features at $E\simeq
0.2$--$2$~keV have recently been detected from at least four sources,
including 1E 1207.4-5209 (0.7 and 1.4~keV, possibly also 2.1,~2.8~keV;
Sanwal et al.~2002; DeLuca et al.~2004; Mori et al.~2005), RX
J1308.6+2127 (0.2-0.3~keV; Haberl et al.~2003), RX J1605.3+3249
(0.45~keV; van Kerkwijk et al.~2004) and RX J0720.4$-$3125 (0.27~keV;
Haberl et al.~2004), and possibly two additional sources (see Zane et
al.~2005).  The identifications of these features, however, remain
uncertain, with suggestions ranging from electron/ion cyclotron lines
to atomic transitions of H, He or mid-Z atoms in a strong magnetic
field (Ho \& Lai 2004; Pavlov \& Bezchastnov 2005; Mori et al.~2005).
These sources also have different X-ray light curves: e.g., RX
J1856.5-3754 and RX J1605+3249 show no variability (pulse fraction
$\lo 1$-$3\%$); RX J0720-3125 shows a single-peaked $P=3.39$~s
pulsation of amplitude $\sim 11\%$, with the spectral hardness and
line width varying with the pulse phase; RX J 1308+2127 has a
double-peaked pulsation with $P=10.3$~s and amplitude $\sim 18\%$.
Another puzzle concerns the optical emission: For at least four of these
sources, the optical counterparts have been identified, but the
optical flux is larger (by a factor of $4$-$10$) than the
extrapolation from the black-body fit to the X-ray spectrum (see
Haberl 2005).

The preceding paragraphs highlight the great 
observational progress made in recent years on the study 
of NS surface emission.
These observations can potentially tell us much about the physics and
astrophysics of NSs.
Future X-ray telescopes ({\it Constellation-X} and {\it XEUS})
will have great capability of studying NS surface emission.
In order to properly interpret the current and future observations,
it is crucial to have a detailed understanding of the radiative 
properties of the outer layers of NSs in the presence of
intense magnetic fields, 
and to model the emergent thermal radiation spectra from the NS.

\subsection{Modeling Neutron Star Surface Radiation}

Thermal radiation from a magnetized NS is mediated by the thin
atmospheric layer (with scale height $0.1$--$10$~cm and density $\sim
10^{-3}$--$10^4$~g~cm$^{-3}$) that covers the stellar surface.  The
physical properties of the atmosphere, such as the chemical
composition, equation of state, and especially the radiative
opacities, directly determine the characteristics of the thermal
emission.  While the surface composition of the NS is unknown, a great
simplification arises due to the efficient gravitational separation of
light and heavy elements (see \S\ref{subsec:outermost}).
The strong magnetic field makes the atmospheric plasma anisotropic and
birefringent. If the surface temperature is not too high, atoms and
molecules
may form in the atmosphere. Moreover, if the magnetic field is
sufficiently strong, the NS envelope may transform into a condensed
phase with very little gas above it (\S\ref{subsec:outermost}).
A superstrong magnetic field will
also make some quantum electrodynamics (QED) effects (e.g., vacuum
polarization) important in calculating the surface radiation 
spectrum (see \S\ref{subsec:vacuum} below).

The first models of zero-field NS atmospheres were constructed
by Romani (1987). Further works used improved opacity and 
equation of state data from the OPAL project
for pure hydrogen, helium and iron compositions
(Rajagopal \& Romani 1996; Zavlin et al.~1996; G\"ansicke et al.~2002;
Pons et al.~2002). These models may be applicable to weakly magnetized
($B\lo 10^8$~G) NSs.  So far most studies of magnetic NS atmospheres
have focused on hydrogen and moderate field strengths of $B\sim
10^{12}$--$10^{13}$~G (e.g., Shibanov et al.~1992;
Zane et al.~2000; Zavlin \& Pavlov 2002).
These models take into account the transport of different photon modes 
through a mostly ionized medium. The opacities adopted in the
models include free-free transitions and electron scattering.
However, since a strong magnetic field greatly increases the binding 
energies of bound species 
(see \S \ref{sec:matter}),
atoms, molecules, and other bound states
may have appreciable abundances in the atmosphere
(see \S \ref{subsec:outermost}).
Thus these magnetic atmosphere models are expected to be valid only 
for relatively high temperatures ($T\go {\rm a~few}\times 10^6$~K) 
where hydrogen is almost completely ionized. 
As the magnetic field increases, we expect these models
to break down at even higher temperatures as bound atoms, molecules 
and condensate become increasingly important. 
Models of magnetic iron atmospheres (with $B\sim 10^{12}$~G)
were studied by Rajagopal et al.~(1997). Because of the complexity
in the atomic physics and radiative transport, these Fe
models are necessarily crude. 

NS atmospheres with superstrong ($B\go 10^{14}$~G) magnetic field have
been studied recently (Ho \& Lai 2001,2003,2004; 
\"Ozel 2001; Zane et al.~2001). Currently most models treat ionized H and He
atmospheres, and include full angle-dependent transport of the photon
polarization modes and ion cyclotron resonance in the opacity. 
The effect of vacuum polarization has been studied, and is found to 
be important for determining the thermal spectra and polarizations 
of magnetized NSs (see \S\ref{subsec:vacuum}).
Recently, NS atmosphere models that include self-consistent treatment of the
thermodynamics and opacities of bound H atoms in strong magnetic
fields were constructed (see \S\ref{subsec:partial}). 

In order to confront theoretical atmosphere models with observational
data (spectra, light curves, and in some cases phase-resolved
spectra), it is necessary to calculate synthetic spectra from the
whole stellar surface, taking into account the effect of gravitational
redshift and light-bending. This can be accomplished using the
standard, well-developed procedure (Pechenick et al.~1983; Beloborodov
2002).  To calculate synthetic spectra, one must know the
distributions of magnetic field (both magnitude and direction) and the
effective temperature over the stellar surface; these are clearly
model-dependent (see Zane \& Turolla 2006).
For a passively cooling NS with a given field
geometry, the surface temperature distribution can be obtained from
the results of NS heat conduction calculations (see
\S\ref{subsec:crust}), provided no extra heating (e.g., associated
with particle bombardment in the polar cap) is present.
Obviously, any spectral feature is expected to be broadened 
if different parts of the NS surface with different field strengths 
contribute similarly to the observed flux 
(e.g., Zane et al.~2001; Ho \& Lai 2004).

\subsection{Effect of Vaccum Polarization on Neutron Star Thermal
Spectrum and Polarization}
\label{subsec:vacuum}

\begin{figure} 
\hskip 2cm
\includegraphics[width=28pc]{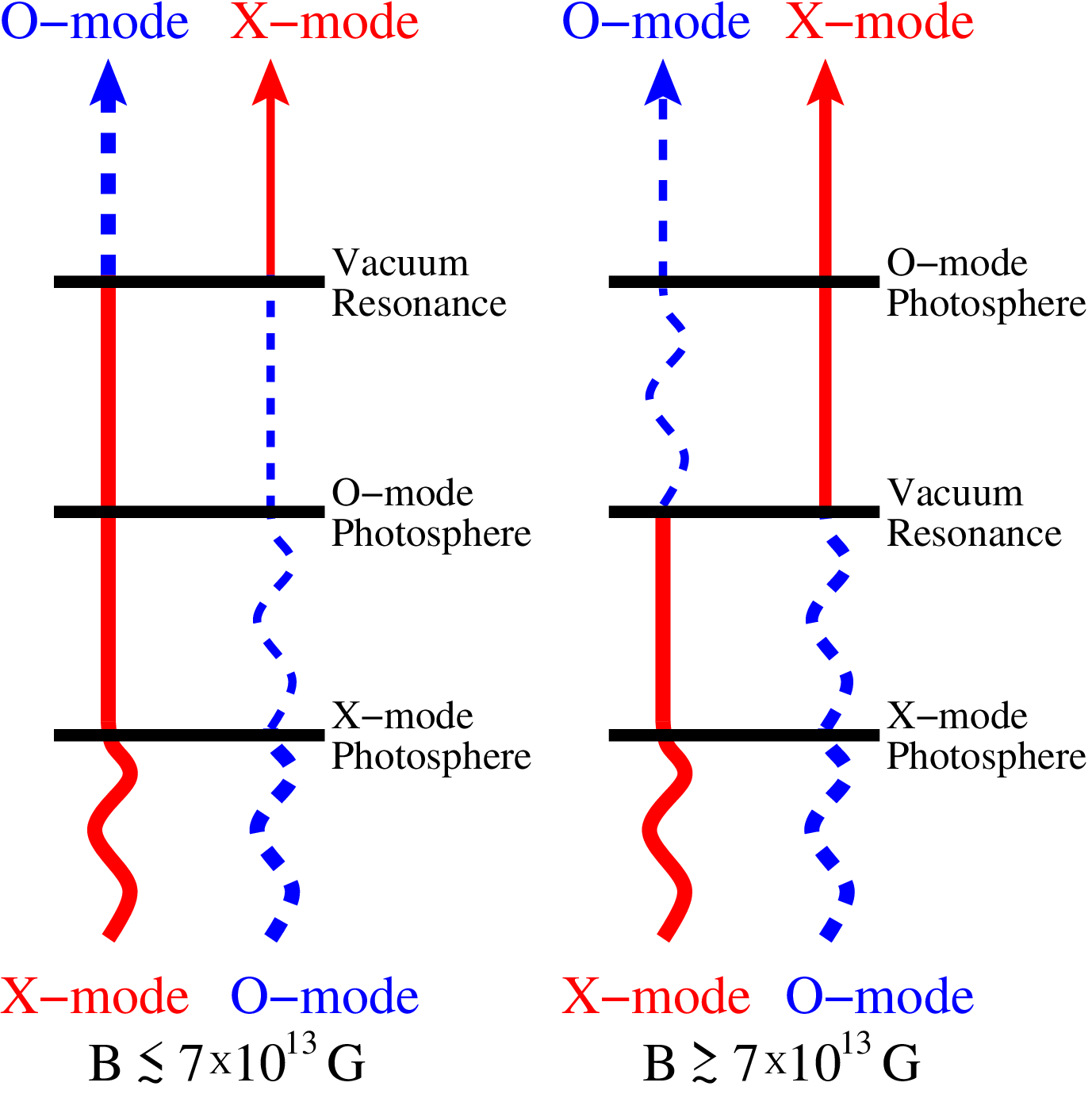}
\caption{A schematic diagram illustrating how vacuum polarization-induced
mode conversion affects the emergent radiation from a magnetized
NS atmosphere.
The photosphere is defined by where the optical depth (measured from
the surface) is 2/3 and is where the photon decouples from the matter.
The left side applies to the ``normal'' field regime
[$B\lo 7\times 10^{13}$~G; see Eq.~(\ref{eq:maglim})], in which the 
vacuum resonance lies outside the photospheres of the two modes.
The right side applies to the ``superstrong'' field regime
($B\go 7\times 10^{13}$~G), in which the vacuum resonance lies
between the two photospheres. In this diagram, complete adiabatic
mode conversion is assumed across the vacuum resonance. 
From Ho \& Lai (2004).}
\label{fig:mc}
\end{figure}

Vacuum polarization can dramatically affect the surface emission from
magnetized NSs (Lai \& Ho 2002,2003a). As discussed in 
\S\ref{subsec:vacuum-p}-\ref{subsec:modevac},
quantum electrodynamics predicts that in strong magnetic fields 
the vacuum becomes birefringent. In a magnetized NS
atmosphere, both the plasma and vacuum polarization contribute to the
dielectric property of the medium, and a ``vacuum resonance'' arises
when these two contributions ``compensate'' each other. When a photon
propagates outward in the NS atmosphere, it may convert from
one mode into another as it traverses the vacuum resonance.
Because the two photon modes have very different opacities,
the vacuum polarization-induced mode conversion can significantly affect
radiative transfer in magnetized atmospheres.
When the vacuum polarization effect is neglected (nv),
the decoupling densities of the O-mode and X-mode photons 
(i.e., the densities of their respective photospheres)
are approximately given by (see Lai \& Ho 2002)
\ba
\rho_{\rm O,nv} & \approx & 0.42\,T_6^{-1/4}E_1^{3/2}S^{-1/2}
\mbox{ g cm$^{-3}$}
\label{eq:denso} \\
\rho_{\rm X,nv} & \approx & 486\,T_6^{-1/4}E_1^{1/2}S^{-1/2}B_{14}
\mbox{ g cm$^{-3}$},
\label{eq:densx}
\ea 
where $T_6=T/(10^6\,{\rm K})$ and $S=1-e^{-E/k_BT}$. Thus the
X-mode photons are produced in deeper, hotter layers in the atmosphere
than the O-mode photons.  When vacuum polarization is taken into
account, the decoupling densities can be altered depending on the
location of the vacuum resonance $\rho_V$ relative to $\rho_{\rm
O,nv}$ and $\rho_{\rm X,nv}$. For ``normal'' magnetic fields, $B<B_l$,
with \be B_l \approx 6.6\times 10^{13}\,
T_6^{-1/8}E_1^{-1/4}S^{-1/4}\mbox{ G},
\label{eq:maglim}
\ee 
the vacuum resonance lies outside both photospheres
($\rho_V<\rho_{\rm O,nv}<\rho_{\rm X,nv}$), and the net thermal spectrum
is not affected by the vacuum resonance. 
For superstrong magnetic fields, $B>B_l$, the vacuum resonance lies
between these two photospheres ($\rho_{\rm O,nv}<\rho_V<\rho_{\rm
X,nv}$), and the effective decoupling depths of the photons are
changed. Indeed, we see from Fig.~\ref{fig:mc} 
that mode conversion makes the
effective decoupling density of X-mode photons (which carry the bulk
of the thermal energy) smaller, thereby depleting the high-energy tail
of the spectrum and making the spectrum closer to black-body (although
the spectrum is still harder than black-body because of nongrey
opacities)\footnote{Note that even when mode conversion is neglected,
the X-mode decoupling depth can still be affected by vacuum
polarization.  This is because the X-mode opacity exhibits a spike
feature near the resonance, and the optical depth across the resonance
region can be significant; see Lai \& Ho (2002).}.  This expectation is
borne out in the atmosphere model calculations (Ho \& Lai
2003). Another important effect of vacuum polarization on the spectrum
is the suppression of proton cyclotron lines and other spectral lines
(Ho \& Lai 2003; Ho et al.~2003).  The physical origin for such line
suppression is related to the depletion of continuum flux, which makes
the decoupling depths inside and outside the line similar.
It was suggested (Ho \& Lai 2003) that the absence of 
cyclotron lines in the quiescent spectra of several magnetars 
(e.g., Juett et al 2002; Patel et al.~2003)
is a natural consequence of the vacuum polarization effect at work in 
these systems, and the detection of lines in several dim
isolated neutron stars (see \S\ref{subsec:obs})
is also consistent with this picture (Ho \& Lai 2004). 

Most studies of radiative transfer in magnetized NS atmospheres 
rely on solving the transfer equations for the
specific intensities of the two photon modes (e.g.
Meszaros 1992; Zavlin \& Pavlov 2002).
These equations cannot properly handle the vacuum-induced mode conversion
phenomenon. This is because mode conversion
intrinsically involves the interference between the modes.
In particular, photons with energies 0.3-2~keV (this is the energy range
in which the bulk of the radiation comes out and spectral lines are
expected for $B\sim 10^{14}$~G) are only partially converted across
the vacuum resonance.
The other problem with the modal description of radiative transport is that
it is valid only in the limit of large Faraday depolarization
(Gnedin \& Pavlov 1974), which is not always satisfied near the vacuum
resonance, especially for superstrong magnetic fields (Lai \& Ho 2003a).
Also, in the presence of dissipation, the two photon modes can collapse
near the resonance (see \S\ref{subsec:modevac}), making the modal 
description meaningless.
To account for the vacuum resonance effect in a quantitative
manner, one must 
solve the transfer equations in terms of the photon intensity matrix 
(Lai \& Ho 2003a) and properly take into account of the probability of
mode conversion (van Adelsberg \& Lai 2006). 
Figure \ref{fig:b50t50} depicts an example of such calculation:
we see that a superstrong magnetic field indeed suppresses the high-energy
tail of the thermal spectrum and reduces width of spectral lines that may be
present.

\begin{figure} 
\includegraphics[width=32pc]{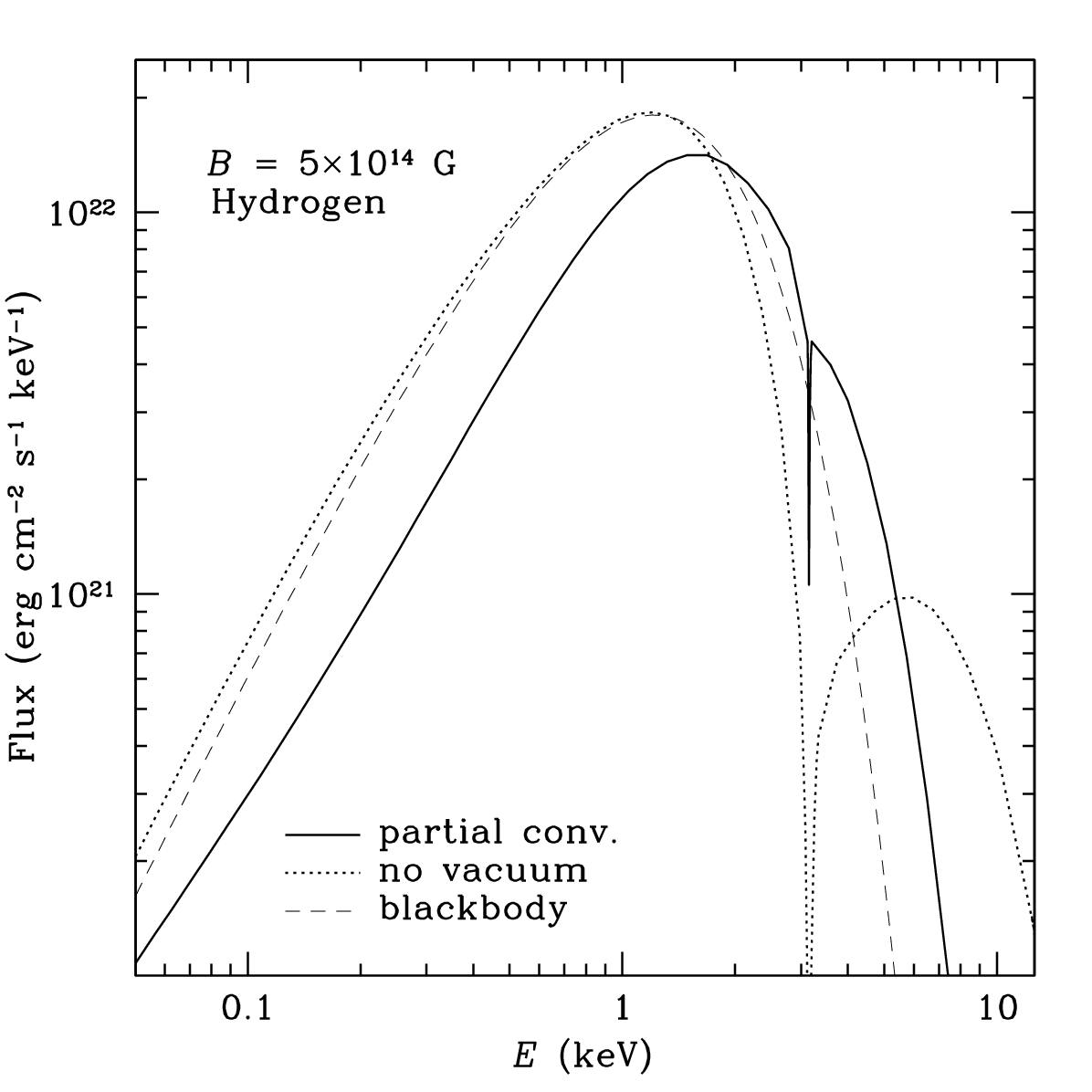}
\caption{The spectrum of a fully ionized hydrogen atmosphere with
$B=5\times 10^{14}$~G and $T_{\rm eff}=5\times 10^6$~K.
The solid line shows the result when the vacuum polarization 
effect is properly taken into account (including partial
mode conversion at the vacuum resonance), the dotted line 
shows the result when the vacuum effect is turned off, 
and the dashed line is for a blackbody with $T = 5\times 10^6$~K.
The $E_{ci}=0.63$~keV ion cyclotron feature is prominent 
in the ``no vacuum'' curve, but is suppressed in the ``partial conv''
curve. From van Adelsberg \& Lai (2006).}
\label{fig:b50t50}
\end{figure}

We note that even for NSs with ``ordinary'' field strengths
($10^{12}\lo B\lo 7\times 10^{13}$~G), vacuum polarization can affect
the X-ray polarization signals of the thermal emission in a
significant way (Lai \& Ho 2003b).  In particular, the vacuum
resonance effect gives rise to an unique energy-dependent polarization
signature: since the mode conversion probability depends on the photon
energy (\S \ref{subsec:inhom}), the plane of linear polarization at
$E\lo 1$~keV is perpendicular to that at $E\go 4$~keV. By contrast,
for superstrong field strengths ($B\go 7\times 10^{13}$~G), the
polarization planes at different energies coincide.  Vacuum
polarization is also important for aligning the polarization vectors
of photons emitted from different patches of the NS, thus ensuring an
appreciable net polarization fraction at the observer (Heyl et
al.~2003; see \S \ref{subsec:inhom}).  The detection of polarized
X-rays from neutron stars can provide a direct probe of strong-field
quantum electrodynamics and constrain the neutron star magnetic field
and geometry.

\subsection{Partially Ionized Atmospheres and Radiation from 
Condensed Surfaces}
\label{subsec:partial}

For sufficiently strong magnetic field and/or low temperature, bound
species have non-negligible abundances in the NS atmosphere.  Early
considerations of partially ionized atmospheres (e.g., Rajagopal et
al.~1997) relied on oversimplified treatments of atomic physics and
nonideal plasma effects in strong magnetic fields.  In recent years,
significant progress has been made in studying partially ionized
hydrogen plasmas: The binding energies and radiative transition rates,
incorporating the center-of-mass motion effects (see
\S\ref{subsec:hatom}), have been calculated for the H atom (Potekhin
1994; Potekhin \& Pavlov 1997); these atomic data have been
implemented in calculations of thermodynamic functions (including
ionization equilibrium; see \S\ref{subsec:outermost}) and radiative
opacities (Potekhin et al.~1999; Potekhin \& Chabrier 2003,2004).  The
bound species contribute to the bound-bound and bound-free opacities,
which, because of the center-of-mass motion effect, do not have a
narrow width or sharp edge.  In addition, the bound species affect the
dielectric tensor of the medium and hence the polarization properties
of the normal modes; these were studied by Potekhin et
al.~(2004) using the Kramers-Kronig relation between the real and
imaginary parts of the plasma polarizability. All these physical
ingredients have been incorporated into NS atmosphere models 
to produce the spectrum of surface emission (Ho et al.~2003;
Potekhin et al.~2004). An example is shown in Fig.~\ref{fig:spec_bw},
clearly demonstrating the importance of the partial-ionization effects.

\begin{figure} 
\includegraphics[width=38pc]{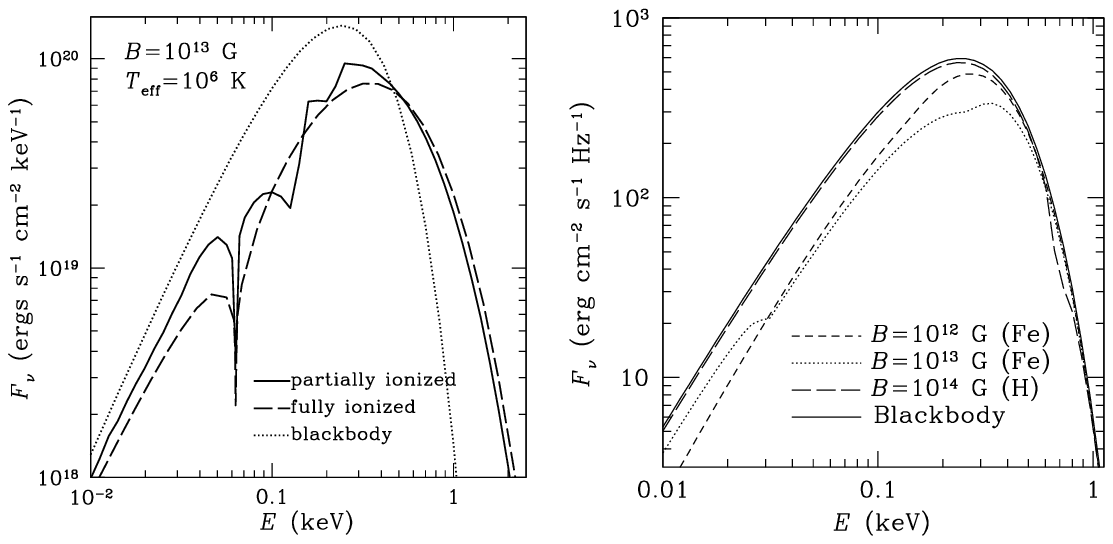}
\caption{Spectral flux as a function of the photon energy $E$.
{\it Left panel:} The case of a partially ionized hydrogen atmosphere model
(solid line) at $B=10^{13}$~G (field normal to the surface) and $T_{\rm eff}
=10^6$~K is compared with the fully ionized model (dashed line) and
with the blackbody spectrum (dotted line). Adapted from
Ho et al.~(2003) and Potekhin et al.~(2004,2006).
{\it Right panel}: The cases of condensed Fe surface
($B=10^{12}$~G, short-dashed line; $10^{13}$~G, dotted line) and
H surface ($B=10^{14}$~G, long-dashed line) at temperature $T=10^6$~K,
compared with the blackbody spectrum (solid line). Adapted from 
van Adelsberg et al.~(2005).}
\label{fig:spec_bw}
\end{figure}

As discussed in \S\ref{subsec:outermost},
for sufficiently low temperatures
and/or strong magnetic fields, the atmosphere may undergo
a phase transition into a condensed state.
Thermal emission from such a condensed (Fe or H) surface was studied 
by van Adelsberg et al.~(2005) (cf. Brinkmann 1980;
Turolla et al.~2004; Perez-Azorin et al.~2005). 
Obviously, the standard radiative transfer equation does not apply 
inside the condensed phase. Instead, we can apply the
Kirchkoff's law for a macroscopic object (generalized for polarized
radiation; see \S 3.1 of van Adelsberg et al.~2005): We 
calculate the (polarization-averaged) reflectivity of the surface 
for a incident ray with incident angle $\theta$; then the emission 
intensity of the surface is simply $I_\nu=\left[1-R_\nu(\theta)\right]
B_\nu(T)$, where $B_\nu(T)$ is the Planck function.
To calculate $R_\nu(\theta)$, we can approximate
the dielectric tensor of the condensed surface using the quasi-electron
gas approximation, with and without the ion response. 
For a smooth condensed surface, the overall emission is 
reduced from the blackbody by less than a factor of 2,
The spectrum exhibits modest deviation from blackbody across 
a wide energy range, and shows mild absorption features associated 
with the ion cyclotron frequency and the electron plasma frequency 
in the condensed matter. Examples of the spectrum for different models 
of the surface are shown on the right panel of Fig.~
\ref{fig:spec_bw}. If the surface is rough (as is likely in the 
Fe case), the surface reflectivity will be close to zero, making 
emission spectrum even closer to blackbody. It is 
possible that the almost perfect blackbody spectrum in the X-ray band
observed from RX J1856.5-3754 (see \S\ref{subsec:obs})
is a result of such condensed surface 
emission, although an expanation for the excess optical emission
is still called for.

\subsection{Resonant Scatterings of Surface Emission}

An interesting issue concerns the possibility that
thermal emission from the NS surface may be modified by scatterings
in the magnetospheric plasma. Since a thermal X-ray photon has energy 
less than the electron cyclotron energy ($E_{ce}=11.6\,B_{12}$~keV) at
the stellar surface, as the photon passes through the magnetosphere,
it goes through the electron cyclotron resonance ($E=E_{ce}$) where
enhanced scattering occurs.  For the Goldreich-Julian charge density
in the closed magnetosphere, the optical depth of such scattering is
negligible, thus the surface emission will not be modified (Rajagopal
\& Romani 1997).  Ruderman (2003) suggested that a larger density of
pair plasmas may be present in the magnetosphere, maintained by
conversion of $\gamma$-rays from the NS's polar-cap and/or out-gap
accelerators. Partly motivated by the observational puzzles
associated with the spectra of isolated NSs, he further
suggested that resonant scattering of thermal photons will result in a
Planck-like spectrum that is modified from the seed surface
spectrum (see also Wang et al.~1998; Lyutikov \& Gavriil 2006).
Also, if the plasma is sufficiently hot, it can Compton upscatter 
the thermal photons.  While this scenario may be plausible for energetic
rotation-powered pulsars, the situation is not clear 
for the dim, radio-quiet isolated NSs, given the absence of any 
nonthermal emission signature from these stars.  
For active magnetars, a corona consisting mainly of
relativistic electron-positron pairs can be generated by 
crustal magnetic field twisting/shearing due to starquakes,
with the plasma density much higher (by a factor of order
$c/\Omega r$) than the Goldreich-Julian value
(Thompson et al.~2002; Thompson \& Beloborodov 2005).
Such a corona plasma can significantly modify 
the surface radiation by multiple scatterings; this may explain the
$2-10$~keV soft tail of the magnetar surface emission.

\section{Non-Thermal Radiation of Rotation-Powered Pulsars} \label{sec:pulsars}

There is abundant evidence that rotation-powered pulsars must be capable of particle acceleration to
energies of at least 10 TeV.  Pulsed $\gamma$-rays above 100 MeV have been detected from seven pulsars 
with high significance and about five others with less significance (Kanbach 2002).  Some 60 pulsars
have been detected at X-ray energies with about 30 having detectable pulsations (Kaspi et al. 2005).  
Both the acceleration voltage and the high-energy luminosities required
can originate from the rotation of a dipole field, if the surface magnetic 
fields $B_s$ are in fact those required
to produce the observed rate of spin-down by magnetic dipole torques
[see Eqn (\ref{eq:B0})].  
The vacuum potential drop across the open field lines is 
\be
V_{\rm pc} \simeq {1\over 2} \left({2\pi \over cP}\right)^2\,B_s R^3 = 6 \times 10^{12}\,{\rm eV}\,
P^{-2}\,B_{12},
\ee
where $B_{12} \equiv B_s/10^{12}$ G and $P$ is the pulsar spin period (in s).
The fraction of this potential available for particle acceleration
varies among different models and from source to source (see Harding
2005 for review), but the overall picture seems energetically
self-consistent (although unsolved problems such as global current
closure still exist).  Two accelerator sites have been studied in most
detail and two main types of model have developed.  Polar cap models
(Daugherty \& Harding 1996, Usov \& Melrose 1996) are based on
particle acceleration beginning near the magnetic poles, where the
magnetic fields are high and one-photon pair creation (\S
\ref{sec:1pho}) usually dominates over two-photon pair creation.
Outer gap models (Cheng et al. 1986, Romani 1996) are based on
acceleration in vacuum gaps in the outer magnetosphere where the
magnetic fields are too low for one-photon pair creation, but pairs
may be created by interactions of $\gamma$-rays produced by
gap-accelerated particles with X-ray photons from either the gap or
the hot neutron star.

\subsection{Electromagnetic cascades}

\subsubsection{Polar cap cascades}

The combination of high particle energies and magnetic field strength approaching, and in some sources exceeding, 
the critical field allow very favorable conditions for pair creation and electromagnetic cascades (Sturrock 1971)
that can greatly enhance the number of particles in a pulsar magnetosphere (Daugherty \& Harding 1982).  
The accelerated particles moving along magnetic field lines with high Lorentz factors radiate $\gamma$-ray 
photons at very small angles to the field ($\theta_0 \sim 1/\gamma$), so the one-photon pair production rate 
for these photons is initially zero.  However, as they propagate through the curved dipole field, their angle 
increases until the threshold condition, $\epsilon\sin\theta = 2$, is
reached and the attenuation coefficient becomes large.  Each member of the pair will be produced in a 
Landau state whose maximum principal quantum number is $n_{\rm max} = 2\epsilon'(\epsilon' - 2)/B'$ 
(Daugherty \& Harding 1983), where $\epsilon' = \epsilon\sin\theta$ is the photon energy in the frame in
which it propagates perpendicular to the local magnetic field.  From the pair production condition,
$\chi \equiv \epsilon B'\sin\theta/2 \gsim 0.1$, an estimate for the maximum pair Landau state is
\be
n_{\rm max} \simeq {0.4\over B'^2}\,\left({0.2\over B'} - 2\right).
\ee
When the local field $B \lsim 0.1 B_Q$, pairs will be created above threshold in highly excited
Landau states and the excitation level is quite sensitive to field strength, $n_{\rm max} \propto 1/B'^3$.  
The pairs will decay through
emission of synchrotron/cyclotron photons, many of which will produce more pairs in excited states.  A
pair cascade can be sustained in such a way through several generations, with pair multiplicities reaching
as high as $10^3 - 10^4$ (pairs per primary electron) (Hibschmann \& Arons 2001, Arendt \& Eilek 2002).  
In high-field pulsars,
where $B \gsim 0.1 B_Q$ near the neutron star surface, the pair creation attenuation coefficient 
is high enough for pair creation near threshold, so pairs are produced in very low Landau states.  In this 
case, cascade pair multiplicities are lower since the number of cyclotron photons drop significantly 
(Baring \& Harding 2001).  Bound pair production (cf. \S \ref{sec:BoundPair}) also becomes important
for $B \gsim 0.1 B_Q$, which will further lower the pair multiplicity.  
When $B \gsim B_Q$, photon splitting dominates over pair production,
as was discussed in \S \ref{sec:PhoSplit}.  Further reduction of the cascade pair yield in this regime
then depends on whether and at what field strength additional splitting modes are possible.  
If all three modes permitted by QED are operating, then a full photon
splitting cascade replaces the pair cascade and a complete suppression of pairs occurs (Baring \& Harding
2001).  If only the $\perp \rightarrow \parallel\parallel$ mode permitted in the non-dispersive limit is
operating, as is more likely, then no suppression of pair yields occurs by photon splitting alone.  
However, the 
$\parallel$ mode photons will create bound rather than free pairs in a strong field 
(Shabad \& Usov 1985, 1986), which at least delays the creation of real pairs until the bound pair 
is dissociated by radiation and/or electric fields (Usov \& Melrose 1996) or moves to higher 
altitude where the field is smaller.  
The pair cascade may then resume, but changes to pair yields in this case have not been studied.  The 
screening (see next paragraph) of the electric field is also delayed by splitting and 
creation of bound pairs, effectively increasing the accelerating potential.  The effect of
bound pair creation on polar cap acceleration has been studied by Usov \& Melrose (1996).  

In the case of young pulsars with high surface fields, the pair cascades are limited by screening of the 
electric field by the pairs as they accelerate in opposite directions (Arons \& Scharlemann 1979).
Such pulsars are capable of accelerating primary particles to high enough energies ($\gamma \sim 10^7$) 
that their curvature
radiation photons can create pairs to initiate the cascade.  The curvature-radiation initiated cascades
produce high multiplicities over a small fraction of a neutron star radius and can screen the $E_{\parallel}$ 
on that length scale (Harding \& Muslimov 2001).  Older pulsars with lower magnetic fields have difficulty
accelerating particles fast enough for their curvature photons to pair produce before the magnetic field
drops.  In this case, inverse-Compton (ICS) photons that are produced by the particles at a lower energy
($\gamma \sim 10^4 - 10^6$), scattering thermal X-rays from the neutron star surface, can create pairs at
lower altitude (Zhang \& Qiao 1996).  However, the ICS-initiated cascades have much lower multiplicity,
since the emission rate decreases as the particles' energy increases (in constrast to the CR emission rate
which increases as with energy), and are not robust enough to screen the $E_{\parallel}$ (Harding \& 
Muslimov 2002).  These cascades are thus limited by the increasing energy of the accelerating particles
and the $E_{\parallel}$ ultimately saturates at several polar cap radii by geometric effects.  In the case
of millisecond pulsars having extremely low magnetic fields, but very short periods, two-photon pair 
creation through interaction of curvature radiation or ICS photons with thermal X-ray photons
from the neutron star surface can dominate over one-photon pair creation (Zhang \& Qiao 1998, 
Harding et al. 2002).  Ultimately, as the pulsar ages the polar cap potential drop becomes too low to 
produce pairs of any kind (Harding \& Mulsimov 2002, Harding et al. 2002).  The resulting ``death line"
in $P$-$\dot P$ space accounts for the abrupt decrease in numbers of radio pulsars at periods longer than 
a few seconds (see Figure {\ref{fig:PPdot}).

\subsubsection{Outer gap cascades}

Pair cascades in the outer magnetosphere of a pulsar, where magnetic field strengths are in the
range $B \sim 10^5 - 10^9$ G, must be sustained by two-photon pair creation.  The cascade-initiating
particles are accelerated to high Lorentz factors in near-vacuum electric fields that develop as a
result of the depletion of charge beyond the null-charge surface $\Omega \cdot B = 0$ (Cheng, Ho 
\& Ruderman 1986).  They radiate curvature, or more accurately, synchro-curvature (Zhang \& Cheng 1996)
$\gamma$ rays tangent to magnetic field lines
within the gap and propagate until they interact with soft photons, either from the hot neutron star
surface or from lower energy gap radiation, to create a pairs.  The pair accelerates in opposite
directions, emitting more synchro-curvature radiation photons which produce more pairs and the 
secondary particles supply current through the gap.  The multiplicity
of outer-gap cascades could be quite high (Hirotani, priv. comm. 2005), with most of
the pairs produced by collisions between downward-going gamma-rays and surface X-rays, with magnetic
pair creation being important near the neutron star surface.   
These cascades are limited by screening of the $E_{\parallel}$ in directions both 
parallel and perpendicular to the magnetic field (Cheng et al. 2000, Hirotani \& Shibata 2001), taking into 
account the curvature of the field lines.  The gaps can also be self-limited if the thermal X rays supplied
by the hot neutron star surface require heating by gap-accelerated particles flowing downward (Cheng 1994,
Zhang \& Cheng 1997).  
However, neutron star cooling that releases the latent heat of formation, provides most of the 
surface X rays in the youngest pulsars.   

Older pulsars cannot sustain pair cascades in an outer gap, because the $E_{\parallel}$ is smaller and
even the potential drop across the entire outer magnetosphere is not sufficient for pair creation.  
An outer-gap death line in $P$-$\dot P$ space is thus predicted (Chen \& Ruderman 1993), which may 
also be a function of pulsar inclination angle (Zhang et al. 2004), and suggests
that pulsars older than $\tau \sim 10^7$ yr do not have outer gaps or the associated high-energy emission. 

\subsection{Observable signatures of high-field physics}

There are several predicted features of non-thermal pulsar emission
that would be signatures of physical processes operating exclusively
in magnetic fields near $B_Q$.  The spectra of polar cap cascade
emission will display high-energy cutoffs due to one-photon pair
creation attenuation (Daugherty \& Harding 1982).  Such cutoffs are
distinguishable by their very sharp, super-exponential shape, so that
the observed spectra will have the form
\be  \label{eq:PCcaspec}
f(\epsilon) = A \epsilon^{-a}\,\exp\left[{-C_{1\gamma}\exp\left(-{\epsilon^{1\gamma}_{\rm esc}/\epsilon}\right)}\right]
\ee
where $a$ is the power-law index, $C_{1\gamma} = 0.2(\alpha/\lambar)(B'R^2/\rho)$ and $\epsilon_{\rm esc}$ is the
escape energy from the magnetosphere for photons of energy $\epsilon$ (Harding et al. 1997, Harding 2001),
\be  \label{eq:Eesc}
\epsilon^{1\gamma}_{\rm esc} \sim 2\,\,{\rm GeV}\,P^{1/2}\,\left({r_0\over R}\right)^{1/2}\, {\rm max}
\left\{0.1, \,B_{s,12}^{-1}\,\left({r_0\over R}\right)^3\right\}
\ee
or the energy above which photons emitted from a radius $r_0$ can
escape the magnetosphere without pair attenuation.  The above estimate
of the escape energy assumes that the condition for pair attenuation
is $\chi \sim 0.2$ for $B' \lsim 0.1$ and $\epsilon\sin\theta = 2$,
the threshold condition, for $B' \gsim 0.1$, that $\sin\theta \sim
R/\rho$, where $\rho$ is the field line radius of curvature at the
polar cap rim and that $\rho$ and $B'$ are evaluated at the photon
emission radius $r$.  The cutoff in the spectrum of polar cap cascades
that are extended over more than a stellar radius will be somewhat
more gradual than Eqn (\ref{eq:PCcaspec}) due to variations in both
the field strength and radius of curvature.  Nevertheless, the cutoff
should still be distinguishable from a simple exponentional.

For pulsars with high magnetic fields $B' \gsim 1$, photon splitting
dominates the attenuation of polar cap cascade photons in the $\perp$
mode (see \S \ref{sec:PhoSplit}), while $\parallel$ mode photons are
attenuated by pair production.  The photon splitting cutoff occurs at
roughly the escape energy for photon splitting,
\be
\epsilon^{sp}_{esc} \sim 93\,{\rm MeV}\,\,P^{3/5}\,\left({r_0\over R}\right)^{3/5}\,
\left\{
\begin{array}{lr}
B'^{-6/5}, &  B' \lsim 1 \\
1,  &  B' \gsim 1  \\
\end{array}
\right.
\ee
where $B' = (r_0/R)^{-3}$ is the local field strength.  At high fields 
the photon splitting escape energy will be lower than that for pair
production.  The photon splitting cutoff will also have a more gradual
shape since the attenuation coefficient is power law rather than
exponential.  Because the $\perp$ mode photons are attenuated at a
lower energy than the $\parallel$ mode photons, the spectrum will be
100\% polarized in the $\parallel$ mode in the energy band between
$\epsilon^{sp}_{esc}$ and $\epsilon^{1\gamma}_{esc}$ (Harding et
al. 1997).  This band will grow as the field strength increases, as
the $\epsilon^{sp}_{esc}$ decreases and the field dependence of
$\epsilon^{1\gamma}_{esc}$ saturates for 
$B' \gsim 1$.  If such
strong polarization in $\gamma$-ray pulsar spectra near the high
energy cutoff is detected by future X-ray and $\gamma$-ray
polarimeters, it would be a clear signature of photon splitting.

The above estimates for pair creation and photon splitting escape
energies assumes flat spacetime.  Near a neutron star, depending on
its compactness (i.e. equation of state), general relativistic effects
including photon red-shift, light bending and changes to the dipole
field are important for photon attenuation (Gonthier \& Harding 1994).
For standard neutron star compactness, curved spacetime effects will
lower the escape energies for both pair creation and photon splitting
by a factor of about 2 (Harding et al. 1997, Baring \& Harding 2001).

\section{Radiation of Magnetars} \label{sec:magnetars}

SGRs and AXPs were both discovered several decades ago, but they have
only recently been recognized as similar objects (for detailed review,
see Woods \& Thompson 2005, Heyl 2005).  SGRs were first detected
around 1979 as $\gamma$-ray transients and were thought to be a type
of classical $\gamma$-ray burst.  They undergo repeated bursts with
several tenths of second duration and average energy
$10^{40}-10^{41}\,\rm erg$, and their bursting often occurs in
episodes spaced years apart.  They more rarely undergo giant
superflares of total energy $10^{45}-10^{47}\,\rm erg$, consisting of
an initial spike of duration several hundred ms followed by a longer
decay of duration several hundred seconds showing pulsations.  Such
superflares have been observed in three SGR sources, SGR0526-66 (the
famous 5th March 1979 event), SGR1900+14 (Hurley et al.~1999) and very
recently in SGR1806-20 
(Hurley et al.~2005; Palmer et al.~2005).
In 1998, SGR1806-20 was
discovered to have 7.47 s pulsations in its quiescent X-ray emission
(Kouveliotou et al. 1998) (and a large $\dot P$ that implied a huge
surface magnetic field of $\simeq 10^{15}$ G if due to dipole
spin-down.  Quiescent periodicities of 8 s and 5.16 s and large $\dot
P$ were subsequently detected in SGR0526-66 and SGR1900+14, implying
similarly high surface magnetic fields.  In all three sources, the
quiescent periods are the same periods seen in the decay phases of
their superflares.  The quiescent pulse profiles are very broad and
undergo dramatic changes just before and after superflares.  The
profiles are often more complex, with multiple peaks before flares,
changing to more simple single peaks profiles following the flares.
All of the SGRs lie near the Galactic plane and are thought to have
distances around 10-15 kpc (except for SGR0526-66, which is in the
LMC).

The quiescent spectra of AXPs and SGRs consist of a thermal component
fit by $\sim$ 0.5-1 keV blackbodies and one or more non-thermal
components, 
with X-ray luminosity $10^{35}-10^{36}$~erg~s$^{-1}$.
Until recently, the non-thermal spectra below 10 keV were
fit with steep power laws having indices 
$\Gamma\sim 2-4$ (spectral flux $\propto \epsilon^{-\Gamma}$). 
When {\sl INTEGRAL} and {\sl RXTE} recently measured the
spectra above 10 keV for the first time hard, non-thermal components
were discovered in three AXPs, and also SGR 1806-20.  In two of the
AXPs, the differential spectra between 10 keV and 50 keV are extremely
flat: 1E 1841-045 (Kuiper et al. 2004) has a power-law index of
$s=0.94$ and 4U 0142+61 (Hartog et al. 2004) displays an index of
$s=0.45$, both much flatter than the steep non-thermal components in
the classic X-ray band.  RXS J1708-40 possesses a slightly steeper
continuum with $s=1.18$.  The non-thermal tail of quiescent emission
in SGR 1806-20 is similarly pronounced (Mereghetti et al. 2005, Molkov
et al. 2005)), but somewhat steeper, with an index of $s=1.6-1.9$
extending beyond 100 keV. 
The luminosity in hard x-rays ($\sim 100$~keV), $\sim 10^{36}$~erg~s$^{-1}$,
even exceeds the thermal luminosity from the stellar surface.
Such hard non-thermal components require
continuous particle acceleration during the quiescent state.

Although it was proposed early on (Ramaty et al. 1980, Katz 1982) that
SGRs were neutron stars with strong magnetic fields in the range $B_S
\sim 10^{11} - 10^{13}$ G to confine the burst radiation, it was
sometime later that surface magnetic fields exceeding $10^{14}$ G were
actually predicted for these objects 
(Duncan \& Thompson 1992; Thompson \& Duncan 1995,1996).
Duncan \& Thompson (DT), exploring the origin of neutron star magnetic
fields, conjectured that in dynamo models magnetic fields $3 \times
10^{17}\,{\rm G}\, (P_i/1\,\rm ms)^{-1}$ could be generated in nascent
neutron stars with initial periods $P_i$ in the ms range (see 
Thompson \& Duncan 1993 and \S\ref{subsec:origin}).
Paczynski (1992) deduced that SGR0526-66 must have a magnetic field of
$5 \times 10^{14}$ G to lower the magnetic scattering cross section
(\S \ref{sec:CompScatt}), and therefore raise the Eddington Limit to
the luminosity of the repeating bursts at a distance of the LMC
(see also Miller 1995).
This is coincidentally the field required if its spin-down age at rotation
period of 8 s is compatible with the age of its associated supernova
remnant N49 
(Duncan \& Thompson 1992). 
Fields of this magnitude will decay of much shorter
timescales due to the onset of ambipolar diffusion (\S
\ref{sec:Bfield}) 
\be 
t_{\rm Amb}^{} \cong 10^4 yr_{} \left(
{\frac{{B_{core} }}{{10^{15} G}}} \right)^{ - 2}.  
\ee 
Diffusion of magnetic flux out of the NS core on these timescales provides the
power to magnetars in the DT model.  Magnetar-strength fields also
apply higher stresses to the stellar crust, so that the yield strain
can exceed the crustal strength.  This property is responsible for the
small SGR and AXP bursts in the DT model (Thompson \& Duncan 1996).
If a toroidal component of the field $B_{core} > 10^{15}$ G develops
in the interior of the star, it can twist the external field 
(Thompson \& Duncan 2001).
Such action can cause the superflares if the twisted field lines reconnect.  
The field line twists following starquakes (and the associated
X-ray burst activities) may last for years and lead to significant heating
of the near vicinity of the star (Thompson et al.~2002; 
Thompson \& Beloborodov 2005). Finally, due to the much faster heat
transport in very strong magnetic fields, there is a modest increase 
in the heat flux through the crust (\S \ref{subsec:crust};
Heyl \& Kulkarni 1998, see Figure \ref{fig:HeylKulkarni1998}).  
All these effects 
may explain the much hotter surface temperatures of magnetars and the 
high quiescent X-ray emission.

\begin{figure} 
\hskip 2cm
\includegraphics[width=12cm]{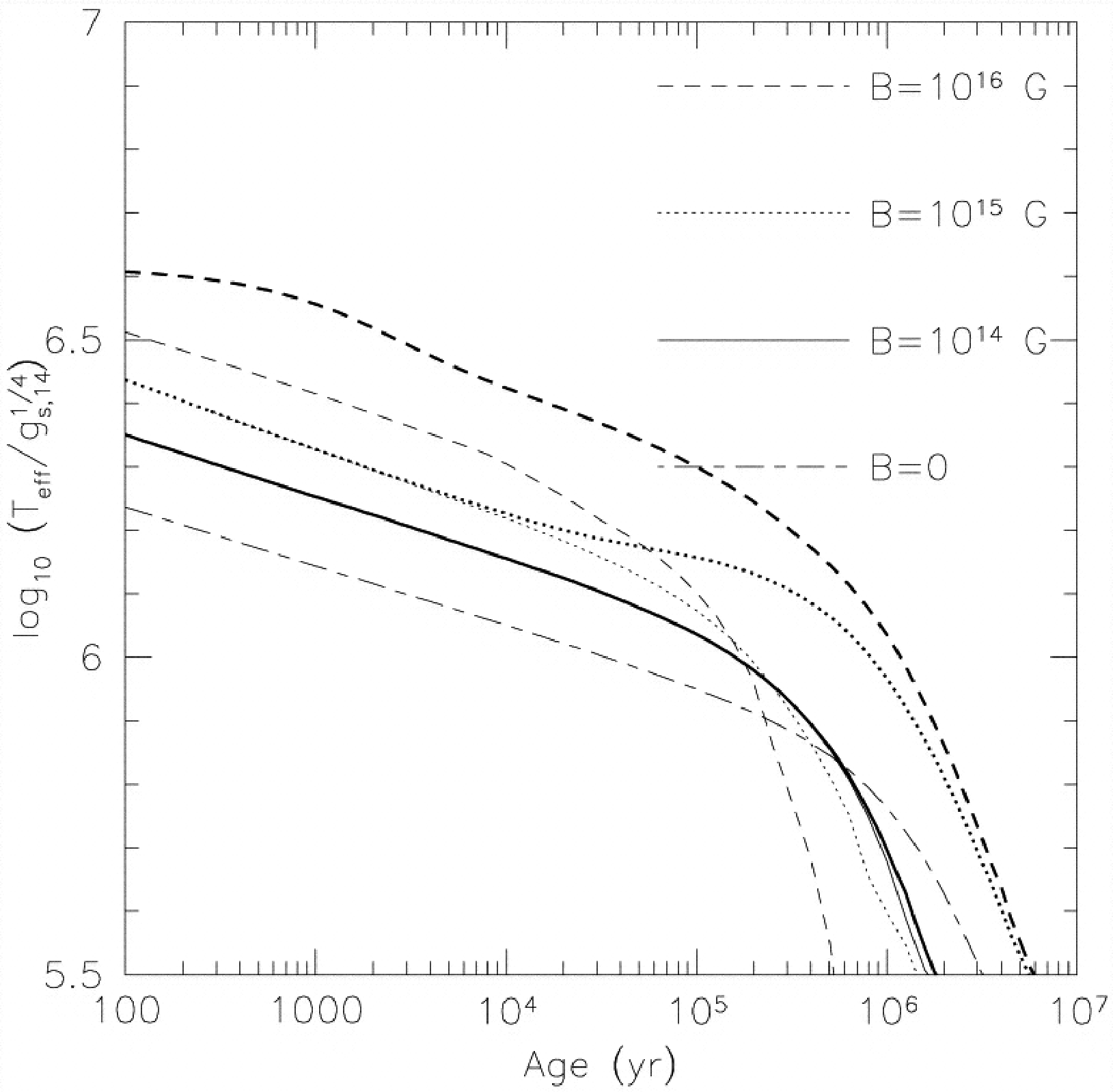}
\caption{Evolution of effective neutron star surface temperature when the magnetic field
decays through the irrotational mode, for different initial magnetic fields.  The light lines
are results for no field decay and the long-short dashed line shows the cooling evolution for
an unmagnetized neutron star.  From Heyl \& Kulkarni (1998).}
\label{fig:HeylKulkarni1998}
\end{figure}

The magnetar model has had success in accounting for the energetics of
the burst and quiescent emission in terms of the response of the
neutron star to the enormous stresses of field decay.  According to
the DT model (Thompson \& Duncan 1995), the magnetar superflares
results from reconnection of sheared or twisted external field lines,
leading to particle acceleration and radiation of hard emission
(see also Lyutikov 2006).
The estimated total energy of such events,
\be
\frac{{B_{core}^2 }}{{8\pi }}R^3  \approx 4 \times 10^{46} {\rm erg} \left( {\frac{{B_{core} }}{{10^{15} G}}} \right)^2, 
\ee
is similar to observed luminosities of superflares.  
The smaller bursts result from cracking of the crust, which is
continually overstressed by diffusion of magnetic flux from the NS
interior.  The shaking of magnetic footpoints then excites Alfven
waves that accelerate particles.  The energy radiated in such events
would be
\be
E_{SGR}  \cong 10^{41} {\rm erg} \left( {\frac{{B_S }}{{10^{15} G}}} \right)^{ - 2} _{} \left( {\frac{l}{{\rm 1 km}}} \right)^2 _{} \left( {\frac{{\theta _{\max } }}{{10^{ - 3} }}} \right)^2,
\ee
where $l$ is the length scale of the displacements, $B_S$ is the crustal field and 
$\theta _{\max }$ is the yield strain of the crust.
The quiescent emission in the DT model is power by magnetic field decay through conduction of heat from
the core.  The NS crust is heated to a temperature of 
\be
T_{crust}  \cong 1.3 \times 10^6 K_{} \left( {\frac{{T_{core} }}{{10^8 K}}} \right)^{5/9} 
\ee
where $T_{core} $ is the core temperature, and luminosity
\be
L_{\rm{x}}  \cong 6 \times 10^{35} {\rm erg s^{-1}} \left( {\frac{{B_{core} }}{{10^{16} G}}} \right)^{4.4}.
\ee
Model atmospheres for the quiescent thermal emission from magnetars
was discussed in \S \ref{sec:atm}.

Magnetar models have so far not had complete success in accounting for
the spectrum of the burst radiation.  In the DT model, $\gamma$ rays
are produced by scattering of X-rays from the neutron star surface by
particles accelerated after the energy released in crust cracking.  In
the equatorial region, the $\gamma$ rays are attenuated by photon
splitting and pair creation (assuming that only the $\perp \rightarrow
\parallel\parallel$ splitting mode operates), forming a pair plasma
that is optically thick to Compton scattering at an equilibrium
temperature of $\sim 100$ keV (Thompson \& Duncan 1995) and confined
by the magnetic field.  Such an estimate results from the deposition of
an energy $\sim 10^{41}$ erg into a volume $\sim R_{\rm max}^3$, with
$R_{\rm max} \sim 10\,R$.  Since the observed burst spectra are fitted
with much lower temperatures ($\sim 10-20$ keV), we cannot be viewing
the pair plasma directly, and perhaps the final spectrum is formed at
small optical depths.  Thompson \& Duncan (1995)
argue that if thermal equilibrium of the pair plasma is maintained
by photon splitting balancing photon merging (its inverse process),
then the photon number is increased while the temperature decreases to
$T_{\rm sp} \sim 11\, {\rm keV}$, which represents a lower limit on
the temperature of the emergent SGR burst spectra.  But this is a
crude estimate of the actual temperature of the observed thermal
spectrum since scattering and further splitting can occur outside the
splitting photosphere, the radius at which $B \sim B_Q$.  The
radiative transfer and spectrum of this hot, highly magnetized, and
probably dynamic, pair plasma has not been fully modeled.  Baring
(1995, Harding \& Baring 1996) modeled spectra resulting from photon
splitting cascades assuming that the three splitting modes allowed by
QED (see \S \ref{sec:PhoSplit}) operate at magnetar field strengths.
The splitting cascade spectra are quasi-Maxwellian in shape and the
spectral peak saturates in fields above $\sim 5-10 B_Q$ at around
20-30 keV due to the saturation of the photon splitting rate in high
fields.  Although this model could explain the magnetar burst spectral
shape, it has not been shown that both photon polarization modes can
split in magnetar fields.

Heyl \& Hernquist (2005a) propose an alternative model of magnetar radiation that appeals to
the non-linearity of electrodynamics due to vacuum polarization in a strong magnetic field.  This model
assumes that the power is generated by crustal fractures or field re-arrangement, as in the DT model,
but postulates that the energy is transported primarily by magnetohydrodynamic fast-mode waves rather than by Alfven waves.  The details of the generation of these waves, whether they are produced directly or through
interaction of Alfven waves, is not addressed.  
The fast MHD waves that travel away from the neutron star surface with sufficiently large amplitude 
and short wavelength can break down into 
electron-positron pairs (Heyl \& Hernquist 1999) if they form discontinuities in the magnetized vacuum 
on the order of a Compton wavelength.  The bulk of the pairs form a hot, optically-thick fireball, 
similar to that produced in the DT model.  Unfortunately, if the burst spectrum is produced
by such an optically-thick pair fireball, the actual mechanisms by which it formed are erased and will not be 
detectable from observation.

Several models have been suggested for explaining the hard component
recently discovered in quiescent emission.  Baring (2004) proposes
that resonant Compton upscattering of thermal X-rays by accelerated
particles in open field region produces the quiescent hard emission.
Thompson \& Beloborodov (2005) have proposed that the hard,
non-thermal quiescent component is due to the creation of a strong
$E_{\parallel}$ induced by twisting of field in closed region;
this leads to synchrotron radiation from electron acceleration at high 
altitude, or free-free emission from a hot ($\sim 100$~keV) transition
layer between the atmosphere and corona (Thompson \& Beloborodov 2005).
In the model of Heyl \& Hernquist (2005b), a small fraction
of the energy generated by the fast MHD wave breakdown goes into a
non-thermal pair component which produces the hard quiescent component
via a pair-synchrotron cascade.

\subsection{Radio emission and pair suppression in magnetars}

Although transient radio emission has been detected following
superflares of SGR1900+14 (Frail et al. 1999) and SGR1806-20 (Gaensler
et al. 2005), 
as well as the outburst of AXP XTE J1810-197 (Halpern et al.~2005),
no persistent, pulsed radio emission has been detected
from any SGRs or AXPs.  The report of radio pulsations from
SGR1900+14 (Shitov et al. 2000) was unconfirmed (Lorimer \& Xilouris
2000) and probably confused with emission from the nearby radio pulsar
PSR J1907+0918.  It was only very recently that radio pulsations were 
detected from one transient AXP, XTE J1810-197 (Camilo et al. 2006), 
after its 2003 X-ray outburst.  The fact that so few magnetars have 
detected radio pulsations is somewhat of a mystery given the dense pair
plasmas invoked by magnetar models.  It is widely believed that the
bright, coherent radio emission seen from rotation-powered pulsars is
produced by plasma instabilities requiring pair plasmas.  One must
conclude that the pair plasma instabilities operating to produce radio
emission in rotation-powered pulsars are not effective in magnetar
fields, that the radio emission is absorbed, or that magnetars do not
produce dense pair plasmas.  The first option is complicated by the
fact that radio pulsars with magnetar fields exist, so that the
inability of pairs to generate the coherent radio emission in
magnetars must be disrupted by the higher energy densities in their
magnetospheres.  During magnetar bursts, the plasma frequency is
almost surely above radio frequencies if even a small fraction of the
total energy density resides in pairs, so that any radio emission
would be absorbed.  However, in quiescence the particle density is
expected to be closer to that of the Goldreich-Julian charge of radio
pulsars, which gives plasma 
frequencies of 1 - 10 GHz near the surface
and falls of $r^{-3/2}$ above the surface.  Even if the particle
densities are several hundred times the Goldreich-Julian density,
radio emission would only be absorbed near the stellar surface.
Another reason which has been given for the absence of radio pulses is
that their small polar caps predict very small radio beams that are
hard to detect (Thompson \& Duncan 1996).  But as the number of known
magnetars increases, we would expect to be in view of a few of the
radio beams, and a number of radio pulsars with similarly small
predicted radio beams have been discovered.

The suppression of pair creation is also a possible explanation for
the absence of radio emission in magnetars.  It was noted that the
field strength $B \sim B_Q$ at which photon splitting begins to
dominate the attenuation of photons near pulsar polar caps occurs at
around the dividing line between radio pulsars and magnetars (Baring
\& Harding 1998) and would impose a death line for pair creation at
high fields.  However, the complete suppression of pair creation
requires that both photon polarization modes undergo splitting (Baring
\& Harding 2001), but only the $\perp$ mode splitting is allowed under
Adler's (1971) kinematic selection rules (\S \ref{sec:PhoSplit}).
Usov (2002) suggested that pair suppression could still occur if bound
pair creation were also taken into account.  Photons in the
$\parallel$ mode will create bound pairs with both particles in the
ground state, which are stable against annihilation (\S
\ref{sec:BoundPair}).  Photons in the $\perp$ mode would create bound
pairs with one or more particles in an excited state, which can be
dissociated by ionization in spin-flip transitions, but these photons
would split before creating the unstable bound pairs.

\section{Accreting X-ray Pulsars} \label{sec:XRPs}

X-ray pulsars are believed to be strongly-magnetized accreting neutron
stars powered by kinetic energy of infalling matter from a companion
star.  The magnetic fields of $10^{11} - 10^{13}$ G are inferred from
the pulsations that require anisotropic infall and radiation, the 
cyclotron lines observed in a number of pulsar spectra (Nagase 1989, dal Fiume et al. 2000), 
and the observed changes in pulse period (Ghosh \& Lamb 1979, Bildsten et al. 1997).  
We can get a fairly good estimate of the effective temperature observed
in X-ray pulsars from $T_{\rm eff} \simeq (L_x/\sigma A_{\rm cap})^{1/4}
\simeq 10$ keV, where $L_x$ is the X-ray luminosity and $A_{\rm cap}$ is
the heated polar cap area.  But clearly, since pulsar spectra are not
blackbody,  we must have an accurate description of the radiating plasma in
order to model the emission.

The conditions in the emission region are closely tied to the dynamics
of the accretion flow and infall, and there are two major regimes that
depend upon whether or not the radiation pressure of the emitting
plasma is capable of decelerating the accretion flow.  The critical
luminosity $L_{\rm crit}$ where this division occurs is not 
accurately determined but can be
estimated by requiring that the outgoing radiation pressure balance the
ram pressure of the infalling matter:
\be  \label{Lcrit}
L_{\rm crit} = {G M m_p c\over \sigma_{\parallel}} {A_{\rm cap}\over R^2}
\simeq 10^{36}\,\rm erg s^{-1} ,                
\ee
when $\sigma_{\parallel} \sim \sigma_T$ and $A_{cap}=0.1 R^2$, 
where $\sigma_{\parallel}$ is the magnetic Thomson scattering cross section
parallel to the field, averaged over the spectrum, polarization and angles
of the radiation.  The value of $L_{\rm crit}$ is about one hundredth of the Eddington
limit for spherical accretion, and is model dependent since the averaged 
cross section should include the radiation produced throughout the atmosphere in a 
self-consistent model.  The high luminosity X-ray pulsars where $L_x > L_{\rm crit}$ 
may radiate several orders of magnitude above $L_{\rm crit}$ and even 
above the spherical Eddington limit.  In low-luminosity sources where 
$L_x \ll L_{\rm crit}$, radiation pressure is not important in decelerating
the accretion flow, in which case it may be decelerated by a collisionless
shock above the neutron star surface or by Coulomb and nuclear collisions
with atmospheric plasma near the surface.  Most (about two thirds) of the 
observed X-ray pulsars fall into the high-luminosity category.

The most important effect of the neutron star magnetic field on the
processes that decelerate the accretion flow and that produce the
observed radiation is the quantization of particle momentum perpendicular
to the field (see \S \ref{sec:elec}).  In sources where $L_x \ll L_{\rm crit}$,
accretion flow can convert its kinetic energy to heat either at a
collisionless shock front (Basko \& Sunyaev 1976) or by Coulomb collisions
with thermal electrons and protons in the atmosphere (Basko \& Sunyaev 1975, Pavlov \& 
Yakovlev 1976).  Self-consistent
models of Coulomb-decelerated acceleration (Harding et al. 1984, Miller et al. 1989) 
result in thin slab-like atmospheres, since
Coulomb stopping lengths are only $\sim 50\,\rm g\,cm^{-2}$ (Kirk \& Galloway 1981, Pakey 1990).
Models for sources with $L_x > L_{\rm crit}$, where radiation pressure dominates
the deceleration, must solve the coupled radiation-hydrodynamic flow to find
the structure of the atmosphere.  In such models, a radiative shock stands off 
several stellar radii above the neutron star surface and the atmosphere has more of
a cylindrical geometry, as radiation escapes from the sides of a decelerating mound
of gas below the shock (Langer \& Rappaport 1982, Kirk 1985).

The dominant cooling process in X-ray pulsar atmospheres is resonant
Bremsstrahlung (or cyclotron cooling), in which an electron in the
ground state is collisionally excited and deexcites through
spontaneous emission, producing a cyclotron photon.  The inverse
process of resonant free-free absorption, where an electron is excited
by absorption of a cyclotron photon and collisionally deexcited, can
in some cases provide significant heating to the atmosphere.  So while
the vast majority of electron excitation-deexcitation events are
resonant Compton scattering at the cyclotron energy the collisions,
although less frequent, control the net production and destruction of
cyclotron photons in these atmospheres.  Another important heating and
cooling process is Compton scattering, which is resonant at the
cyclotron harmonics in a magnetic field.  While scattering does not
produce nearly as much heating and cooling as Bremsstrahlung since it
does not create and destroy cyclotron photons, it exchanges energy
between photons and electrons in the atmosphere (Meszaros \& Ventura
1979).  The presence of a strong magnetic field introduces substantial
changes in the stopping length due to Coulomb scattering.  A accreting
proton moving along the magnetic field experiences a much smaller drag
force than in the non-magnetic case because the electrons are limited
to momentum transfers along the field unless they can be excited to
higher states.  The infalling protons must diffuse in momentum space
through many small-angle collisions, gradually veering away from the
magnetic field direction where their drag force will increase and they
can decelerate faster.  As a result, the stopping lengths are very
dependent on the proton momentum diffusion and significantly larger
than in an un-magnetized plasma 
(see, e.g., Nelson, Salpeter \& Wasserman 1993).

\subsection{Cyclotron lines}

The cyclotron lines seen in the spectra of accreting X-ray pulsars are
formed by resonant scattering (\S \ref{sec:CompScatt}) of photons with
electrons that occupy discrete Landau states.  If the plasma
temperature is less than $T_B = \epsilon_B/k$, then the electrons
occupy primarily the ground Landau state [regime (i) of \S
\ref{sec:NS}] and have a one-dimensional thermal distribution of
momenta along the magnetic field.  At magnetic field strengths typical
of neutron stars, the cyclotron decay rates from excited Landau states
($n > 0$), $\Gamma_n = 4 \times 10^{15} n B_{12}^2 \,\rm s^{-1}$, are
much greater than the rate of collisional excitation.  Unless there
are other processes that can populate the excited states on faster
timescales, the bulk of the electrons are expected to occupy the
ground state in a one-dimensional distribution.  Under these
conditions, absorption of a cyclotron photon is always followed by the
emission of another photon, so resonant scattering is more important
than absorption.  As discussed in \S \ref{sec:CompScatt} , the Compton
scattering cross section has sharp maxima at the cyclotron resonance
(Eqn [\ref{eq:eps_abs}]) and its harmonics, so that photons scatter
preferentially at these energies.  Since the scattering cross section
has a strong dependence on angle to the magnetic field, the cyclotron
resonance scattering features (CRSFs) provide excellent diagnostics of
the emission geometry as well the physical conditions (temperature and
density) of the radiating plasma.

The formation of CRSFs at the fundamental and at the harmonics
proceeds by very different routes.  When a photon at the fundamental
resonance frequency (Eqn [\ref{eq:eps_abs}] with 
$N=1$) scatters with
an electron in the ground state the scattered photon also appears at
the fundamental.  The scattered photon may have a different angle,
but unless the continuum photons are beamed, resonant photons from
other angles scatter into view.  Absorption-like features can only
form in the fundamental by repeated scattering, causing diffusion of
photons in energy and angle through the combined effects of electron
recoil and the natural line width of the resonance (Wasserman \&
Salpeter 1980, Wang et al. 1989).  Resonant photons can then diffuse
from the line core into the line wings where the cross section drops
to near $\sigma_T$ and they can then escape, primarily in the red
wing.  When a photon at the first harmonic 
($N=2$) scatters with an
electron from the same distribution, the scattered photon will appear
most of the time at the fundamental in fields $B' \lsim 0.2$, exciting
the electron to the 
$n=1$ state.  The electron then decays through
cyclotron emission, spawning an additional photon at the fundamental.
Although this is technically still a resonant scattering ($n= 0
\rightarrow 2 \rightarrow 1$ followed by $n=1 \rightarrow 0$),
analogous to Raman scattering in atomic physics, incident photons are
destroyed in the first harmonic and the line formation here is well
approximated by absorption (Fenimore et al. 1988, Wang et al. 1988,
Harding \& Daugherty 1991).  Thus, the second and higher harmonics are
formed in a similar way, with photons in the 
$N$th resonance exciting the electron from the 
ground state to the 
$n-1$ state, with the
subsequent spawning of $n-1$ additional photons at the fundamental
through single harmonic number transitions.  Photon spawning partially
fills in the fundamental (Wang et al. 1988), so that the line features
at the cyclotron fundamental and higher harmonics will not appear in
the same ratios in the observed spectrum as they do in the cross
section.  In magnetic fields $B' \gsim 0.2$, the dominance of
cyclotron decays to the ground state (\S \ref{sec:CycRad}) reduce the
rate of photon spawning in the fundamental (Araya \& Harding 1999)
since scattering in the $N$th harmonic is more likely to produce a
scattered photon in the $N$th resonance ($n= 0 \rightarrow N
\rightarrow 0$).  Formation of the harmonics in these high fields
therefore proceeds more by diffusion out of the line core, as the
fundamental is formed, rather than by excitation of the electron.

As in most line-formation scenarios, the broadening of the CRSFs
results from a combination of natural line width, which gives a
Lorentzian profile, and the Doppler effect, which gives 
a Gaussian profile.  Due the one-dimensional momentum distribution of the
electrons, the line broadening is asymmetric and dependent on viewing
angle to the magnetic field, $\theta$.  First order (non-relativistic)
Doppler broadening alone gives a line width that is approximately
\be
\Delta\epsilon_D \approx \epsilon_N(2T)^{1/2}\cos\theta,
\ee
where $T$ is the plasma temperature in units of $mc^2$ and
$\epsilon_N$ is the resonant energy (Eqn [\ref{eq:eps_abs}]).  At
small viewing angles, the lines are broadened by the full Doppler
motion, while in the limit of large viewing angles, only transverse
(second-order) Doppler broadening contributes to the line width.
Since the transverse Doppler broadening is a red shift (from time
dilation), the line profiles become increasingly asymmetric at large
incidence angles.  The natural line width, due to the finite lifetime
of excited Landau states, is approximately equal to the cyclotron
decay rate from state $N$ which, in the non-relativistic limit, is
\be
\Delta\epsilon_N \approx {4\over 3}\alpha B'^2 N
\ee
At large angles, $\Delta\epsilon_N$ provides the main contribution to
the broadening on the blue side of the line.  In a relativistic
treatment of line broadening, the cyclotron decay rate is dependent on
the electron spin, with the rates of spin-flip transitions being
smaller (Herold et al. 1982).

\begin{figure} 
\hskip -0.5cm
\includegraphics[width=16cm]{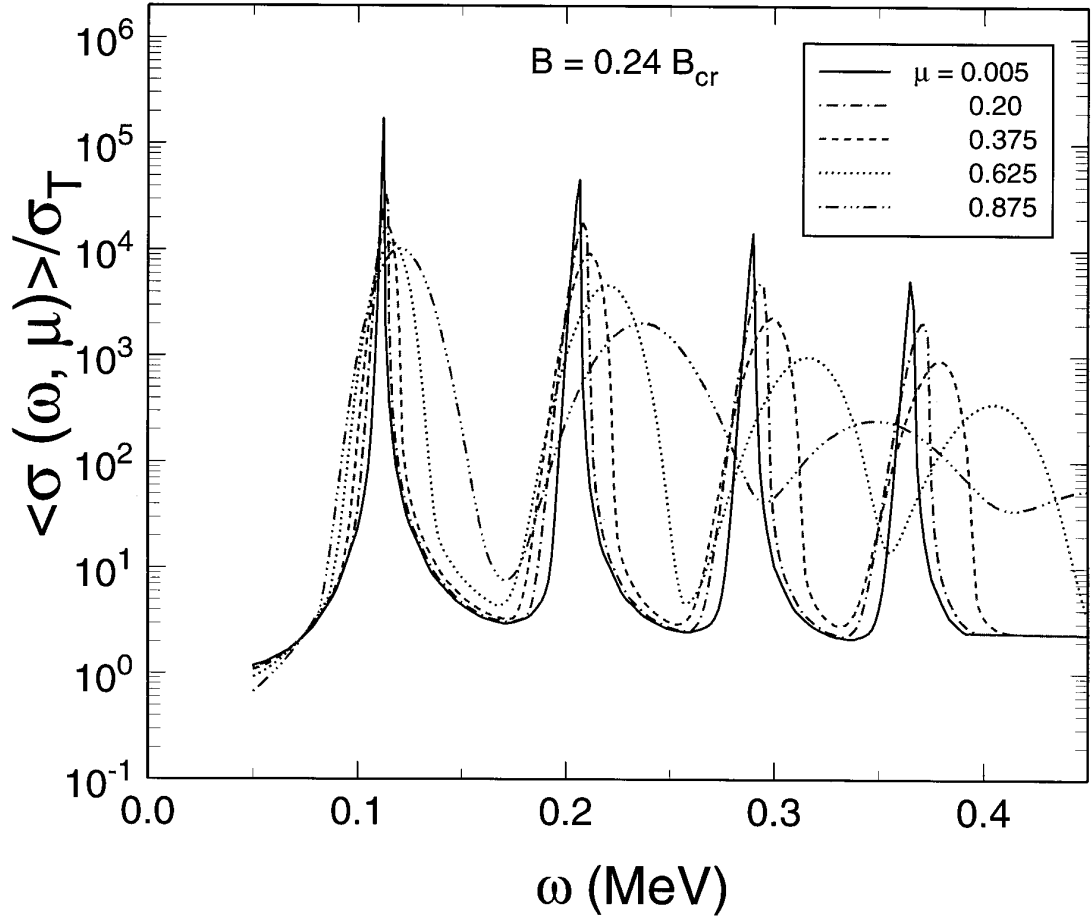}
\caption{Relativistic cyclotron scattering line profiles for $B'$ = 0.24 and thermal electron 
temperature of 31 keV, for different viewing angles, $\mu = \cos\theta$.}
\label{fig:ArayaHarding1999}
\end{figure}

Techniques for modeling CRSFs in XRP atmospheres have included either
solving the equations of radiative transfer using a Feautrier method
(Meszaros \& Nagel 1985, Nagel 1981, Alexander \& Meszaros 1991b,
Burnard et al. 1988) or Monte Carlo simulation (Wang et al. 1988,
Araya \& Harding 1999, 2000).  Given the complexities of the line
transfer, these models usually assume a static atmosphere and
simplified geometries, in particular homogeneous slabs, with the
magnetic field perpendicular to the slab normal, or cylinders parallel
to the field.  An exception is the work of Arons et al. (1987) and
Burnard et al. (1988) who treat a time-dependent accretion column.
Such studies have generally confirmed that the shape and relative
depths of the fundamental and higher harmonics of CRSF are very
dependent on viewing geometry as well as geometry of the scattering
plasma.  In particular, CRSFs seen at small angles to the magnetic
field tend to be broader with weak harmonics, whereas CRSFs seen at
large angles are narrower, more asymmetric and have stronger features
at higher harmonics.  In slab geometry the depth of the fundamental is 
largest at large angles to the field, where the optical depth is
largest, but in cylinder geometry the fundamental feature is largest at 
small angles (Araya \& Harding 1999).  In many cases the
fundamental is flanked by emission wings, from photons escaping the
line core, and may even appear completely in emission, at large angles
in cylinder geometry, due to photons diffusing from smaller angles.

Given the sensitive dependence of CRSF characteristics with angle to
the field, one would expect a great deal of variation in the features
with rotational phase of the neutron star.  This is in fact what is
observed in phase resolved spectra of accreting XRPs (Heindl et
al. 2004, Orlandini \& dal Fiume 2001).  The line depths as well as
their energies change with phase, but the harmonic relationships are
preserved, as would be expected of CRSFs.

\section{Conclusion}

In this article we have reviewed the physics that applies in extremely
strong magnetic fields, the properties of strongly magnetized neutron
stars, atmospheres and magnetospheres, and the present status of our
understanding and models of the astrophysical sources that are their
manifestation.  While the use of non-relativistic quantum mechanics is
adequate (even for $B>B_Q$) for describing bound states or any
processes where the electron stays in the ground Landau level, one
must use relativistic quantum mechanics (QED) in computing the rates
and cross sections for many free particle and photon processes.
Nearly all of the cross sections for the first and second-order free
particle processes in strong magnetic fields, such as cyclotron
radiation and absorption, one photon and two-photon pair creation and
annihilation and Compton scattering, have been calculated and studied.
Only a few of the higher-order processes, such as Bremsstrahlung and
photon splitting, have been investigated since they become
increasingly complex.  The unusual and interesting properties of
photon propagation in strong magnetic fields has been a topic of
intensive study in recent years, especially after the discovery of
magnetars.  Among all the processes that have been investigated,
vacuum polarization has been found to be of particular importance.

A variety of astrophysical sources, that include rotation-powered and
accretion-powered neutron stars and the magnetars, SGRs and AXPs, are
believed to be very strongly magnetized neutron stars.  The processes
that operate only in such strong fields may be fundamental to the
functioning of these sources.  For rotation-powered pulsars,
one-photon pair creation is thought to be the primary process
attenuating high-energy photons, and the created pairs may be critical
in the production of the observed radio pulsations.  We do not yet
fully understand though how and why the radio emission seems to turn
off well before the pulsar spin-down has ceased.  Theoretical models
show that pair production decreases as the pulsar ages, but does it
turn off suddenly, or are there some threshold properties
(e.g. multiplicity or energy spectrum) for radio emission?  In the
case of the magnetars, power levels far exceeding their spin-down
power requires fast magnetic field decay on timescales only possible
through processes such as ambipolar diffusion, that become effective
in fields above $10^{14}$ G.  We do not yet understand, however, why
these sources are radio quiet.  Is pair production suppressed by the
strong field or are the collective plasma processes disrupted?
Theoretical studies have so far not been able to find a convincing
mechanism for the suppression of the radio emission in magnetars.
Even more puzzling is the growing number of magnetars and radio
pulsars having very similar spin properties, and therefore similar
implied surface magnetic field strengths.  Aside from their radio
emission properties, radio pulsars and magnetars also have very
different X-ray emission levels, transient emission and glitching
behavior.  What are the hidden characteristics of these sources that
distinguish them?  Also of interest are the emerging population of
``dim'' isolated neutron stars with apparently pure thermal
emission. The spectral features detected in some of these sources are
exciting, but their identifications remain unclear. The nature of
these sources is unknown. Could they be descendants of magnetars?

To find answers to these questions, new ideas and more theoretical
work are surely needed.  But new and more sensitive instruments are on
the horizon that will also provide some clues as well.  The ALFA
pulsar survey (Cordes et al. 2005) began operation last year at
Arecibo, one of the world's most powerful radio telescopes, and this
survey is expected to discover at least 1000 new radio pulsars, nearly
doubling the current number.  Among the newly discovered pulsars will
be many more radio pulsars with magnetar field strengths.  The
Gamma-Ray Large Area Telescope (GLAST) (McEnery et al. 2004), due to
launch in 2007, will have a point source detection threshold about 20
times below that of EGRET and will discover hundreds of new
$\gamma$-ray pulsars.  GLAST will also have sensitivity up to 200 GeV
and will be able to make very sensitive measurements of spectral
high-energy cutoffs.  Third-generation ground-based Air Cherenkov
detectors, such as H.E.S.S.  (Hinton et al. 2004) in Namibia, have
begun operation.  They are sensitive to $\gamma$-rays in the range 50
GeV to 50 TeV and may discover or put important constraints on pulsar
and magnetar spectra and their nebulae.  Further into the future are
planned X-ray telescopes, such as Constellation X, XEUS, and X-ray
polarimeters, such as AXP, POGO and ACT.  Polarimeters in particular will be
extremely important in looking for some of the signatures of very
strong magnetic fields that have been discussed in this article, such
as the vacuum polarization resonance and photon splitting cutoffs.

\ack
We thank Matthew Baring for comments on the manuscript. This work has been supported 
in part by NSF grant AST 0307252 (DL).

\section*{References}

\begin{harvard}

\item Adler, S. L., Bahcall, J. N., Callan, C. G. \& Rosenbluth, M. N. 1970, {\it Phys. Rev. Lett.} {\bf 25} 1061.

\item Adler, S.~L. 1971, {\it Ann. Phys.} {\bf 67} 599.

\item Akgun, T., Link, B., \& Wasserman, I. 2006, {\it MNRAS}, {\bf 365}, 653.

\item Alcock, C., Farhi, E., \& Olinto, A. 1986, {\it ApJ}, {\bf 310}, 261.

\item Alford, M., Rajagopal, K., Reddy, S., \& Wilczek, F. 2001, {\it Phys. Rev.}, {\bf D64}, 075017.

\item Alford, M., et al.~2005, {\it ApJ}, {\bf 629}, 969.

\item Alexander, S. G. \& Meszaros, P. 1991a, {\it ApJ} {\bf 372} 554.

\item Alexander, S. G. \& Meszaros, P. 1991b, {\it ApJ} {\bf 372} 565.

\item Al-Hujaj, O.-A., \& Schmelcher, P. 2003a, {\it Phys. Rev.} {\bf A67}, 023403

\item Al-Hujaj, O.-A., \& Schmelcher, P. 2003b, {\it Phys. Rev.} {\bf A68}, 053403

\item Alpar, M. A.; Cheng, A. F.; Ruderman, M. A.; Shaham, J. 1982, {\it Nature} {\bf 300} 28.

\item Araya, R. \& Harding, A. K. 1999, {\it ApJ} {\bf 517} 334.

\item Araya-Góchez, R. A. \& Harding, A. K. 2000, {\it ApJ} {\bf 544 } 1067.

\item Arendt, P. N. \& Eilek, J. A. 2002, {\it ApJ} {\bf 581} 451.

\item Arons, J., \& Barnard, J.J. 1986, {\it ApJ}, {\bf 302}, 120

\item Arons, J., \& Scharlemann, E.~T. 1979, {\it ApJ} {\bf 231} 854.

\item Arons, J., Klein, R. \& Lea, S. 1987, {\it ApJ} {\bf 312} 666.

\item Arras, P., Cumming, A., \& Thompson, C. 2004, {\it ApJ}, {\bf 608}, L49

\item Ashcroft, N.W., and N.D. Mermin, 1976, Solid State Physics (Saunders
College: Philadelphia)

\item Asseo, E., \& Riazuelo, A. 2000, {\it MNRAS}, {\bf 318}, 983

\item Avron, J.E., Herbst, I.B., \& Simon, B. 1978, {\it Ann. Phys. (NY)} {\bf 114}, 431

\item Baade, W. \& Zwicky, F. 1934, {\it PhRv} {\bf 46} 76.

\item Barnard, J.J. 1986, {\it ApJ}, {\bf 303}, 280

\item Backer, D. C., Kulkarni, S. R., Heiles, C., Davis, M. M. \& Goss, W. M. 1982, {\it Nature} {\bf 300} 615.

\item Baring, M. G. 1988, {\it MNRAS} {\bf 235} 51.

\item Baring, M. G. 1991, {\it  A \& A} {\bf 249} 581.

\item Baring, M.~G. 1995, {\it ApJ} {\bf 440} L69.

\item Baring, M.~G. 2004, in  {\it Young Neutron Stars and Their Environments}
    eds F. Camilo and B. M. Gaensler
    (IAU Symposium 218, Astronomical Society of the Pacific, San Francisco), p. 267.

\item Baring, M. G. \& Harding, A. K., 1992, in Proc. of Second GRO Science Workshop, ed. C.R. Schrader, N. Gehrels \& B. Dennis (NASA CP-3137, Washington, DC), 245.

\item Baring, M.~G. \& Harding, A. K., 1998, {\it ApJ} {\bf 507} L55.

\item Baring, M.~G. \& Harding, A. K., 2001, {\it ApJ} {\bf 547} 929.

\item Baring, M.~G., Gonthier, P. L. \& Harding, A. K., 2005, {\it ApJ} {\bf 630} 430.

\item Basko, M. M. \& Sunyaev, R. A. 1975, {\it  A \& A} {\bf 42} 311.

\item Basko, M. M. \& Sunyaev, R. A. 1976, {\it MNRAS} {\bf 175} 395. 

\item Bayes, D., \& Vincke, M. 1998, in {\it Atoms and Molecules in Strong Magnetic
Fields}, edited by P. Schmelcher and W. Schweizer (Plenum Press, New York),
p.~141

\item Baym, G., Pethick, C., \& Pines, D. 1969, {\it Nature}, {\bf 224}, 674

\item Baym, G., \& Pethick, C. 1979, {\it Ann. Rev. Astron. Astrophys.}, {\bf 17}, 415

\item Baym, G., Pethick, C., \& Sutherland, P. 1971, {\it ApJ} {\bf 170}, 299. 

\item Becker, W., \& Pavlov, G.G. 2002, in The Century of Space Science,
ed. J. Bleeker et al. (Kluwer) (astro-ph/0208356)

\item Bekefi, G. 1966, {\it Radiation Processes in Plasmas} (Wiley: New York).

\item Beloborodov, A.M. 2002, {\it ApJ}, {\bf 566}, L85

\item Bezchastnov, V.G., et al.~1998, {\it Phys. Rev.} {\bf A58}, 180

\item Bhatia, V. B., Chopra, N. \& Panchapakesan, N. 1992, {\it ApJ} {\bf 388} 131.

\item Bhattacharya, D., et al. 1992, {\it A\&A}, {\bf 254}, 198

\item Bhattacharya, D., \& Srinivasan, G. 1995, in X-ray Binaries, ed.
W.H.G.Lewin et al. (Cambridge: Cambridge Univ. Press), p495

\item Bialynicka-Birula, Z. \& Bialynicki-Birula, I. 1970, {\it PhRvD } {\bf 2} 2341.

\item Bildsten, L. et al. 1997, {\it ApJ} {\bf 113} 367.

\item Bisnovati-Kogan, G.S., \& Komberg, B.V. 1975, {\it Sov. Astron.}, 
{\bf 18}, 217

\item Blaes, O., Blandford, R., Madau, P., \& Koonin, S. 1990, {\it ApJ}, {\bf 363}, 612

\item Blandford, R., et al. 1983, MNRAS, 204, 1025

\item Blandford, R.D., \& Hernquist, L. 1982, {\it J. Phys. C} {\bf 15}, 6233

\item Blandford, R. D. \& Scharlemann, E. T. 1976, {\it MNRAS} {\bf 174} 59.

\item Bonazzola, S., Heyvaerts, J., \& Puget, J. 1979, {\it A \& A}, {\bf 78}, 53.

\item Brainerd, J. J. 1989, {\it ApJ} {\bf 341} L67.

\item Brainerd, J. J. \& Lamb, D. Q. 1987, {\it ApJ} {\bf 313} 231.

\item Braithwaite, J., \& Spruit, H.C. 2006, A\&A, submitted (astro-ph/0510287)

\item Brinkmann, W. 1980, A\&A, 82, 352

\item Broderick, A.M., Prakash, M., \& Lattimer, J.M. 2000, {\it ApJ}, {\bf 537}, 351

\item Budden, K.G. 1961, Radio Waves in the Ionosphere (Cambridge: Cambridge Univ. Press)


\item Bulik, T. 1998, {\it Acta Astronomica} {\bf 48} 695.

\item Burnard, D. J., Klein, R. I. \& Arons, J. 1988, {\it ApJ} {\bf 324} 1001.

\item Burns, M. L. \& Harding, A. K. 1984, {\it ApJ} {\bf 285} 747.

\item Burwitz, V., et al. 2003, {\it A\&A}, {\bf 399}, 1109

\item Bussard, R. W., Alexander, S. B. and Meszaros, P. 1986, {\it Phys. Rev. D} {\bf 34} 440.

\item Camilo, F. et al. 2006, Nature, submitted.

\item Canuto, V., Lodenquai, J. and Ruderman, M. 1971, {\it Phys. Rev. D} {\bf 3} 2303.

\item Canuto, V. and Ventura, J. 1977, {\it Fundamentals of Cosmic Physics} {\bf Vol. 2} p. 203.

\item Canuto, V., \& Ventura, J. 1972, {\it Ap. Space Sci.}, {\bf 18}, 104

\item Caraveo, P.A., et al.~2004, {\it Science}, {\bf 305}, 376

\item Ceperley, D., \& Alder, B. 1980, {\it Phys. Rev. Lett.} {\bf 45}, 566

\item Chabrier, G., 1993, {\it ApJ}, {\bf 414}, 695

\item Chabrier, G., Ashcroft, N.W., \& DeWitt, H.E. 1992, {\it Nature}, {\bf 360}, 6399

\item Chabrier, G., Saumon, D.S., \& Potekhin, A.Y. 2006, {\it Journal of Phys. A}, in press

\item Chakrabarty, D. 2005, in Binary Radio Pulsars (ASP Conf. Ser.), eds.
F.A. Rasio \& I.H. Stairs (astro-ph/0408004), p. 279.

\item Chang, P., Arras, P., \& Bildsten, L. 2004, {\it ApJ}, {\bf 616}, L147

\item Chen, K. \& Ruderman, M. A., 1993, {\it ApJ} {\bf 402} 264.

\item Cheng, A.F., \& Ruderman, M. 1979, {\it ApJ}, {\bf 229}, 348

\item Cheng, K. S. 1994, in {\it Toward a Major Atmospheric Cherenkov Detector} ed. T. Kifune (Tokyo: Universal
Academy), 25.

\item Cheng, K.~S., Ho, C., \& Ruderman, M.~A. 1986, {\it ApJ} {\bf 300} 500.

\item Cheng, K.~S., Ruderman, M.~A. \& Zhang, L. 2000, {\it ApJ} {\bf 537} 964.

\item Cho, J., \& Lazarian, A. 2004, ApJ, 615, L41

\item Cordes, J. M et al. 2006, {\it ApJ}, 637, 446.

\item Cumming, A., Zweibel, E., \& Bildsten, L. 2001, ApJ, 557, 958

\item Cumming, A., Arras, P., \& Zweibel, E. 2004, ApJ, 609, 999

\item dal Fiume, D et al. 2000, {\it AdSpR} {\bf 25} 399.

\item Daugherty, J. K. and Bussard, R. W. 1980, {\it ApJ} {\bf 238} 296.

\item Daugherty, J.~K., \& Harding, A.~K. 1982, {\it ApJ} {\bf 252} 337.

\item Daugherty, J. K. \& Harding, A. K. 1983, {\it ApJ} {\bf 273} 761.

\item Daugherty, J.~K., \& Harding, A.~K. 1986, {\it ApJ} {\bf 309} 362.

\item Daugherty, J.~K., \& Harding A.~K. 1996, {\it ApJ} {\bf 458} 278.

\item Daugherty, J. K. \& Lerche, I. 1975, {\it ApSS} {\bf 38} 437.

\item Daugherty, J. K. \& Lerche, I. 1976, {\it PhRvD } {\bf 14} 340.

\item Daugherty, J. K. \& Ventura, J. 1978, {\it PhRvD } {\bf 18} 1053.

\item De Luca, A., et al. 2004, {\it A\&A}, {\bf 418}, 625

\item De Luca, A., et al. 2005, {\it ApJ}, {\bf 623}, 1051 (astro-ph/0412662)

\item Demeur, M., Heenen, P.-H., \& Godefroid, M. 1994, {\it Phys. Rev. A} {\bf 49}, 176

\item den Hartog, P.~R., Kuiper, L., Hermsen, W. \& Vink, J 2004, Astron. Tel. 293.

\item Donati, J.-F., et al.~2006, MNRAS, 365, L6

\item Douchin, F., \& Haensel, P. 2001, {\it A\&A}, {\bf 380}, 151

\item Drake, J.J. et al. 2002, {\it ApJ}, {\bf 572}, 996

\item Duncan, R. C. 2000, Fifth Huntsville Gamma-Ray Burst Symposium, Eds: R. Marc Kippen, Robert S. Mallozzi, Gerald J. Fishman. AIP Vol. 526 (American Institute of Physics, Melville, New York) p.830 [astro-ph/0002442].

\item Duncan, R. C. \& Thompson, C. 1992, {\it ApJ} {\bf 392} 9.



\item Erber, T. 1966, {\it Rev. Mod. Phys} {\bf 38} 626.

\item Faber, J.A., et al. 2002, {\it Phys. Rev. Lett.}, {\bf 89}, 231102

\item Faucher-Gigu\'er\'e, C., \& Kaspi, V.M. 2006, ApJ, submitted
(astro-ph/0512585)

\item Ferrario, L., \& Wichramasinghe, D.T. 2005, MNRAS, 356, 615

\item Flowers, E.G., et al.
1977, {\it ApJ}, {\bf 215}, 291

\item Fenimore, E. E. et al. 1988, {\it ApJ} {\bf 335} L71.

\item Frail, D. A.,  Kulkarni, S. R. \& Bloom, J. S  1999, {\it Nature} {\bf 398} 127.

\item Furry, W. H. 1937, {\it Phys. Rev.} {\bf 51} 125.

\item Fushiki, I., Gudmundsson, E.H., \& Pethick, C.J. 1989, {\it ApJ}, {\bf 342}, 958

\item Fushiki, I., Gudmundsson, E.H., Pethick, C.J., \& Yngvason, J. 1992, {\it Ann. Phys. (NY)}, {\bf 216}, 29

\item Gaensler, B. et al. 2005, {\it Nature} {\bf 434} 1104.

\item G\"ansicke, B. T., Braje, T. M., \& Romani, R.W. 2002, {\it A\&A}, {\bf 386}, 1001

\item  Garstang, R. H. 1977, RPPh, 40, 105.

\item  Gavriil, F.P., Kaspi, V.M., \& Woods, P.M. 2002, Nature, 419, 142

\item Ghosh, P. \& Lamb, F. K. 1979, {\it ApJ} {\bf 234} 296.

\item Ginzburg, V.~L. 1970,
{\it The Propagation of Electromagnetic Waves in Plasmas},
2nd ed. (London: Pergamon)

\item Glendenning, N.K. 1992, {\it Phys. Rev. D} {\bf 46}, 1274

\item Glendenning, N.K. 2000, {\it Compact Stars: Nuclear Physics, Particle Physics
and General Relativity} (Springer: New York)

\item Gnedin, Yu.N. \& Pavlov, G.G. 1974, {\it Sov. Phys. JETP}, {\bf 38}, 903


\item Gnedin, Yu.N., Pavlov, G.G., \& Shibanov, Yu.A. 1978, {\it JETP Lett.}, {\bf 27}, 305

\item Gonthier, P.~L., \& Harding A.~K. 1994, {\it ApJ} {\bf 425} 747.

\item Gonthier, P.~L., Harding A.~K., Baring, M. G., Costello, R. M. 
   \& Mercer, C. L. 2000, {\it ApJ} {\bf 540} 907.

\item Gonthier, P. et al. 2004, ApJ, 604, 775

\item Gold, T. 1968, {\it Nature} {\bf 218} 731.

\item Goldreich, P. \& Reisenegger, A. 1992, ApJ, 395, 250.

\item Graziani, C. 1993, {\it ApJ} {\bf 412} 351.

\item Haberl, F., et al. 2003, {\it A\&A}, {\bf 403}, L19


\item Haberl, F. 2005, Proceedings of the 2005 EPIC XMM-Newton Consortium
   Meeting, Ringberg Castle, Germany, April 11-13 2005, Edt. U.G. Briel, S.
   Sembay and A. Read, MPE Report 288 (astro-ph/0510480)

\item Haberl, F., et al. 2004, {\it A\&A}, {\bf 424}, 635 

\item Haensel, P., \& Zdunik, J.L. 1990, {\it A\&A}, {\bf 222}, 353


\item 
Halpern, J.P., et al.~2005, {\it ApJ}, {\bf 632}, L29

\item Han, J.L., et al.~1998, {\it MNRAS}, {\bf 300}, 373

\item Harding, A. K. 1986, {\it ApJ} {\bf 300} 167 (462).

\item Harding, A. K. 2001, in {\it High Energy Gamma-Ray Astromony} ed. F. A. Aharonian \& H. J. Volk, AIP Conf. Series Vol. 558 (AIP: New York), 115.

\item Harding, A. K. 2003, in in Pulsars, AXPs and SGRs Observed with BeppoSAX, ed. G. Cusumano, E. Massaro, 
T. Mineo (Italy: Aracne Editrice), p. 127 [astro-ph/0304120].

\item Harding, A. K. 2005, in {\it Astrophysical Sources of High Energy Particles and Radiation} eds. T. Bulik, B. Rudak, G. Madejski, AIP Conference Proceedings, 
vol 801, 241.

\item Harding, A. K. \& Baring, M. G. 1996, in {\it Gamma-ray bursts}, Proceedings of the 3rd Huntsville Symposium, (Woodbury, NY: American Institute of Physics), AIP Conference Proceedings Series vol. 384, ed. C. Kouveliotou, M. F. Briggs, and G. J. Fishman, p.941.

\item Harding, A. K. \& Daugherty, J. K.: 1991, {\it ApJ} {\bf 374} 687.

\item Harding, A.~K. \& Muslimov, A.~G. 2001, {\it ApJ} {\bf 556} 987. 

\item Harding, A.~K. \& Muslimov, A.~G. 2002, {\it ApJ} {\bf 568} 862.

\item Harding, A. K., Muslimov, A. \& Zhang, B. 2002, {\it ApJ} {\bf 576} 366.

\item Harding, A. K., Meszaros, P., Kirk, J. G. \& Galloway, D. J. 1984, {\it ApJ},
{\bf 278} 369.

\item Harding, A. K. and Preece, R. D. 1987, {\it ApJ} {\bf 319} 939.

\item Harding, A.~K., Baring, M.~G. \& Gonthier, P.~L. 1997, {\it ApJ} {\bf 476} 246.

\item Haxton, W.C. 1995, {\it ARAA}, {\bf 33}, 459

\item Heger, A., Woosley, S.E., \& Spruit, H.C. 2005, ApJ, 626, 350

\item Heindl, W. A. et al. 2004, 
{\it X-ray Timing 2003: Rossi and Beyond.} AIP Conference Proceedings, Vol. 714, held 3-5 November, 2003 in Cambridge, MA. Edited by Philip Kaaret, Frederick K. Lamb, and Jean H. Swank. Melville, NY: American Institute of Physics, 2004., p.323-330

\item Heinke, C.O., et al. 2003, {\it ApJ}, {\bf 588}, 452

\item Heiselberg, H. 2002, astro-ph/0201465

\item Heisenberg, W. \& Euler, H. 1936, 
{\it Z. Physik}, {\bf 98}, 714

\item Hernquist, L. 1984, {\it ApJ Suppl. Ser.}, {\bf 56}, 325

\item Hernquist, L. 1985, {\it MNRAS}, {\bf 213}, 313

\item Herold, H., Ruder, H., \& Wunner, J. 1981, {\it J.~Phys. B} {\bf 14}, 751

\item Herold, H. 1979, {\it Phys. Rev. D} {\bf 19} 2868.

\item Herold, H., Ruder, H. and Wunner, G. 1982, {\it  A \& A} {\bf 115} 90.

\item Herold, H., Ruder, H. \& Wunner, G. 1985, {\it PhRvL} {\bf 54} 1452.

\item Heyl, J. 2005, in {\it Proc. of 22nd Texas Symposium on Relativistic Astrophysics at Stanford University}, Stanford, 2004, ed. P. Chen et al., eConf C041213.

\item Heyl, J.S. \& Hernquist, L. 1997, {\it J. Phys. A}, {\bf 30}, 6485-6492

\item Heyl, J.S., \& Hernquist, L. 1998, {\it MNRAS}, {\bf 300}, 599

\item Heyl, J. \& Hernquist, L. 1999, {\it PhRvD } {\bf 59} 045005.

\item Heyl, J.S., \& Hernquist, L. 2001, {\it MNRAS}, {\bf 324}, 292

\item Heyl, J. \& Hernquist, L. 2005a, {\it ApJ} {\bf 618} 463.

\item Heyl, J. \& Hernquist, L.  2005b, {\it MNRAS} {\bf 362} 777.

\item Heyl, J. S. \& Kulkarni, S. R. 1998, {\it ApJ} {\bf 506} L61.

\item Heyl, J.S., Shaviv, N.J., \& Lloyd, D. 2003, {\it MNRAS}, {\bf 342}, 134

\item Hewish, A., Bell, S. J., Pilkington, J. D., Scott, P. F. \& Collins, R. A. 1968, {\it Nature} {\bf 217} 709.

\item Hibschman, J. A. \& Arons, J. 2001, {\it ApJ} {\bf 554} 624.

\item Hinton, J. A. et al. 2004, New Astron.Rev., 48, 331. 

\item Hirotani, K. \& Shibata, S. 2001, {\it MNRAS} {\bf 325} 1228.

\item Ho, W.~C.~G., \& Lai, D. 2001, {\it MNRAS}, {\bf 327}, 1081

\item Ho, W.~C.~G., \& Lai, D. 2003, {\it MNRAS}, {\bf 338}, 233

\item Ho, W.~C.~G., \& Lai, D. 2004, {\it ApJ}, {\bf 607}, 420

\item Ho, W.~C.~G., Lai, D., Potekhin, A.~Y., \& Chabrier, G. 2003, {\it ApJ}, {\bf 599}, 1293

\item Hollerbach, R., \& R\"udiger, G. 2002, MNRAS, 337, 216

\item Hurley, K.  et al. 1999, {\it ApJ} {\bf 510} L110.

\item Hurley, K.  et al. 2005, {\it Nature} {\bf 434}, 1098

\item Ibrahim, A.L., Swank, J.H., \& Parke, W. 2003, ApJ, 584, L17

\item Jackson, P. 1975, {\it Classical Electrodynamics} (Wiley, New York).

\item Jackson, M.S., \& Halpern, J.P. 2005, {\it ApJ}, {\bf 633}, 1114

\item Jauch, M. M. \& Rohrlich, F. 1980, {\it The Theory of Photons and Electrons} (2nd Ed. Springer, Berlin).

\item Johnson, B.R., Hirschfelder, J.O., \& Yang, K.-H. 1983, {\it Rev. Mod. Phys.} {\bf 55}, 109

\item Johnson, M.H., \& Lippmann, B.A. 1949, {\it Phys. Rev.}, {\bf 76}, 828

\item Jones, M.D., \& Ceperley, D.M. 1996, {\it Phys. Rev. Lett.}, {\bf 76}, 4572

\item Jones, M.D., Ortiz, G., \& Ceperley, D.M. 1999, {\it Phys. Rev. A.} {\bf 59}, 2875

\item Jones, P.B., 1985, {\it MNRAS}, {\bf 216}, 503

\item Jones, P.B., 1986, {\it MNRAS}, {\bf 218}, 477

\item Jones, P.B. 1999, {\it Phys. Rev. Lett.}, {\bf 83}, 3589

\item Jones, P.B. 2004, {\it Phys. Rev. Lett.}, {\bf 93}, 221101

\item Juett, A.M., Marshall, H.L., Chakrabarty, D., \& Schulz, N.S. 2002, {\it ApJ}, {\bf 568}, L31

\item Kadomtsev, B.B., 1970, Zh. Eksp. Teor. Fiz. {\bf 58}, 1765 
[Sov. Phys. JETP {\bf 31}, 945 (1970)]

\item Kadomtsev, B.B., \& Kudryavtsev, V.S. 1971, {\it Pis'ma Zh. Eksp. Teor. Fiz.} 
{\bf 13}, 61 [{\it Sov. Phys. JETP Lett.} {\bf 13}, 42 (1971)]

\item Kaminker, A. D., Pavlov, G. G. \& Mamradze, P. G. 1987, {\it ApSS} {\bf 138} 1.

\item Kanbach, G. 2002, in {\it Neutron Stars, Pulsars, and Supernova Remnants} ed. W. Becker 
and H. Lesch, and J. Trümper (Max-Plank-Institut für extraterrestrische Physik:
Garching bei München), 91.

\item Kaspi, V.M. et al. 2003, {\it ApJ} {\bf 588} L93.

\item Kaspi, V.M. \& McLaughlin, M.A. 2005, {\it ApJ} {\bf 618}, L41

\item Kaspi, V.M., Roberts, M.S.E., \& Harding, A.K. 2005, in Compact Stellar X-ray Sources, ed. W.H.G. Lewin \& M. van der Klis, ( (astro-ph/0402135).

\item Katz, J. I. 1982, {\it ApJ} {\bf 260} 371.

\item Khersonskii, V.K., 1987, {\it Sov. Astron.} {\bf 31}, 225

\item Kirk, J.G. 1980, {\it Plasma Phys.}, {\bf 22}, 639

\item Kirk, J. G. 1985, {\it  A \& A} {\bf 142} 430.

\item Kirk, J. G. \& Galloway, D. J. 1981, {\it MNRAS} {\bf 195} 45.

\item Klepikov, N. P. 1954, {\it Zh. Eksp. Teor. Fiz.} {\bf 26} 19.

\item Kohri, K., \& Yamada, S. 2002, {|it Phys.\ Rev. D}, {\bf 65}, 043006

\item Konar, S., \& Bhattacharya, D. 1998, MNRAS, 303, 588

\item K\"ossl, D., Wolff, R.G., M\"uller, E., \& Hillebrandt, W. 1988,
{\it Astron. Astrophys.} {\bf 205}, 347

\item Kouveliotou, C.  et al. 1998, {\it Nature} {\bf 393} 235.

\item Kozlenkov, A. A. and Mitrofanov, I. G. 1987, {\it Sov. Phys. JETP} {\bf 64} 1173.

\item Kretschmar, P. et al. 2005, Atel \#601.

\item Kuiper,ÊL., Hermsen,ÊW. \& Mendez, 2004, {\it ApJ} {\bf 613} L1173.  

\item Kulkarni, S.R., et al. 2003, {\it ApJ}, {\bf 585}, 948

\item Lai, D. 2001, {\it Rev.\ Mod.\ Phys.}, {\bf 73}, 629

\item 
Lai, D., \& Ho, W.~C.~G. 2002, {\it ApJ}, {\bf 566}, 373

\item 
Lai, D., \& Ho, W.~C.~G. 2003a, {\it ApJ}, {\bf 588}, 962

\item 
Lai, D., \& Ho, W.~C.~G. 2003b, {\it Phys. Rev. Lett.} {\bf 91}, 071101

\item  
Lai, D., \& Salpeter, E.E. 1995, {\it Phys. Rev.} {\bf A52}, 261

\item  
Lai, D., \& Salpeter, E.E. 1996, {\it Phys. Rev.} {\bf A53}, 152

\item 
Lai, D., \& Salpeter, E.~E. 1997, {\it ApJ}, {\bf 491}, 270

\item 
Lai, D., Salpeter, E.~E., \& Shapiro, S.~L. 1992, {\it Phys. Rev. A.} {\bf 45}, 4832 

\item  
Lai, D., \& Shapiro, S.L. 1991, {\it ApJ}, {\bf 383}, 745

\item  
Lai, D., \& Wiseman, A.G. 1996, {\it Phys. Rev. D} {\bf 54}, 3958

\item  
Landau, L.D., \& Lifshitz, E.M. 1977, {\it Quantum Mechanics} (Pergamon Press:
New York)

\item  
Landau, L.D., \& Lifshitz, E.M. 1980, {\it Statistical Physics} (Pergamon Press:
New York)

\item Langer, S. H. \& Rappaport, S. 1982, {\it ApJ} {\bf 257} 733.

\item Latal, H. G. 1986, {\it ApJ} {\bf 309} 372.

\item  
Lattimer, J.M., \& Prakash, M. 2004, {\it Science}, {\bf 304}, 536

\item Lattimer, J.M., \& Prakash, M. 2005, PhRvL, 94, 1101.

\item 
Lieb, E.H., Solovej, J.P., \& Yngvason, J. 1994a, {\it Commun. Pure and Applied Math.} {\bf 47}, 513

\item 
Lieb, E.H., Solovej, J.P., \& Yngvason, J. 1994b, {\it Commun. Math. Phys.}, {\bf 161}, 77

\item Lin, J.R., \& Zhang, S.N. 2004, ApJ, 615, L133

\item Link, B. 2003, {\it Phys. Rev. Lett.}, {\bf 91}, 101101

\item Lorimer, D.R., et al. 1997, MNRAS, 289, 592

\item Lorimer, D. R.\& Xilouris, K. M., 2000, {\it ApJ} {\bf 545} 385.

\item Lovelace, R.V.E., et al. 2005, ApJ, 625, 957

\item Lyne, A.G. 2004, in Young Neutron Stars and Their Environments
(IAU Symposium No.~218), eds. F. Camilo \& B.M. Gaensler. p.257

\item 
Lyne, A. G., et al. 2004, {\it Science}, {\bf 303}, 1153

\item  
Lyutikov, M. 1998, {\it MNRAS}, {\bf 293}, 447

\item  
Lyutikov, M. 2006, {\it MNRAS}, {\bf 367}, 1594

\item  
Lyutikov, M., \& Gavriil, F. 2006, {\it MNRAS}, {\bf 368}, 690

\item Manchester, R.N. 2004, Science, 567, 542

\item McEnery, J.~E., Moskalenko, I.~V., \& Ormes, J.~F. 2004, in {\it Cosmic Gamma Ray Sources}, 
Kluwer ASSL Series, edited by K.S. Cheng and G.E. Romero (Kluwer: Dordrecht), p. 361.

\item  
McLaughlin, M.A., et al. 2003, {\it ApJ} {\bf 591}, L135

\item  
Medin, Z., \& Lai, D. 2006, {\it Phys. Rev. A}, submitted

\item Melatos, A. 1997, {\it MNRAS} {\bf 288}, 1049

\item Melatos, A., \& Phinney, E.S. 2001, {\it Pub. Astron. Soc. Aust.} 
{\bf 18}, 421

\item  
Melrose, D.B. 1974, {\it Plasma Phys.}, {\bf 16}, 845

\item Melrose, D. B. \& Kirk, J. G.  1986, A \& A.,156, 268.

\item  
Melrose, D.B, \& Stoneham, R.J. 1977, {\it PASA}, {\bf 3}, 120

\item  
Melrose, D.B., Gelalin, M.E., Kennett, M.P., \& Fletcher, C.S. 1999, {\it J. Plasma Phys.}, {\bf 62}, 233

\item  
Melrose, D.B., \& Luo, Q. 2004, {\it MNRAS}, {\bf 352}, 915

\item Melrose, D. B. \& Parle, A. J. 1983, {\it Aust. J. Phys.} {\bf 36} 775.

\item Melrose, D. B. \& Zheleznyakov, V. V. 1981, A \& A, 95, 86.

\item Mereghetti, S., et al. 2005, {\it  A \& A Lett.} {\bf 433} L9.

\item Mestel, L. 1999, {\it Stellar Magnetism} (Clarendon Press, Oxford).

\item M\'{e}sz\'{a}ros, P. 1992, {\it High-Energy Radiation From Magnetized Neutron Stars} 
(Chicago: University of Chicago Press).

\item M\'{e}sz\'{a}ros, P. \& Nagel, W. 1985, {\it ApJ} {\bf 299} 138.

\item M\'{e}sz\'{a}ros, P. \& Ventura, J. 1979, {\it PhRvD } {\bf 19} 3565.

\item  
Miller, M.~C. 1995, {\it ApJ}, 488, L29

\item  
Miller, M.~C., \& Neuhauser, D. 1991, {\it MNRAS}, 253, 107

\item Miller, G. S., Wasserman, I. \& Salpeter, E. E. 1989, {\it ApJ} {\bf 346} 405.

\item Molkov, S., et al. 2005, {\it A \& A  Lett.} {\bf 433} L13.

\item  
Mori, K., \& Hailey, C.J. 2002, {\it ApJ}, 564, 914

\item  
Mori, K., Chonko, J.C., \& Hailey, C.J. 2005, {\it ApJ}, 631, 1082

\item  
Mori, K., \& Ruderman, M. 2003, {\it ApJ}, {\bf 592}, L95

\item Morris, D. J. et al. 2002, {\it MNRAS} {\bf 335} 275.

\item  
Mueller, R.O., Rau, A.R.P., \& Spruch, L. 1971, {\it Phys. Rev. Lett.}, 
{\bf 26}, 1136

\item Nagase, F. 1989, {\it Publ. Astron. Soc. Japan} {\bf 41} 1.

\item  
Nagara, H., Nagata, Y., \& Nakamura, T. 1987, {\it Phys. Rev. A} {\bf 36}, 1859

\item Nagel, W. 1981, {\it ApJ} {\bf 251} 278 (288).

\item Narayan, R., \& Ostriker, J.P. 2000, {\it ApJ} {\bf 352}, 222

\item 
Nelson, R.W., Salpeter, E.E. \& Wasserman, I. 1993, {\it ApJ}
{\bf 418}, {874}

\item 
Neuhauser, D., Koonin, S.E., \& Langanke, K. 1987, {\it Phys. Rev. A}
{\bf 36}, 4163

\item 
Nice, D.J., et al.~2005, {\it ApJ}, submitted.

\item Orlandini, M. \& dal Fiume, D. 2001, in X-Ray Astronomy : Stellar Endpoints, AGN, and the Diffuse X-Ray Background, Ed. by N. E. White et al. (Melville, NY: AIP conference proceedings, Vol. 599), p.283.

\item 
Ortiz, G., Jones, M.D., \& Ceperley, D.M. 1995, {\it Phys. Rev. A.} {\bf 52},
R3405

\item Ostriker, J.~P., \& Gunn, J.~E. 1969, {\it ApJ} {\bf 157} 1395.

\item 
{\"O}zel, F. 2001, {\it ApJ}, {\bf 563}, 276.

\item Pacini, F. 1967, Nature, 216, 567.

\item Paczynski, B. 1992, {\it Acta Astronomica} {\bf 42} 145.

\item Pakey, D. D. 1990, PhD Thesis, University of Illinois.

\item Palmer, D. M., et al. 2005, {\it Nature} {\bf 434} 1107.

\item Parmar, A. 1994, in {\it The Evolution of X-ray Binaries} Eds. Steve Holt and Charles S. Day. (New York: AIP Press)AIP Conference Proceedings, Vol. 308, p.415.

\item  
Patel, S.K., et al. 2003, {\it ApJ}, {\bf 587}, 367

\item  
Pavlov, G. G., Shibanov, Yu. A, \& Yakovlev, D. G. 1980, {\it Ap. Space Sci.},
{\bf 73}, 33

\item 
Pavlov, G.~G., \& M\'esz\'aros, P. 1993, {\it ApJ}, {\bf 416}, 752

\item  
Pavlov, G.G., \& Bezchastnov, V.G. 2005, {\it ApJ}, {\bf 635}, L61





\item  
Pavlov, G.G., Sanwal, D., \& Teter, M.A. 2003, in ``Young Neutron Stars 
and Their Environments'' (IAU Symp.218, ASP Conf. Proc.), eds. F. Camilo
\& B.M. Gaensler

\item Pavlov, G. G. \& Yakovlev, D. G. 1976, {\it Sov. Phys. JETP} {\bf 43} 389.

\item Payne, D.J.B., \& Melatos, A. 2004, MNRAS, 351, 569

\item  
Pechenick, K.R., Ftaclas, C., \& Cohen, J.M. 1983, {\it ApJ}, {\bf 274}, 848.

\item 
Perez-Azorin, J.F., Miralles, J.A., \& Pons, J.A. 2005, {\it A \& A}, {\bf 433}, 275

\item  
Pethick, C.J., \& Ravenhall, D.G. 1995, {\it Ann. Rev. Nucl. Part. Sci.}, {\bf 45}, 429

\item 
Petrova, S.A. 2006, {\it MNRAS}, 366, 1539.

\item 
Petrova, S.A., \& Lyubarskii, Y.E. 2000, {\it ApJ}, {\bf 355}, 1168

\item 
Pons, J.A., Walter, F.M., Lattimer, J.M., et al. 2002, {\it ApJ}, 564, 981

\item 
Potekhin, A.~Y. 1994, {\it J.\ Phys. B}, {\bf 27}, 1073

\item 
Potekhin, A.~Y. 1998, {\bf J.\ Phys. B}, {\bf 31}, 49

\item  
Potekhin, A.Y. 1999, {\it A\&A}, {\bf 351}, 787


\item  
Potekhin, A.Y., \& Chabrier, G. 2003, {\it ApJ}, {\bf 585}, 955

\item  
Potekhin, A.Y., \& Chabrier, G. 2004, {\it ApJ}, {\bf 600}, 317

\item  
Potekhin, A.Y., Chabrier, G., Lai, D., et al. 2006, {\it Journal of Phys. A}, 39, 4453

\item  
Potekhin, A.Y., Chabrier, G., \& Shibanov, Y.A. 1999, {\it Phys. Rev. E} {\bf 60}, 2193
(Erratum: {\it Phys. Rev. E} {\bf 63}, 01990)

\item  
Potekhin, A.Y., Lai, D., Chabrier, G., \& Ho, W.C.G. 2004, {\it ApJ}, {\bf 612}, 1034


\item 
Potekhin, A.~Y., \& Pavlov, G.~G. 1997, {\it ApJ}, {\bf 483}, 414

\item  
Potekhin, A.Y., \& Yakovlev, D.G. 2001, {\it A\&A}, {\bf 374}, 213

\item  
Potekhin, A.Y., et al. 2003, {\it ApJ}, {\bf 594}, 404

\item  
Prakash, M., et al. 2001, in Physics of Neutron Star Interiors, 
eds. D. Blaschke, N.K. Glendenning \& A. Sedrakian (Springer) 
(astro-ph/0012136)

\item  
Pr\"oschel, P., R\"osner, W., Wunner, G., Ruder, H., \& Herold, H.
1982, {\it J. Phys. B} {\bf 15}, 1959

\item  
Radhakrishnan, V., \& Rankin, J.M. 1990, {\it ApJ}, {\bf 352}, 528

\item  
Rajagopal, M., \& Romani, R.W. 1996, {\it ApJ}, {\bf 461}, 327

\item  
Rajagopal, M., \& Romani, R.W. 1997, {\it ApJ}, {\bf 491}, 296

\item  
Rajagopal, M., Romani, R.W., \& Miller, M.C. 1997, {\it ApJ}, {\bf 479}, 347

\item Ramaty, R., Bonazzola, S., Cline, T. L., Kazanas, D., M\'{e}sz\'{a}ros, P. \& Lingenfelter, R. E. 1980, {\it Nature} {\bf 287} 122.

\item Reisenegger, A. 2001, in Magnetic Fields across the H-R Diagram (ASP
Conf. Series), eds. G. Mathys et al. (astro-ph/0103010)

\item
Reisenegger, A., Prieto, J.P., Benguria, R., Lai, D., \& Araya, P.A.
2005, in Magnetic Fields in the Universe: From
Laboratory and Stars to Primordial Structures, AIP Conf. Porc., Vol.~784, 
p.263 (astro-ph/0503047)

\item  
Relovsky, B.M., and H. Ruder, 1996, {\it Phys. Rev. A} {\bf 53}, 4068

\item
Rheinhardt, M., \& Geppert, U. 2002, Phys. Rev. Lett., 88, 101103

\item
Rheinhardt, M., et al. 2004, A\&A, 420, 631

\item
Romani, R.W. 1990, Nature, 347, 741

\item  
Romani, R.W. 1987, {\it ApJ}, {\bf 313}, 718

\item Romani, R.~W. 1996, {\it ApJ} {\bf 470} 469.

\item  
Ruder, H. et al. 1994, {\it Atoms in Strong Magnetic Fields} (Springer-Verlag)

\item  
Ruderman, M. 1974, in {\it Physics of Dense Matter} (I.A.U. Symp. No. 53),
edited by C.J. Hansen (Dordrecht-Holland: Boston), p.117

\item  
Ruderman, M. 1991a, ApJ, 366, 261

\item
Ruderman, M. 1991b, ApJ, 382, 576

\item  
Ruderman, M. 2003, in ``X-ray and Gamma-ray Astrophysics of Galactic Sources'' (ESRIN, Rome)
(astro-ph/0310777)

\item
Ruderman, M. 2004, in 
The Electromagnetic Spectrum of Neutron Stars (NATO-ASI Proceedings) 
eds. A. Baykal et al. (astro-ph/0410607)

\item Ruderman, M. A. \& Sutherland, P. G. 1975, ApJ, 195, 19.

\item 
Rutledge, R.E. et al. 2002, {\it ApJ}, {\bf 478}, 405

\item Rybicki, G. B. and Lightman, A. P. 1979, {\it Radiative Processes in Astrophysics} (Wiley: New York).

\item 
Salpeter, E.E. 1961, {\it ApJ}, {\bf 134}, 669

\item 
Sang, Y., \& Chanmugam, G. 1987, ApJ, 323, L61

\item 
Sanwal, D., et al.~2002, {\it ApJ}, {\bf 574}, L61

\item 
Schaaf, M.E. 1990, {\it A\&A}, {\bf 227}, 61

\item 
Schatz, H., Bildsten, L., Cumming, A., \& Wiescher, M. 1999, {\it ApJ}, {\bf 524}, 1014.

\item 
Schmelcher, P., Cederbaum, L.S., \& Meyer, H.-D. 1988, {\it Phys. Rev. A.} {\bf 38}, 6066 

\item 
Schmelcher, P., Cederbaum, L.S., \& Kappers, U. 1994, 
in {\it Conceptual Trends in Quantum Chemistry}, edited by E.S. Kryachko
(Kluwe Academic Publishers)

\item 
Schmelcher, P., Ivanov, M.V., \& Becken, W. 1999, {\it Phys. Rev. A} {\bf 59}, 3424

\item Schott, G. A. 1912, {\it Electromagnetic Radiation} (Cambridge Univ. Press:Cambridge)

\item
Schwinger, J. 1951, {\it Phys. Rev.}, {\bf 82}, 664

\item 
Schwinger, J., 1988, {\it Particles, Sources and Fields} (Addison-Wesley, 
Redwood City)

\item 
Schubert, C. 2000, {\it Nucl.\ Phys. B}, {\bf 585}, 407

\item Shabad, A. E. \& Usov, V. V.: 1982, {\it Nature} {\bf 295} 215.

\item Shabad, A. E. \& Usov, V. V. 1985, {\it ApSS} {\bf 117} 309.

\item Shabad, A. E. \& Usov, V. V. 1986, {\it ApSS} {\bf 128} 377.

\item 
Shapiro, S.L., \& Teukolsky, S.A. 1983, {\it Black Holes, White Dwarfs
and Neutron Stars} (Wiley, New York)

\item 
Shaviv, N., Heyl, J.S., \& Lithwick, Y. 1999, {\it MNRAS}, {\bf 306}, 333

\item 
Shibanov, Y.A., Pavlov, G.G., \& Zavlin, V.E., \& Ventura, J. 1992, {\it A\&A}, {\bf 266}, 313

\item 
Shibazaki, N., et al. 1989, Nature, 342, 656

\item 
Shibata, M., Tangiguchi, K., \& Uryu, K. 2005, {\it Phys. Rev. D} {\bf 71}, 084021

\item Shitov, Yu. P.; Pugachev, V. D.; Kutuzov, S. M. 2000, in {\it Pulsar Astronomy - 2000 and Beyond}, ASP Conference Series, Vol. 202; (San Francisco: ASP). Ed by M. Kramer, N. Wex, and N. Wielebinski, p. 685.

\item Semionova, L. \& Leahy, D. 1999, {\it PhRvD } {\bf 60} 3011.

\item Semionova \& Leahy  2000, {\it A \& AS} {\bf 144} 307.

\item Semionova, L. \& Leahy, D. 2001, {\it A \& A} {\bf 373} 272.

\item 
Slane, P.O., Helfand, D.J., \& Murray, S.S. 2002, {\it ApJ}, {\bf 571}, L45

\item 
Soffel, M., Ventura, J., Herold, H., Ruder, H., \& Nagel, W. 1983, {\it A\&A}, {\bf 126},
251

\item Sokolov, A. A. and Ternov, I. M. 1968, {\it Synchrotron Radiation} (Pergamon: New York).

\item Sokolov, A. A. \& Ternov, I. M. 1986, {\it Radiation From Relativistic Electrons} (AIP: New York)

\item 
Stairs, I.H. 2004, {\it Science}, {\bf 304}, 547

\item Stoneham, R. J. 1979, {\it J. Phys. A.} {\bf 12} 2187.

\item Sturrock, P. A. 1971, {\it ApJ} {\bf 164} 529.

\item Srinivasan, G., et al. 1990, Curr. Sci., 59, 31

\item Svensson, R. 1982, {\it ApJ} {\bf 258} 335.

\item Svetozarova, G.I. and Tsytovich, V.N. 1962, Radiofizika, 5, 658.

\item Thompson, C. \& Beloborodov, A.~M. 2005, {\it ApJ} {\bf 634} 565.

\item Thompson, C. \& Duncan, R. C. 1993, {\it ApJ} {\bf 408} 194.

\item Thompson, C. \& Duncan, R. C. 1995, {\it MNRAS} {\bf 275} 255.

\item Thompson, C. \& Duncan, R. C. 1996, {\it ApJ} {\bf 473} 332.

\item Thompson, C. \& Duncan, R. C. 2001, {\it ApJ} {\bf 561} 980.

\item Thompson, C., Lyutikov, M. \& Kulkarni, S.R. 2002, {\it ApJ} 
{\bf 574}, 332

\item  
Thorolfsson, A., et al.
1998, {\it ApJ}, {\bf 502}, 847

\item  
Tiengo, A., Goehler, E., Staubert, R., \& Mereghetti, S. 2002, {\it A\&A}, {\bf 383}, 182

\item Toll, J. S. 1952, PhD Thesis, Princeton University.

\item
Tsai, W.Y. \& Erber, T. 1975, 
Phys. Rev. D, 12, 1132

\item  
Turolla, R., Zane, S., \& Drake, J.J. 2004, {\it ApJ}, 603, 265

\item  
Urpin, V.A., \& Geppert, U. 1995, MNRAS, 275, 1117

\item  
Urpin, V.A., \& Shalybkov, D. 1999, MNRAS, 304, 451

\item  
Urpin, V.A., \& Yakovlev, D.G. 1980, Sov. Astron., 24, 425

\item Usov, V. V. 2002, {\it ApJ} {\bf 572} L87.

\item  
Usov, N.A., Grebenshchikov, Yu.B., \& Ulinich, F.R. 1980, {\it Sov.  Phys. JETP}, {\bf 51}, 148

\item Usov, V. V. 1992, {\it Nature} {\bf 357} 472.

\item Usov, V. V. \& Melrose, D. B. 1995, {\it AuJPh} {\bf 48} 571.

\item Usov, V. V. \& Melrose, D. B. 1996, {\it ApJ} {\bf 464} 306.

\item Vainstein, S.I., Chitre, S.M., \& Olinto, A.V. 2000, Phys. Rev. E 61, 4422

\item  
van Adelsberg, M., \& Lai, D. 2006, {\it MNRAS}, submitted

\item  
van Adelsberg, M., Lai, D., Potekhin, A.Y., \& Arras, P. 2005, {\it ApJ}, {\bf 628}, 902

\item  
van Kerkwijk, M.H., \& Kulkarni, S.R. 2001, {\it A\&A}, {\bf 378}, 986

\item  
van Kerkwijk, M.H., et al. 2004, {\it ApJ}, {\bf 608}, 432

\item  
Van Riper, K.A., 1988, {\it ApJ}, {\bf 329}, {\bf 339}

\item Vasisht, G. \& Gotthelf, E. V. 1997, ApJ, 486, L129.

\item Ventura, J. 1979, {\it PhRvD } {\bf 19} 1684.

\item  
Vignale, G., \& Rasolt, M. 1987, {\it Phys. Rev. Lett.} {\bf 59}, 2360

\item  
Vignale, G., \& Rasolt, M. 1988, {\it Phys. Rev. B} {\bf 37}, 10685


\item  
Virtamo, J., 1976, {\it J. Phys. B} {\bf 9}, 751


\item  
Wang, C., \& Lai, D. 2006, {\it ApJ}, submitted

\item  
Wang, F.Y.-H., Ruderman, M., Halpern, J.P., \& Zhu, T. 1998,
{\it ApJ}, {\bf 498}, 373

\item Wang, J.C.L., Wasserman, I. and Salpeter, E. E. 1988, {\it ApJ Supp} {\bf 68} 735.

\item Wang, J. C. L., Lamb, D. Q., Loredo, T. J., Wasserman, I. M., Salpeter, E. E. 1989, {\it PhRvL} {\bf 63} 1550.

\item Wasserman, I. \& Salpeter, E. E.  1980, {\it ApJ} {\bf 241} 1107.

\item  
Weber, F. 2005, {\it Prog. Part. Nucl. Phys.}, {\bf 54}, 193

\item White, D. 1974, {\it PhRvD } {\bf 9} 868.

\item
Wiebicke, H.-J., \& Geppert, U. 1996, A\&A,309, 203

\item Woods, P. M. \& Thompson, C. 2005, in {\it Compact Stellar X-ray Sources} 
eds. W.H.G. Lewin and M. van der Klis, in press [astro-ph/0406133].

\item Wunner, G. 1979, {\it PhRvL} {\bf 42} 79.

\item Wunner, G., Ruder, H. \& Herold, H. 1981, {\it JPhB} {\bf 14} 765.

\item Wunner, G., Paez, J., Herold, H. \& Ruder, H. 1986, {\it A \& A} {\bf 170} 179.

\item  
Yakovlev, D.G., \& Kaminker, A.D. 1994, in The Equation of State in
Astrophysics, edited by G. Chabrier and E. Schatzman 
(Cambridge Univ. Press: Cambridge), p.214

\item  
Yakovlev, D.G., \& Pethick, C.J. 2004, {\it ARAA}, {\bf 42}, 169

\item 
Zane, S., et al. 2005, {\it ApJ}, {\bf 627}, 397

\item 
Zane, S., \& Turolla, R. 2006, {\it MNRAS}, 366, 727

\item 
Zane, S., Turolla, R., \& Treves, A. 2000, {\it ApJ}, {\bf 537}, 387

\item{}
Zane, S., Turolla, R., Stella, L., \& Treves, A. 2001, {\it ApJ}, {\bf 560}, 384

\item 
Zavlin, V.E., Pavlov, G.G., \& Shibanov, Y.A. 1996, {\it A\&A}, {\bf 315}, 141

\item 
Zavlin, V.E. \& Pavlov, G.G. 2002, in Proc. 270 WE-Heraeus Seminar on {\it Neutron
Stars, Pulsars, and Supernova Remnants}, eds. W. Becker et al.(Bad Honnef, Germany)
(astro-ph/0206025)

\item Zhang, J. L. \& Cheng, K. S. 1996, {\it Ch A \& A} {\bf 20} 239.

\item Zhang, J. L. \& Cheng, K. S. 1997, {\it ApJ} {\bf 487} 370.

\item Zhang, B. \& Harding, A. K. 2000, {\it ApJ} {\bf 535} L51.

\item Zhang, B. \& Qiao, G. J. 1998, {\it A \& A} {\bf 338} 62.

\item Zhang, L., Cheng, K. S., Jiang, Z. J. \& Leung, P. 2004, {\it ApJ} {\bf 604} 317.

\item Zhang, B. \& Qiao, G. J. 1996, {\it A \& A} {\bf 310} 135.

\item 
Zheleznyakov, V.V., Kocharovski\u{i}, V.V., \& Kocharovski\u{i}, Vl.V. 1983, {\it Sov. Phys. Usp.}, {\bf 26}, 877

\end{harvard}
           
\end{document}